\documentclass[prd,
a4paper,
nofootinbib,
onecolumn
]{revtex4}
\pdfoutput=1 

 \usepackage{braket}
\usepackage{blkarray}
\usepackage{graphicx}
\usepackage{amsfonts}
\usepackage{soul}
\usepackage{amssymb}
\usepackage{amsmath}
\usepackage{capt-of} 
\usepackage{pifont}
\usepackage{slashed} 
\usepackage{enumerate}
\usepackage{color}
\usepackage{comment}
\usepackage{bbold}
\usepackage{tikz}
\usepackage{amsmath}
\usepackage{tikz}
\usepackage{empheq}
\usepackage{adjustbox}
\usepackage{multirow}
\usepackage{hyperref}
\usepackage[utf8]{inputenc}
\usepackage{adjustbox}
\usepackage{erewhon}
\usepackage{tikz-feynman}
\usepackage{tcolorbox}

\definecolor{oucrimsonred}{rgb}{0.6, 0.0, 0.0}
\definecolor{persianblue}{rgb}{0.11, 0.22, 0.73}
\definecolor{forestgreen}{rgb}{0.13,0.35,0.13}
\definecolor{lightgray}{rgb}{0.83, 0.83, 0.83}
 \hypersetup{colorlinks, citecolor=oucrimsonred, linkcolor=persianblue, urlcolor=oucrimsonred}
\def\hhref#1{\href{http://arxiv.org/abs/#1}{#1}} 
\definecolor{cornellred}{rgb}{0.7, 0.11, 0.11}
\definecolor{navyblue}{rgb}{0.0, 0.0, 0.5}
\definecolor{amethyst}{rgb}{0.6, 0.4, 0.8}
\definecolor{yellow}{rgb}{1.0, 1.0, 0.0}
\definecolor{firebrick}{rgb}{0.7, 0.13, 0.13}
\definecolor{tangerineyellow}{rgb}{1.0, 0.8, 0.0}
\definecolor{deepfuchsia}{rgb}{0.76, 0.33, 0.76}
\definecolor{amber}{rgb}{1.0, 0.75, 0.0}
\definecolor{VioletRed4}{rgb}{0.55, 0.13, .32}
\definecolor{indiagreen}{rgb}{0.07, 0.53, 0.03}
\definecolor{aucolor}{rgb}{0.13, 0.55, 0.13}

\newcommand{\be}{\begin{equation}}
\newcommand{\de}{\partial}
\newcommand{\ee}{\end{equation}}
\newcommand{\bea}{\begin{eqnarray}}
\newcommand{\eea}{\end{eqnarray}}
\newcommand{\nn}{\nonumber}

\definecolor{oucrimsonred}{rgb}{0.6, 0.0, 0.0}

\definecolor{violachiaro}{rgb}{1,0.6,1}
\definecolor{gbcolor}{rgb}{.43,.22,.12}
 
\definecolor{gbcolor2}{rgb}{.9,.2,.6}
\definecolor{gbcolor3}{rgb}{.3,.2,.6}

\definecolor{verdechiaro}{rgb}{0.6,1,0.6}
\definecolor{giallochiaro}{rgb}{1,1,0.6}
\definecolor{bluscuro}{rgb}{0.15, 0.2, 0.9}
\definecolor{verdes}{rgb}{0.1, 0.5, 0.1}%
\definecolor{tangerineyellow}{rgb}{1.0, 0.8, 0.0}

  \hypersetup{colorlinks, citecolor=persianblue, linkcolor=firebrick, urlcolor=verdes}

\begin{document}

\title[]{Power suppressed corrections show new features of infrared cancellations} 

\date{\today}
\author{Paolo Ciafaloni$^{a}$}
\email{paolo.ciafaloni@le.infn.it}
\author{Denis  Comelli$^{b}$}
\email{comelli@fe.infn.it}
\author{Alfredo Urbano$^{c,d}$}
\email{alfredo.urbano@uniroma1.it}
\affiliation{$^a$INFN Sezione di Lecce e Universita` del Salento - Lecce, Italy}
\affiliation{$^b$INFN Sezione di Ferrara, I-44122 Ferrara, Italy}
\affiliation{$^c$Dipartimento di Fisica, ``Sapienza'' Universit\`a di Roma, Piazzale Aldo Moro 5, 00185, Roma, Italy}
\affiliation{$^d$I.F.P.U., Institute  for  Fundamental Physics  of  the  Universe, via  Beirut  2, I-34014 Trieste, Italy.}

\begin{abstract}
\noindent 
The cancellation of infrared (IR) divergences is an old topic in quantum field theory whose main results are condensed into the 
celebrated Kinoshita-Lee-Nauenberg (KLN) theorem. 
In this paper we consider mass-suppressed corrections to the leading (i.e. double-logarithmic) IR divergences in the context of 
spontaneously broken gauge theories. 
We work in a simplified theoretical set-up 
based on the spontaneously broken $U'(1)\otimes U(1)$ gauge group.
We analyze, at the one-loop level and including mass-suppressed terms, 
the double-logarithmic corrections to the decay channels of an hypothetical  heavy $Z'$ gauge boson coupled to light chiral fermions and mixed with a light massive $Z$ gauge boson.  {Limited to this theoretical framework, only final state IR corrections are relevant.}
We find that full exploitation of the KLN theorem requires non-trivial combinations of various decay channels in order to get rid of the mass-suppressed IR corrections. 
Based on this observation we show that, starting from any two-body decay of the heavy $Z'$ gauge boson,
 the cancellation of the mass-suppressed double-logarithmic corrections requires the sum 
 over the {\it full} decay width 
 (thus enforcing the inclusion of final states which are na\"{\i}vely unrelated to the starting one). 
En route, we prove a number of technical results that are relevant for the computation of mass-suppressed double-logarithms of IR origin.
Our results are relevant for models that enlarge the Standard Model by adding a  heavy $Z'$.
\end{abstract}
\maketitle
\tikzset{->-/.style={decoration={
  markings,
  mark=at position #1 with {\arrow{>}}},postaction={decorate}}}
\tikzset{-<-/.style={decoration={
  markings,
  mark=at position #1 with {\arrow{<}}},postaction={decorate}}} 

\section{Introduction}\label{sec:Intro}

A proper treatment of radiative corrections of infrared (IR) origin is of paramount importance in order to calculate theoretical predictions at the level of accuracy demanded by  experiments. In first place, the very presence of divergences requires to define suitable observables - obtained by adding virtual corrections and real emissions diagrams - that are finite. In second place, 
finite observables feature, as a remnant of IR divergences, large logarithms that threaten to spoil the convergence of the perturbative series. It is then necessary to calculate higher orders contributions and/or perform an all-order resummation
of  leading terms 
(see \cite{Agarwal:2021ais} for a collection of the main references).
In the Standard Model (SM), in the case of QED and QCD, one loop leading IR radiative correction are proportional to 
$\alpha \log^2(Q/\lambda)$, $Q$ being a typical energy scale of the observable under consideration and $\lambda$ an IR cutoff (for instance a fictitious photon mass). After defining suitable inclusive observables, a cancellation takes place between virtual and real corrections and the divergence for $\lambda\to 0$ is cancelled \cite{Bloch:1937pw}. In the case of weak interactions, the mass of the weak gauge bosons $M_Z\approx M_W$ acts as an IR cutoff and one loop corrections of IR origin are proportional 
to $\log^2(Q/M_W)$ \cite{cc1}; therefore there is no need to add real emission in order to obtain a finite result. Nevertheless, electroweak corrections feature, in the case of inclusive observables, a striking phenomenon: real emissions and virtual corrections do not cancel each other and the  the full dependence on the infrared cutoff $M_W$ is retained; this effect has been called
 Block-Nordsieck violation \cite{BNV}.

 In this paper we focus on power-suppressed contributions, that at one loop are
proportional to $(\mu^2/Q^2)\log^2(Q/\lambda)$. If $\mu$ coincides with $\lambda$, there are no true divergences since 
$\lambda^2\log^2 \lambda\to 0$ as $\lambda\to 0$. The divergence, and the need to add real emissions, is instead retained if $\mu$ is some fixed mass like a fermion mass for instance. This kind of contributions has been considered in the context of QED and QCD \cite{MS}; a more general approach with focus on the cancellation between real and virtual corrections can be found in  \cite{Ak}. For the virtual effects in a spontaneously broken theory see \cite{Ciafaloni:2009tf}.

The purpose of this paper is to elucidate the mechanism of KLN-like cancellations of power-suppressed radiative corrections of IR origin; in order to do so we choose   the simplest possible Model: 
a $U'(1)\otimes U(1)$ gauge model, with spontaneous symmetry breaking. 
The spectrum of the Model also includes fermions and scalars as discussed in detail in appendix\,\ref{app:Coupl}.
Our observable is the decay width of an heavy $Z'$ gauge boson, the final state being composed  of various light particles.
Our computations concern  one loop results and retain the full mass  dependent  structure for the coefficients that multiply the double log corrections generically of the form $f(m_i/Q)\log^2Q^2$, where  $m_i$ are the masses of the various light particles  and $Q$ is the heavy $Z'$ mass. Our results are relevant for models that enlarge the Standard Model by adding a  heavy U(1) Z'. As a byproduct of our computations we discover some interesting technical points that concern the mechanism of cancellation of power suppressed terms and a general recipe for their computations (discussed in appendix\,\ref{p+k}) and 
the analytical properties of the Passarino-Veltman  $\mathcal{C}_0$ function\,\cite{Passarino:1978jh} (discussed in appendix\,\ref{app:Vir}).
The paper is organized  as follows.
\\
In section\,\ref{sec:Notation} we collect, for ease of reading, the key relations which play crucial 
importance throughout our work.
\\
 In section\,\ref{sec:WarmUp}, we introduce the power suppressed leading log  corrections analysing the SM   electroweak  decay process $Z\to e^+e^-$ plus $Z\to e^+e^-\,\gamma$. We get the expressions both for virtual and real photon emission and we elucidate the way the cancellation of the IR divergence works including also the power suppressed terms.
 \\
In section \ref{sponte}, we use our toy model based on $U'(1)\otimes U(1)$ to study the interplay between various decay channels needed to ensure the cancellation of the power suppressed leading log corrections. 
The chiral character of the abelian gauge groups results important to emphasize the full generality of the mechanism of the various cancellations.
\\
In section\,\ref{kln} we return to the original Kinoshita-Lee-Nauenberg (KLN) cancellation theorem and revisit 
from a deeper perspective its assumption in light of the results obtained in the previous section.
\\
Finally, we summarize our findings in section\,\ref{sec:Conclusions}.

\section{Notations}\label{sec:Notation}
In this paper we calculate the decay rate of a heavy particle, that decays at tree level into two light particles. We denote by $Q$ the heavy particle mass and by $m_i$ the light particles masses. The tree level width $\Gamma_{\rm tree}$ receives a one loop correction that we call $\Gamma^V$ in which we identify the various terms on the basis of the scaling behaviour with powers of $\log Q$:
\be
\Gamma=\Gamma_{\rm tree}+\Gamma^V;\qquad
\Gamma^V=\Gamma_{LL}^V\,\log^2 Q^2+\Gamma_{NL}^V\, \log Q^2+\delta\Gamma^V
\ee 
We are interested only in the $\log^2$ term $\Gamma_{LL}^V$. Notice that the argument of the double log will also depend on the light masses but we are interested in  the high energy limit, i.e.  the scaling behavior with $Q$. 
Therefore terms like $\log^2(Q/m_1)$
and $\log(Q/m_1)\,\log(Q/m_2)$ produce the same $\log^2 Q$ behavior while differing by single $\log Q$ terms that we neglect. 
Among the double-log terms there are leading terms that truly scale with $Q$, like $\log^2 Q$, 
but also power-suppressed terms that are proportional to some power of the light masses, like  for instance   
$(m_i^2/Q^2)\log^2 Q$. 
We label the first kind of terms terms ``$LL$'' and the second kind ``$LL_{PS}$'' where $PS$ stands for Power Suppressed. These  terms have attracted a lot of interest in the literature in various contexts, see refs.\,\cite{MS,Ak,Ciafaloni:2009tf}. The power-suppressed double-log terms are the main topic of the present work.

In the calculation of the one loop width appears the relevant two-body phase space $\Phi_2$ and the Passarino-Veltman scalar integral
$\mathcal{C}_0$\,\cite{Passarino:1978jh} (see our definitions in eq.\,(\ref{eq:totalDR}) and eq.\,(\ref{eq:ScalarC0})). We  define:
\be\label{gvrw}
\Gamma_{LL}^V=Q\,V(\epsilon_i) \, {\cal I}^V\,,\qquad {\cal I}^V=-i \, Q^3 \,  \Phi_2 \,  \mathcal{C}_0\,.
\ee
The a-dimensional combination ${\cal I}^V$   behaves as $\propto \log^2 Q$ with no power-suppressed terms, 
see eq.\,(\ref{eq:ExpressionIV}) and eq.\,(\ref{eq:AsiIV}). All the dependence on the mass terms $\epsilon_i\equiv m_i/Q$ is therefore encoded  in the a-dimensional function $V(\epsilon_i)$. It turns out that the latter is a polynomial of degree 6 in the variables $\epsilon_i$ (see appendix\,\ref{app:2b} where our results are shown).

Finally, for real corrections we adopt the same notations as for virtual corrections, making obvious modifications. 
For instance, eq.\,(\ref{gvrw}) still holds in the form
\be\label{oiuh}
\Gamma_{LL}^R=Q\,R(\epsilon_i) \, {\cal I}^R\,,\qquad {\cal I}^R=Q^2\int \frac{d\Phi_3}{D_1D_2}\,.
\ee
where  the integral over the three body of the two relevant propagator denominators, dubbed 
$D_{1,2}$ in the above equation, is involved.  
We discuss in more detail the structure of the above integral in appendix\,\ref{app:Re} 
(see in particular eq.\,(\ref{eq:Main3Body}) and eq.\,(\ref{eq:ExpressionIR}) for a more explicit form of ${\cal I}^R$).
A quite interesting property is that   ${\cal I}^R$ and 
${\cal I}^V$ have the same asymptotic high-energy behavior, that is
\be\label{final}
{\cal I}^V={\cal I}^R=\frac{1}{512\,\pi^3} \log^2 Q^2+{\cal O}(\log Q)\,.
\ee
{
In our analysis is important to verify the validity of eq.\,(\ref{final}) when power-suppressed double logarithms are taken into 
account.
As far as virtual corrections are concerned, we demonstrate analytically in appendix\,\ref{app:Vir} 
that ${\cal I}^V$ in eq.\,(\ref{final}) is actually free from any power-suppressed double logarithm. 
As far as real corrections are concerned, and considering
the most generic mass configuration,
we were not able to find solid-rock analytical proof that the same statement holds true for $\mathcal{I}^R$. 
However, in appendix\,\ref{app:Vir} we verify that the QED limits (both with massless and massive fermions) enjoy this property. 
For these reasons, we do not consider too rash the conclusion that eq.\,(\ref{final}) remains valid also when 
power corrections are taken into account, and we shall stick to this assumptions throughout the rest of our work.
}

From now on we drop the subscript $LL$ from $\Gamma^{V,R}_{LL}$; by 
$\Gamma^{V,R}$ we therefore indicate only the double log component of the (virtual or real) one loop correction to the width.

Let us now illustrate another very general property, that we call ``{\it IR unitarity theorem}'' and is demonstrated in 
appendix\,\ref{p+k}.  Consider the   decay of an initial particle  into final particles, {\it in a generic theory}. The relevant diagrams are depicted in fig.\,\ref{fig:Bubble} and symbolize virtual corrections (when the cut encounters two particle lines) and real emissions (when the cut encounters three particle lines). Notice that diagrams that involve one loop self-energies are not drawn, since they do not give rise to double logs.

The {\it IR unitarity theorem} can be written as:
\be\label{pkth}
V^{(X)}_{CD}+V^{(X)}_{AB}+R^{(X)}_{AD}+R^{(X)}_{BC}= 0
\ee
Note that the $X$ state, present in the ``$t$-channel'' of the bubble diagram, corresponds to the soft (infrared and collinear) particle emitted from (for real corrections) or exchanged between (for virtual corrections),  the two hard states, described with the lower indices of the virtual $V$ and real $R$ processes.
The above results are valid for any theory and it concerns only decay processes. 
Notice, however, that the relationship with the observables is not a priori clear. 
For instance $A,B$ may be different particles from $C,D$. 
Consequently, 
$V^{(X)}_{CD}$ is related to the $Z'\to CD$ decay while
  $V^{(X)}_{AB}$ is related to the different $Z'\to AB$ decay. 
  In next section we show how the IR unitarity theorem is useful in order to define IR safe observables.

\vspace{0.5cm}	
\begin{adjustbox}{max width=0.94\textwidth}\hspace{.5cm}    
    \raisebox{-22.5mm}{
	\begin{tikzpicture}
	\draw[thick] (-0.5,0)--(0,0);
	\draw[thick] (0,0)--(1.25/2,1.25/2); 
	\draw[thick] (1.25/2,1.25/2)--(1.25,1.25); 
	\draw[thick] (1.25,-1.25)--(1.25/2,-1.25/2);
	\draw[thick] (1.25/2,-1.25/2)--(0,0);
	\draw[thick] (1.25,1.25)--(1.25,-1.25);
	\draw[thick] (1.25,1.25)--(2.5,0); 
	\draw[thick] (2.5,0)--(1.25,-1.25);
	\draw[thick,dash dot,persianblue][thick] (1.25/2,1.25)--(1.25/2,-1.25);  	
	\draw[thick] (2.5,0)--(2.5+0.5,0);
	\draw[black,fill=tangerineyellow,thick] (0,0)circle(2.75pt); 
	\draw[black,fill=tangerineyellow,thick] (2.5,0)circle(2.75pt); 
	\draw[black,fill=tangerineyellow,thick] (1.25,1.25)circle(2.75pt); 
	\draw[black,fill=tangerineyellow,thick] (1.25,-1.25)circle(2.75pt);
	\node at (1.5,0.2) {\scalebox{1}{$X$}};
	\node at (0.2,0.6) {\scalebox{1}{$A$}};
	\node at (0.2,-0.6) {\scalebox{1}{$B$}};
	\node at (2.3,0.6) {\scalebox{1}{$C$}};
	\node at (2.3,-0.6) {\scalebox{1}{$D$}};
	\node at (1.25,-2) {\scalebox{1}{$V^{(X)}_{AB}$}};	
	\end{tikzpicture}}	~~~~~~~
    \raisebox{-22.5mm}{
	\begin{tikzpicture}
	\draw[thick] (-0.5,0)--(0,0);
	\draw[thick] (0,0)--(1.25,1.25); 
	\draw[thick] (1.25,-1.25)--(0,0);
	\draw[thick] (1.25,1.25)--(1.25,-1.25);
	\draw[thick] (1.25,1.25)--(1.25+1.25/2,1.25-1.25/2);
    \draw[thick] (1.25+1.25/2,1.25-1.25/2)--(2.5,0); 
	\draw[thick] (2.5,0)--(1.25+1.25/2,-1.25/2);
	\draw[thick] (1.25+1.25/2,-1.25/2)--(1.25,-1.25);  	
	\draw[thick,dash dot,persianblue][thick] (1.25+1.25/2,1.25)--(1.25+1.25/2,-1.25); 
	\draw[thick] (2.5,0)--(2.5+0.5,0);
	\draw[black,fill=tangerineyellow,thick] (0,0)circle(2.75pt); 
	\draw[black,fill=tangerineyellow,thick] (2.5,0)circle(2.75pt); 
	\draw[black,fill=tangerineyellow,thick] (1.25,1.25)circle(2.75pt); 
	\draw[black,fill=tangerineyellow,thick] (1.25,-1.25)circle(2.75pt); 	
	\node at (1,0.2) {\scalebox{1}{$X$}};
	\node at (0.2,0.6) {\scalebox{1}{$A$}};
	\node at (0.2,-0.6) {\scalebox{1}{$B$}};
	\node at (2.3,0.6) {\scalebox{1}{$C$}};
	\node at (2.3,-0.6) {\scalebox{1}{$D$}};
	\node at (1.25,-2) {\scalebox{1}{$V^{(X)}_{CD}$}};			
	\end{tikzpicture}}	~~~~~~~
    \raisebox{-22.5mm}{
	\begin{tikzpicture}
	\draw[thick] (-0.5,0)--(0,0);
	\draw[thick] (0,0)--(1.25/2,1.25/2); 
	\draw[thick] (1.25/2,1.25/2)--(1.25,1.25); 
	\draw[thick] (1.25,-1.25)--(0,0);
		\draw[thick] (1.25,1.25)--(1.25,-1.25);
	\draw[thick] (1.25,1.25)--(2.5,0); 
	\draw[thick] (2.5,0)--(1.25+1.25/2,-1.25/2);
	\draw[thick] (1.25+1.25/2,-1.25/2)--(1.25,-1.25); 	
	\draw[thick] (2.5,0)--(2.5+0.5,0);
	\draw[thick,dash dot,persianblue][thick] (0,1.25)--(2.5,-1.25); 	
	\draw[black,fill=tangerineyellow,thick] (0,0)circle(2.75pt); 
	\draw[black,fill=tangerineyellow,thick] (2.5,0)circle(2.75pt); 
	\draw[black,fill=tangerineyellow,thick] (1.25,1.25)circle(2.75pt); 
	\draw[black,fill=tangerineyellow,thick] (1.25,-1.25)circle(2.75pt); 
	\node at (1.25,-2) {\scalebox{1}{$R^{(X)}_{AD}$}};		
	\node at (1.5,0.2) {\scalebox{1}{$X$}};	
	\node at (0.2,0.6) {\scalebox{1}{$A$}};
	\node at (0.2,-0.6) {\scalebox{1}{$B$}};
	\node at (2.3,0.6) {\scalebox{1}{$C$}};
	\node at (2.3,-0.6) {\scalebox{1}{$D$}};		
	\end{tikzpicture}}	~~~~~~~
    \raisebox{-22.5mm}{
	\begin{tikzpicture}
	\draw[thick] (-0.5,0)--(0,0);
	\draw[thick] (0,0)--(1.25,1.25); 
	\draw[thick] (1.25,-1.25)--(1.25/2,-1.25/2);
	\draw[thick] (1.25/2,-1.25/2)--(0,0);
		\draw[thick] (1.25,1.25)--(1.25,-1.25);
	\draw[thick] (1.25,1.25)--(1.25+1.25/2,1.25-1.25/2);
    \draw[thick] (1.25+1.25/2,1.25-1.25/2)--(2.5,0);  
	\draw[thick] (2.5,0)--(1.25,-1.25); 	
	\draw[thick] (2.5,0)--(2.5+0.5,0);
    \draw[thick,dash dot,persianblue][thick] (2.5,1.25)--(0,-1.25); 
	\draw[black,fill=tangerineyellow,thick] (0,0)circle(2.75pt); 
	\draw[black,fill=tangerineyellow,thick] (2.5,0)circle(2.75pt); 
	\draw[black,fill=tangerineyellow,thick] (1.25,1.25)circle(2.75pt); 
	\draw[black,fill=tangerineyellow,thick] (1.25,-1.25)circle(2.75pt); 	
	\node at (1,0.2) {\scalebox{1}{$X$}};
	\node at (0.2,0.6) {\scalebox{1}{$A$}};
	\node at (0.2,-0.6) {\scalebox{1}{$B$}};
	\node at (2.3,0.6) {\scalebox{1}{$C$}};
	\node at (2.3,-0.6) {\scalebox{1}{$D$}};
	\node at (1.25,-2) {\scalebox{1}{$R^{(X)}_{BC}$}};			
	\end{tikzpicture}}  
	\end{adjustbox}
	\label{fig:Bubble}
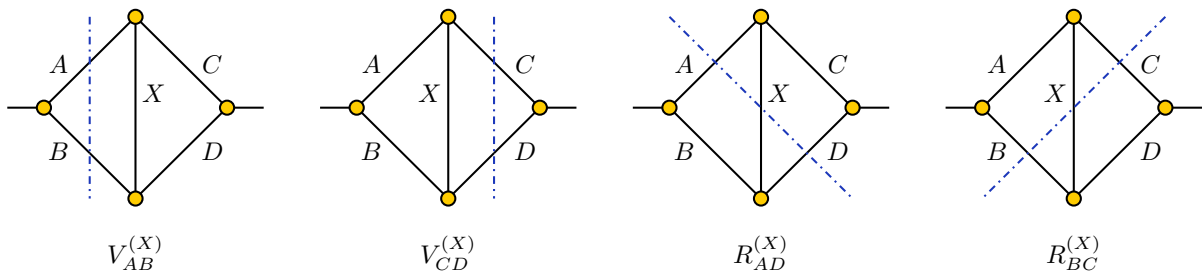
\captionof{figure}{{\em 
Diagrams contributing to 
one-loop two-body decay (the first two diagrams with the cut that encounters two particle lines) and three-body decay	
(the last two diagrams with the cut that encounters three particle lines).
	}}\vspace{0.5cm}

 
\section{Warm-up discussion: power-suppressed infrared divergences in abelian gauge theories}\label{sec:WarmUp}
To warm up we consider, in the SM, the electroweak  decay process $Z\to e^+e^-$, and compute QED radiative corrections; the photon is given a fictitious mass $\lambda$ to regularize IR divergencies, while $m_\psi$ is the electron mass. 
At the tree level, the above decay process is described by the total decay rate (averaged over the three polarizations of the decaying $Z$ boson, and taking the sum over the final-state spins) 
\begin{align}\label{eq:TreeLevelDecayZ}
\Gamma_{Z,\,{\rm tree}} = \frac{g^2\,Q}{96 \, \pi \,  c_W^2} 
\left[(1-4\,s_W^2 + 8\, s_W^4)
-\epsilon_{\psi}^2(1+ 8\,s_W^2 - 16\,s_W^4)
\right]
\sqrt{1-4\,\epsilon_{\psi}^2}
~~\overset{\epsilon_{\psi}\to 0}{~=~}~~
\frac{g^2\,Q}{96\,\pi\, c_W^2}\,(1-4\,s_W^2 + 8\, s_W^4)\,,
\end{align}
 where the hard scale $Q$ is set by the $Z$ boson mass, $Q = m_Z$, $g$ is the SU(2) weak coupling constant while for the Weinberg angle $\theta_W$ we define the shorthand notations $c_W \equiv \cos\theta_W$
 and $s_W \equiv \sin\theta_W$
and we define $\epsilon_{\psi} \equiv m_{\psi}/Q$.  
We now compute, at one-loop, infrared QED radiative corrections.  
This exercise will serve as a warm-up for the more complicated computation that we shall do in the context of 
a spontaneously broken abelian gauge theory. 
We start from one-loop virtual corrections to the decay $Z(q)\to e^-(p_a)e^+(p_b)$ and  we work in the Z rest frame, where $q=(Q,0,0,0)$.
At the level of the modulus squared of the amplitude, 
the relevant diagram (together with its complex conjugate) is given by
\begin{align}
   |M_{vir}|^2=& \int\frac{d^4k}{(2\pi)^4}
    \raisebox{-12mm}{
	\begin{tikzpicture}
	\draw[thick,style={decorate, decoration={snake}}] (-0.75,0)--(0,0);
	\draw[->-=.7,>=Latex][thick] (0,0)--(1.25/2,1.25/2); 
	\draw[->-=.7,>=Latex][thick] (1.25/2,1.25/2)--(1.25,1.25); 
	\draw[->-=.7,>=Latex][thick] (1.25/2,-1.25/2)--(0,0); 
	\draw[->-=.7,>=Latex][thick] (1.25,-1.25)--(1.25/2,-1.25/2); 
	\draw[->-=.6,>=Latex][thick] (0,0)--(1.25,+1.25);
	\draw[->,>=Latex][thick] (0.85,0.25)--(0.85,-0.25);
	\draw[thick,style={decorate, decoration={snake}}] (1.25/2,1.25/2)--(1.25/2,-1.25/2);
	\draw[black,fill=tangerineyellow,thick] (0,0)circle(2.75pt);
	\draw[black,fill=tangerineyellow,thick] (1.25/2,1.25/2)circle(2.75pt); 
	\draw[black,fill=tangerineyellow,thick] (1.25/2,-1.25/2)circle(2.75pt); 
	\node at (1.75,1.25) {\scalebox{0.85}{$e^-(p_a)$}};
	\node at (1.75,-1.) {\scalebox{0.85}{$e^+(p_b)$}};
	\node at (-0.6, 0.3) {\scalebox{0.85}{$Z(q)$}};
	\node at (1.1,0) {\scalebox{0.85}{$k$}};
	\end{tikzpicture}}~\times~
	\underbrace{
	\raisebox{-12mm}{
	\begin{tikzpicture}
	\draw[->-=.6,>=Latex][thick] (1.25,1.25)--(2.5,0); 
	\draw[->-=.6,>=Latex][thick] (2.5,0)--(1.25,-1.25); 	
	\draw[thick,style={decorate, decoration={snake}}] (2.5,0)--(2.5+0.75,0);
	\draw[black,fill=tangerineyellow,thick] (2.5,0)circle(2.75pt);
	\end{tikzpicture}}}_{{\rm tree\,level\,ampl}^*}	~~~ = \label{eq:VirtualStructure} \\ 	
&\frac{g^2}{4 \, c_W^2}\int\frac{d^4k}{(2\pi)^4}\frac{ie^2}{[(p_a + k)^2-m_{\psi}^2]
[(-p_b + k)^2-m_{\psi}^2](k^2-\lambda^2)}
\bigg[\frac{1}{3}\sum_{{\rm pol\,}Z}\epsilon_{\rho}(q)\epsilon_{\sigma}^*(q)\bigg]
T^{\rho\sigma}\,.\nn
\end{align}
\begin{equation}
T^{\rho\sigma}=g_{\mu\nu}{\rm tr}[(\slashed{p}_a+m_{\psi})\gamma^{\mu}(\slashed{p}_a+\slashed{k}+m_{\psi})\gamma^{\rho}A_{LR}
(\slashed{p}_b - \slashed{k}-m_{\psi})\gamma^{\nu}(\slashed{p}_b-m_{\psi}) A_{RL}\gamma^{\sigma}
]
\end{equation}
where we use the shorthand notation $A_{LR}\equiv (1-2 \, s_W^2) \, P_L -2 \, s_W^2 \, P_R$ with 
$P_{R,L} = (1\pm \gamma^5)/2$ the chiral projectors. 
We adopt the Feynman gauge propagator for gauge bosons. 
In eq.\,(\ref{eq:VirtualStructure}), we use the  
fictitious photon mass $\lambda \equiv \,\epsilon \,Q$ while the electron mass is $m_\psi \equiv \,\epsilon_\psi \,Q$.\footnote{
We remark that in our approximation, and within our gauge-fixing choice, 
one-loop self-energy diagrams vanish. 
This is because these diagrams only contain single-logarithmic infrared divergences while we are only interested 
in double-logarithmic corrections. It is worth emphasising  that this is a gauge-dependent statement. 
In the Coulomb or in the axial gauge, for instance, infrared double-logarithms originate from the self-energy corrections to
the
external lines\,\cite{Frenkel:1976bj}.}

After doing the trace, the structure $[\sum_{{\rm pol\,}Z}\epsilon_{\rho}(q)\epsilon_{\sigma}^*(q)]
T^{\rho\sigma}$ contains two kind of terms. 
The first kind 
does not depend on the integral variable $k$; $Q^4$ is one of those terms for instance. This kind of terms sums up to a combination of invariants multiplied by the scalar integral with three propagators at the denominator; this integral is known as the $\mathcal{C}_0$ Passarino-Veltman function in the literature and is defined in eq. (\ref{eq:ScalarC0}). 
The second kind of terms does depend on the integration variable $k$, like for instance terms proportional to $Q^2(k\cdot p_a)$. In this case we use the trick explained in appendix\,\ref{app:Vir}, which entails the substitutions:
\begin{equation}
p_a\cdot k\to -\frac{\lambda^2}{2},\quad p_b\cdot k\to -\frac{\lambda^2}{2},\quad k^2\to\lambda^2
\end{equation}
Eventually therefore, all of the terms we are interested in (i.e. at the double log level) turn out to be proportional to the $\mathcal{C}_0$ function
\footnote{This trick is equivalent to the Passarino-Veltman reduction mechanism where, at the end, we keep only the full coefficient in front of the $\mathcal{C}_0$ function}. Taking into account the two body phase space, we obtain the final result: 
\begin{equation}\label{eq:VirtualZdecayMassiveE}
\Gamma_{Z\to\psi\bar{\psi}}^V = Q \, 
V_{\psi\bar{\psi}}  \, {\cal I}^V;\qquad
V_{\psi\bar{\psi}}=-\frac{2\, g^4 }{   3\,c_W^2\,s_W^2} 
\left[
(1-4 s_W^2+8 s_W^4)(1+\epsilon^2)^2 - \epsilon_{\psi}^2 (3+2 \epsilon^2)
 + 2 (1+8 s_W^2-16 s_W^4) \epsilon_{\psi}^4
\right]  
\end{equation}
where $ {\cal I}^V$ is defined in eq. (\ref{gvrw}).

 The calculation of real corrections proceeds along the same lines of virtual ones.
 Of course kinematics is different since in the case of real corrections we have $p_a+p_b+k=q$ instead of $p_a+p_b=q$ and three body phase space instead of the two body one. Moreover we have two propagator denominators instead of the three present in the case of virtual corrections.
The final result is:
\begin{equation}\label{eq:RealCorrectionZdecayEle}
\Gamma_{Z\to\psi\bar{\psi}}^R = 
Q\,R_{\psi\bar{\psi}} {\cal I}^R;\qquad
R_{\psi\bar{\psi}}=\frac{2\, g^4 }{3\,c_W^2\,s_W^2} 
\left[
(1-4 s_W^2+8 s_W^4)(1+\epsilon^2)^2 - \epsilon_{\psi}^2 (3+2 \epsilon^2)
 + 2 (1+8 s_W^2-16 s_W^4) \epsilon_{\psi}^4
\right]  
\end{equation}
where  ${\mathcal I}^R $ is   the integral of the two propagators over the three body phase space $d\Phi_3$:
\be\label{pokcd}
{\mathcal I}^R(Q^2,m_\psi^2,\lambda^2)=Q^2\int\frac{d\Phi_3(m_\psi,m_\psi,k)}{[(p_a+k)^2-m_\psi^2] [(p_b+k)^2-m_\psi^2]}
\ee

\subsection{Z decay into massless fermions}\label{sec:MasslessQED}
Let us now discuss our results for QED corrections to Z decay, beginning with the case $m_\psi=0$ (massless QED). In this case the $\mathcal{C}_0$ function and the three-body phase space can be exactly calculated. We find 
\bea
&&{\mathcal I}^V=\frac{1}{256\,\pi^3}\,\left[{\rm Li}_2(1+\epsilon^2)+\frac{1}{2}\log^2(-\epsilon^2)+\frac{\pi^2}{6}\right]\,,
\\
&&{\mathcal I}^R= \frac{1}{256\,\pi^3}\,\left[-{\rm Li}_2\left(\frac{1}{1+\epsilon^2}\right) + {\rm Li}_2\left(\frac{\epsilon^2}{1+\epsilon^2}\right)+
2\;\log(\epsilon)\;\log\left(\frac{\epsilon}{1+\epsilon^2}\right)\right]\,.
\eea
At the double-log level, we find
\begin{equation}\label{eq:nonso}
{\mathcal I}^R(Q^2,0,\lambda^2)=\frac{1}{512\,\pi^3}\,\log^2\epsilon^2+ 0\times \epsilon^2\log^2\epsilon^2\,,\qquad 
{\mathcal I}^V(Q^2,0,\lambda^2)=\frac{1}{512\,\pi^3}\,\log^2\epsilon^2+0\times \epsilon^2\log^2\epsilon^2\,,
\end{equation}
where it is emphasized that there are no power suppressed terms at any order in $\epsilon$.
We obtain here an interesting result: the integrals which are relevant for virtual and for real corrections have the same double-log behavior; moreover. these integrals are free from power suppressed terms.  
This is rather surprising since ${\mathcal I}^V$ depends, as discussed in eq.\,(\ref{gvrw}), 
on the $C_0$ function and the two-body phase 
space  while ${\mathcal I}^R$ is an integral over the three body phase space, as discussed in eq.\,(\ref{pokcd}). This result holds more in general than in QED: indeed, it holds for any combination of internal and external masses, as is shown in appendix\,\ref{app:Vir}. With this in mind, in order to calculate the various observables and in order to compare real and virtual corrections, we can concentrate on the modulus squared of the amplitudes rather than on details of the relevant phase space. 

  Specializing eqs.\,(\ref{eq:VirtualZdecayMassiveE},\,\ref{eq:RealCorrectionZdecayEle}) to the case $\epsilon_\psi=0$ and using eq.\,(\ref{eq:nonso})
	we obtain that radiative corrections are factorized with respect to the tree level width in eq.\,(\ref{eq:TreeLevelDecayZ}):
\begin{align}\label{eq:DecayWidthZ}
 \Gamma^R    = -\Gamma^V = \Gamma_{Z,\,{\rm tree}}\,
  \frac{\alpha}{2\pi} \mathcal{R}(\epsilon)\log^2\epsilon^2\ \,,
~~~~~~~~~~~~
\mathcal{R} (\epsilon)  = (1+\epsilon^2)^2.
\end{align}
The $\epsilon$-independent term in $\mathcal{R} (\epsilon)$ corresponds to the so called eikonal approximation  which features a factorization at the amplitude level.  The eikonal  approximation is a semiclassical one, and features the emission of one 
(or more) soft particles along a hard line as illustrated in fig.\,\ref{eikonal}, panel {\it (a)}. Because of its classical nature, the soft particle emission does not change the spin of the emitting particle and it is universal, i.e. it does not depend neither on the spin of the emitting particle
   nor on the form of the vertex ${\cal M}_{0}$ describing the two body decay; in this case, double-log unsuppressed terms are generated in the three body width.
   
\vspace{0.5cm}
\begin{adjustbox}{max width=0.94\textwidth}    
    \raisebox{0mm}{
	\begin{tikzpicture} 
	  \draw[->-=.7,>=Latex][thick] (-0.5,-0.5)--(0.3,-0.2); 
	  \draw[->-=.7,>=Latex][thick] (0.3,-0.2)--(1.2,0.15); 	
	  \draw[thick,style={decorate, decoration={snake}}] (0.3,-0.2)--(1,-1); 
	  \draw[->,>=Latex][thick,double] (-0.5,-0.5 + 0.7)--(0.7,0.15 + 0.5);
	  \draw[black,fill=green,thick,opacity=0.3] (-1,-1)circle(20pt);
	  \draw[black] (-1,-1)circle(20pt);
	  \draw[black,fill=tangerineyellow,thick] (0.3,-0.2)circle(2.75pt);
	\node at (-1,-1) {\scalebox{1}{$\mathcal{M}_0$}};
	\node[rotate=20] at (-0.2,0.6) {\scalebox{1}{${\rm \textbf{hard}}$}};
	\node[rotate=-47] at (0.6,-0.9) {\scalebox{1}{${\rm soft}$}};
	\node at (1.2,-0.7) {\scalebox{1}{$k$}};
	\node[rotate=20] at (1,0.35) {\scalebox{1}{$p$}};
	\node[rotate=20] at (-0.2,-0.05) {\scalebox{1}{$p+k$}};
	\node at (0,-2.25) {\scalebox{1}{{\it (a)}}};
	\end{tikzpicture}}\hspace{2.cm}
    \raisebox{0mm}{
	\begin{tikzpicture} 
	  \draw[->-=.7,>=Latex][thick] (-0.5,-0.5)--(0.3,-0.2); 
	  \draw[->-=.7,>=Latex][thick] (0.3,-0.2)--(1,-1); 	
	  \draw[thick,style={decorate, decoration={snake}}] (0.3,-0.2)--(1.2,0.15); 
	  \draw[->,>=Latex][thick,double] (-0.5,-0.5 + 0.7)--(0.7,0.15 + 0.5);
	  \draw[black,fill=green,thick,opacity=0.3] (-1,-1)circle(20pt);
	  \draw[black] (-1,-1)circle(20pt);
	  \draw[black,fill=tangerineyellow,thick] (0.3,-0.2)circle(2.75pt);
	\node at (-1,-1) {\scalebox{1}{$\mathcal{M}_0$}};
    \node[rotate=20] at (-0.2,0.6) {\scalebox{1}{${\rm \textbf{hard}}$}};	
    \node[rotate=-47] at (0.6,-0.9) {\scalebox{1}{${\rm soft}$}};
    \node at (1.2,-0.7) {\scalebox{1}{$k$}};
	\node[rotate=20] at (1,0.35) {\scalebox{1}{$p$}};
	\node[rotate=20] at (-0.2,-0.05) {\scalebox{1}{$p+k$}}; 
	\node at (0,-2.25) {\scalebox{1}{{\it (b)}}};
	\end{tikzpicture}}\hspace{2.cm}
    \raisebox{0.3mm}{
	\begin{tikzpicture} 
	  \draw[thick,dashed] (-0.5,-0.5)--(0.3,-0.2); 
	  \draw[thick,style={decorate, decoration={snake}}] (0.3,-0.2)--(1,-1); 	
	  \draw[thick,style={decorate, decoration={snake}}] (0.3,-0.2)--(1.2,0.15); 
	  \draw[->,>=Latex][thick,double] (-0.5,-0.5 + 0.7)--(0.7,0.15 + 0.5);
	   \draw[black,fill=tangerineyellow,thick] (0.3,-0.2)circle(2.75pt);
	  \draw[thick,dashed] (-0.5+4,-0.5)--(0.3+4,-0.2); 
	  \draw[thick,style={decorate, decoration={snake}}] (0.3+4,-0.2)--(1+4,-1); 	
	  \draw[thick,dashed] (0.3+4,-0.2)--(1.2+4,0.15); 
	  \draw[->,>=Latex][thick,double] (-0.5+4,-0.5 + 0.7)--(0.7+4,0.15 + 0.5);
	  \draw[black,fill=tangerineyellow,thick] (0.3+4,-0.2)circle(2.75pt);
	\node[rotate=20] at (-0.2,0.6) {\scalebox{1}{${\rm \textbf{hard}}$}};
	\node[rotate=-47] at (0.6,-0.9) {\scalebox{1}{${\rm soft}$}};	
	\node[rotate=20] at (1,0.35) {\scalebox{0.9}{${\rm gauge\,boson}$}};
	\node[rotate=20] at (-0.35,-0.15) {\scalebox{0.9}{${\rm Higgs}$}}; 
	\node[rotate=20] at (-0.2+4,0.6) {\scalebox{1}{${\rm \textbf{hard}}$}};
	\node[rotate=-47] at (0.6+4,-0.9) {\scalebox{1}{${\rm soft}$}};	
		\node[rotate=20] at (1+4,0.35) {\scalebox{0.9}{${\rm Goldstone}$}};
	\node[rotate=20] at (-0.35+4,-0.15) {\scalebox{0.9}{${\rm Higgs}$}}; 
	\node at (2.3,-0.1) {\scalebox{1}{$\equiv$}}; 
	\node at (2.3,-1.75) {\scalebox{1}{{\it (c)}}};
	\end{tikzpicture}}		
	\end{adjustbox}
	
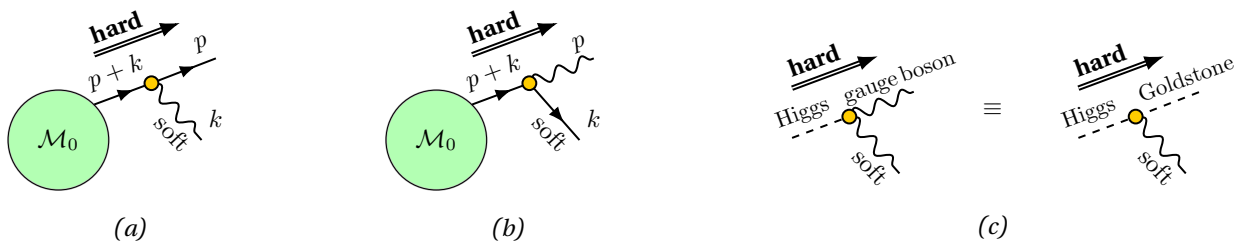
\captionof{figure}{{\em Pictorial representation of eikonal (a) and non-eikonal (b) contributions to a photon emission amplitude. It is understood that $k$ is soft and collinear with $p$, i.e. $k=\epsilon p$ with $\epsilon \ll 1$. In the eikonal case spin is unchanged along the hard line. (c): a Higgs boson emits a soft gauge boson transforming into a gauge boson along the hard line. This is equivalent to a Higgs transforming into a Goldstone boson \cite{g/h}.}}\label{eikonal}\vspace{0.5cm}
	
On the other hand, in the case of a non-eikonal contribution of the kind shown in 
fig.\,\ref{eikonal}, panel {\it (b)}, the spin of the particle is changed along the hard line; in this case no double log unsuppressed terms are generated.\footnote{Note that the eikonal and non-eikonal emissions belong to the same Feynman diagram, albeit evaluated in two different kinematical configurations}

Since the eikonal emission is the only one producing unsuppresed double-logs terms and since it is universal,  whatever the form of the hard vertex,  it is always true 
that $\mathcal{R} (\epsilon)=1+\dots$ where the dots represent the $\epsilon$-dependent terms that depend on the particular process considered.  For instance, in the case of the decay of
a heavy neutral scalar $\Phi \to e^{+}e^-$, mediated 
by the interaction vertex $\Phi\bar{\psi}\psi$,  we obtain $\mathcal{R} (\epsilon)=1$ exactly. So the leading IR behavior, governed by the eikonal,  is universal while the power suppressed terms are process dependents.
Limiting the analysis to the eikonal current is precisely what separates, in the standard approach, 
  the hard part of the scattering process from the infrared physics. 
  However, this picture is no longer valid when power-suppressed corrections are included 
  precisely because, as explained above, one is forced to go beyond the eikonal approximation to 
  compute higher twists. 
  This is worrisome since it seems to imply that power-suppressed corrections may jeopardize factorization. 

Let us now come the pattern of cancellations between real and virtual corrections. Eq. (\ref{eq:DecayWidthZ}) implies that $\Gamma^{V}
+\Gamma^R= 0$. At the level of leading (eikonal) IR corrections, the cancellation of $\log^2\epsilon^2$ terms is guaranteed by the KLN \cite{ Kinoshita:1962ur, Lee:1964is} theorem: the sum of virtual and real corrections must be free of IR divergences, i.e. terms that diverge in the limit of vanishing  photon mass $\epsilon\to 0$. 
However, power suppressed terms $\propto \epsilon^2\log^2\epsilon^2$ are zero in the $\epsilon\to 0$ limit and the KLN theorem says nothing about them. 
Nevertheless, these terms also cancel. 
While there is no apriori reason for this cancellation, there is  a technical reason. 
On the one hand, the fact that ${\mathcal I}^V = {\mathcal I}^R$ including power corrections -- as discussed in 
eq.\,(\ref{eq:nonso}) -- implies that comparing real corrections with virtual ones one can forget about phase space details and consider only  squared amplitudes. 
On the other hand, the squared amplitudes for real and virtual corrections are related by what we call the
 ``$p+k$ theorem'' (see appendix\,\ref{p+k}) where it turns out that are  exactly equal also for  double-log 
 power-suppressed corrections. 
Hence the cancellation. 

\subsection{Z decay into massive fermion\label{QEDmassive}}
Let us now turn our attention to the massive case $m_\psi\neq 0$. 
As a first observation, we note that the virtual correction $\Gamma_{Z\to\psi\bar{\psi}}^V$ in eq.\,(\ref{eq:VirtualZdecayMassiveE}) 
does not factorize the tree-level result in eq.\,(\ref{eq:TreeLevelDecayZ}).
Furthermore, the scalar three-point integral diverges logarithmically in the limit $\lambda\to 0$ (but finite $m_{\psi}$). 
This is the soft infrared divergence associated to the massless photon exchange. Because of the presence of terms proportional to $\epsilon_{\psi}^2$ in $V_{\psi\bar{\psi}}$, eq. (\ref{eq:VirtualZdecayMassiveE}) features   power suppressed terms proportional to $\epsilon_\psi^2\log^2\epsilon^2$ that are divergent in the IR $\epsilon\to 0$ limit; 
the same holds for the real corrections, see eq.\,(\ref{eq:RealCorrectionZdecayEle}).
This   shows that power-suppressed corrections (in this case those proportional to 
$m_{\psi}$) are  infinities in the physical limit $\lambda\to 0$.
  We conclude that in gauge theories, power-suppressed infrared corrections may give rise to actual 
divergences. Let us now come to the pattern of cancellations. Just as in the massless cases, we have ${\cal I}^V={\cal I}^R$; in fact this is a general (and by no means obvious) statement, that we discuss in appendix\,\ref{p+k}. Moreover we see from eqs (\ref{eq:VirtualZdecayMassiveE},\ref{eq:RealCorrectionZdecayEle}) that $V_{\psi\bar{\psi}}=-R_{\psi\bar{\psi}}$. This is also a general relationship 
that comes from the ``$p+k$ theorem'' discussed in appendix\,\ref{p+k}. We conclude that $\Gamma^R+\Gamma^V=0$. 
Just like in the massless case, 
we have a complete cancellation of all leading and power-suppressed terms at one loop. 
This cancellation is only in part related to the KLN theorem: for instance, $\epsilon_\psi^2\log^2\epsilon$ terms are IR divergent and must cancel due to the KLN theorem, while $\epsilon^2\log^2\epsilon$ terms are not IR divergent  and are unrelated to KLN; they cancel due to the ``IR unitarity theorem''  that we prove in appendix\,\ref{p+k}.

All the above results can be summarized in the following rule.
 {\it In QED, power-suppressed double-logarithmic corrections at one loop cancel out, 
 if we sum over all possible cut diagrams}.
Diagrammatically, this statement reads
\begin{align}
    \sum_{{\rm all\,cuts}}
\begin{adjustbox}{max width=0.94\textwidth}    
    \raisebox{-13mm}{
	\begin{tikzpicture}
	\draw[thick,style={decorate, decoration={snake}}] (-0.5,0)--(0,0);
	\draw[->-=.6,>=Latex][thick] (0,0)--(1.25,1.25); 
	\draw[->-=.6,>=Latex][thick] (1.25,-1.25)--(0,0);
		\draw[thick,style={decorate, decoration={snake}}] (1.25,1.25)--(1.25,-1.25);
	\draw[->-=.6,>=Latex][thick] (1.25,1.25)--(2.5,0); 
	\draw[->-=.6,>=Latex][thick] (2.5,0)--(1.25,-1.25); 	
	\draw[thick,style={decorate, decoration={snake}}] (2.5,0)--(2.5+0.5,0);
	\draw[black,fill=tangerineyellow,thick] (0,0)circle(2.75pt); 
	\draw[black,fill=tangerineyellow,thick] (2.5,0)circle(2.75pt); 
	\draw[black,fill=tangerineyellow,thick] (1.25,1.25)circle(2.75pt); 
	\draw[black,fill=tangerineyellow,thick] (1.25,-1.25)circle(2.75pt); 
	\end{tikzpicture}}~~~=~~
    \raisebox{-13mm}{
	\begin{tikzpicture}
	\draw[thick,style={decorate, decoration={snake}}] (-0.5,0)--(0,0);
	\draw[->-=.8,>=Latex][thick] (0,0)--(1.25/2,1.25/2); 
	\draw[->-=.6,>=Latex][thick] (1.25/2,1.25/2)--(1.25,1.25); 
	\draw[->-=.8,>=Latex][thick] (1.25,-1.25)--(1.25/2,-1.25/2);
	\draw[->-=.6,>=Latex][thick] (1.25/2,-1.25/2)--(0,0);
	\draw[thick,style={decorate, decoration={snake}}] (1.25,1.25)--(1.25,-1.25);
	\draw[->-=.6,>=Latex][thick] (1.25,1.25)--(2.5,0); 
	\draw[->-=.6,>=Latex][thick] (2.5,0)--(1.25,-1.25);
	\draw[thick,dash dot,persianblue][thick] (1.25/2,1.25)--(1.25/2,-1.25);  	
	\draw[thick,style={decorate, decoration={snake}}] (2.5,0)--(2.5+0.5,0);
	\draw[black,fill=tangerineyellow,thick] (0,0)circle(2.75pt); 
	\draw[black,fill=tangerineyellow,thick] (2.5,0)circle(2.75pt); 
	\draw[black,fill=tangerineyellow,thick] (1.25,1.25)circle(2.75pt); 
	\draw[black,fill=tangerineyellow,thick] (1.25,-1.25)circle(2.75pt); 
	\end{tikzpicture}}	~~~+~~
    \raisebox{-13mm}{
	\begin{tikzpicture}
	\draw[thick,style={decorate, decoration={snake}}] (-0.5,0)--(0,0);
	\draw[->-=.6,>=Latex][thick] (0,0)--(1.25,1.25); 
	\draw[->-=.6,>=Latex][thick] (1.25,-1.25)--(0,0);
	\draw[thick,style={decorate, decoration={snake}}] (1.25,1.25)--(1.25,-1.25);
	\draw[->-=.8,>=Latex][thick] (1.25,1.25)--(1.25+1.25/2,1.25-1.25/2);
    \draw[->-=.6,>=Latex][thick] (1.25+1.25/2,1.25-1.25/2)--(2.5,0); 
	\draw[->-=.8,>=Latex][thick] (2.5,0)--(1.25+1.25/2,-1.25/2);
	\draw[->-=.6,>=Latex][thick] (1.25+1.25/2,-1.25/2)--(1.25,-1.25);  	
	\draw[thick,dash dot,persianblue][thick] (1.25+1.25/2,1.25)--(1.25+1.25/2,-1.25); 
	\draw[thick,style={decorate, decoration={snake}}] (2.5,0)--(2.5+0.5,0);
	\draw[black,fill=tangerineyellow,thick] (0,0)circle(2.75pt); 
	\draw[black,fill=tangerineyellow,thick] (2.5,0)circle(2.75pt); 
	\draw[black,fill=tangerineyellow,thick] (1.25,1.25)circle(2.75pt); 
	\draw[black,fill=tangerineyellow,thick] (1.25,-1.25)circle(2.75pt); 
	\end{tikzpicture}}	~~~+~~
    \raisebox{-13mm}{
	\begin{tikzpicture}
	\draw[thick,style={decorate, decoration={snake}}] (-0.5,0)--(0,0);
	\draw[->-=.8,>=Latex][thick] (0,0)--(1.25/2,1.25/2); 
	\draw[->-=.6,>=Latex][thick] (1.25/2,1.25/2)--(1.25,1.25); 
	\draw[->-=.6,>=Latex][thick] (1.25,-1.25)--(0,0);
		\draw[thick,style={decorate, decoration={snake}}] (1.25,1.25)--(1.25,-1.25);
	\draw[->-=.6,>=Latex][thick] (1.25,1.25)--(2.5,0); 
	\draw[->-=.8,>=Latex][thick] (2.5,0)--(1.25+1.25/2,-1.25/2);
	\draw[->-=.6,>=Latex][thick] (1.25+1.25/2,-1.25/2)--(1.25,-1.25); 	
	\draw[thick,style={decorate, decoration={snake}}] (2.5,0)--(2.5+0.5,0);
	\draw[thick,dash dot,persianblue][thick] (0,1.25)--(2.5,-1.25); 	
	\draw[black,fill=tangerineyellow,thick] (0,0)circle(2.75pt); 
	\draw[black,fill=tangerineyellow,thick] (2.5,0)circle(2.75pt); 
	\draw[black,fill=tangerineyellow,thick] (1.25,1.25)circle(2.75pt); 
	\draw[black,fill=tangerineyellow,thick] (1.25,-1.25)circle(2.75pt); 
	\end{tikzpicture}}	~~~+~~
    \raisebox{-13mm}{
	\begin{tikzpicture}
	\draw[thick,style={decorate, decoration={snake}}] (-0.5,0)--(0,0);
	\draw[->-=.6,>=Latex][thick] (0,0)--(1.25,1.25); 
	\draw[->-=.8,>=Latex][thick] (1.25,-1.25)--(1.25/2,-1.25/2);
	\draw[->-=.6,>=Latex][thick] (1.25/2,-1.25/2)--(0,0);
		\draw[thick,style={decorate, decoration={snake}}] (1.25,1.25)--(1.25,-1.25);
	\draw[->-=.8,>=Latex][thick] (1.25,1.25)--(1.25+1.25/2,1.25-1.25/2);
    \draw[->-=.6,>=Latex][thick] (1.25+1.25/2,1.25-1.25/2)--(2.5,0);  
	\draw[->-=.6,>=Latex][thick] (2.5,0)--(1.25,-1.25); 	
	\draw[thick,style={decorate, decoration={snake}}] (2.5,0)--(2.5+0.5,0);
    \draw[thick,dash dot,persianblue][thick] (2.5,1.25)--(0,-1.25); 
	\draw[black,fill=tangerineyellow,thick] (0,0)circle(2.75pt); 
	\draw[black,fill=tangerineyellow,thick] (2.5,0)circle(2.75pt); 
	\draw[black,fill=tangerineyellow,thick] (1.25,1.25)circle(2.75pt); 
	\draw[black,fill=tangerineyellow,thick] (1.25,-1.25)circle(2.75pt); 
	\end{tikzpicture}}  ~~~= ~~ 0
	\end{adjustbox}
\end{align}
The first two cuts correspond to virtual corrections while that last two cuts to real emission. 
This is the conventional unitarity rule (KLN theorem) that guarantees infrared finiteness. 
Our computation, however, shows that this cancellation is more general, and involves 
power-suppressed corrections that do not diverge in the IR. This happens by virtue of a combination of the ``$p+k$ theorem'' and of the relationship in eq.\,(\ref{final}). 
 
\section{The case of spontaneously broken abelian gauge theories
\label{sponte}}

In order to clarify  the main features  related to the cancellation of the higher twist double logs
 we choose the simplest (but not so simple) model able to match the main points. 

Our  model  is based on the gauge group
$U'(1)\otimes U(1)$ where an heavy $Z'$ gauge boson (related to the spontaneous breaking of $U'(1)$) is
decaying into lighter particles: light $Z$ gauge bosons (due to the spontaneous breaking of $U(1)$), a light Higgs,    light massive chiral fermions, both charged under the full group $U'(1)\otimes U(1)$.
The   Lagrangian of the model can be divided into three pieces: ${\cal L}_f$ describes the gauge interaction of the fermionic fields 
$\psi$, ${\cal L}_s$ describes the gauge interaction of the scalar fields $\varphi$ and $\varphi^{\prime}$ 
plus the Higgs potentials $V(\varphi)$ and ${\cal V}(\varphi')$ generating a specific spontaneous breaking pattern as a consequence of 
which the mass of the $Z'$ gauge boson is much higher than the mass of the $Z$ boson; finally ${\cal L}_m$ gives the Yukawa interactions between  fermions and  scalars. 
Finally the gauge fixing Lagrangian is discussed in appendix\,\ref{app:Coupl}.
\bea\label{lag}
 {\cal L}_f&=&\bar\psi\;\left[\partial_{\mu} + 
 i g \left(y_L P_L + y_R P_R  \right)Z_{\mu} + ig' \left(f_L P_L + f_R P_R\right)
 Z_{\mu}'\right]\gamma^{\mu}\psi\,,
 \\
\label{scalar}
{\cal L}_s &=& \left| 
\left(\partial_{\mu} + i g' f_{\varphi'}Z_{\mu}'\right)\varphi'\right|^2
+
\left| \left(\partial_{\mu}+i g'f_{\phi}Z'+i g y_{\phi}Z_{\mu} \right)\varphi\right|^2 + 
V(\varphi)+{\cal V}(\varphi')\,,
\\
{\cal L}_m &=&\left(h_f \varphi\,\bar\psi\,P_L\,\psi + h.c.\right)\,.\label{eq:Lm}
\eea
where $P_{R/L}=(1\pm\gamma_5)/2$, $f_{R/L}\;(y_{R/L})$ are the $U'(1)$ ($U(1)$) 
fermionic gauge charges of the right-handed fermions $\psi_R=P_R\;\psi$ and left-handed $\psi_L=P_L\;\psi$, respectively.
To implement, in a natural way, 
the spontaneous breaking of the gauge groups $U'(1)\otimes U(1)$ 
the two complex Higgs fields get a vev and can be decomposed as
$\varphi'= (h'+v'+i\;\phi')/\sqrt{2}$,
with $v'$ the vev breaking $U'(1)$,  and  $\varphi=(h+v+i\;\phi)/\sqrt{2}$,
with $v$ the vev  for the breaking of $U(1)$. 
The scalar fields are charged with respect $U'(1)$ with hypercharge $f_{\phi'}$ and $f_{\phi}=f_R-f_L$ and with respect $U(1)$ with hypercharge $y_{\phi'}=0$ and $y_{\phi}=y_R-y_L$.
The hierarchy  requirements of a much higher   mass for
 the $Z'$ gauge boson over the $Z$  implies $v'\gg v$ (assuming that the gauge coupling are of the same order).
We summarize in table\,\ref{tab:Spectrum} below the main quantum numbers and the mass spectrum of the various fields.
\begin{table}[htp]
\begin{center}
\begin{tabular}{|c|c|c|}
\hline
\textbf{field} & $\boldsymbol{U(1)}$ \textbf{charge}& $\boldsymbol{U'(1)}$ \textbf{charge}\\
\hline\hline
$\psi_{L/R}=P_{L/R}\;\psi$ & $y_{L/R}$ &$f_{L/R}$\\
$\varphi$ & $y_{\phi}=y_R-y_L$ &$f_{\phi}=f_R-f_L$\\
$\varphi'$ & $0$ &$f_{\phi'}$ \\
\hline
\end{tabular}
\qquad \begin{tabular}{|c|c|}
\hline
\textbf{fields} & \textbf{mass spectrum} \\
\hline\hline
$Z,\;\phi $ & $M
$  \\
$Z',\;\phi'$ & $Q
$  \\
$\psi$ & $m_{\psi} 
$   \\
$h$ & $m_{h} $   \\
\hline
\end{tabular} 
$\qquad \epsilon=\epsilon_\phi\equiv\frac{M}{Q},\;\;\epsilon_\psi\equiv\frac{m_\psi}{Q},\;\;\epsilon_h\equiv\frac{m_h}{Q} $
\end{center}\vspace{-0.4cm}
\caption{{\it 
Summary of the main quantum numbers and the mass spectrum for the model given by the Lagrangian density in 
eqs.\,(\ref{lag},\,\ref{scalar},\,\ref{eq:Lm}).
}}\label{tab:Spectrum}
\end{table}%
Note that the gauge fields $Z$ and $Z'$ mix in a mass matrix that needs a diagonalization process describe in the appendix. The diagonal eigenstates   $M\ll Q$   result non trivial function of the vevs $v$ and $v'$, see eq.\,(\ref{massesMQ}). All the Feynman rules are given in 
appendix\,\ref{app:Coupl}.
Our  computations are done in Feynman gauge. In order to shorten our  notation, in virtual corrections, when we write a $Z$ internal line we mean  that the contributions  of the Feynman   $Z$ propagator  and   of the Goldstone $\phi$ propagator are summed over. In the case of  real $Z$ emission we use the sum over   transverse and longitudinal polarizations.

 In order to understand the pattern of double logs of IR origin, let us consider first the $Z'$ decay into fermion-antifermion. 
 
 Virtual corrections are affected by IR sensitive terms; the usual procedure consists in adding real $Z$ emission, i.e.  summing  the probabilities for $Z'\to\psi\bar{\psi}$
and $Z'\to\psi\bar{\psi}Z$, $Z$ being the light gauge boson. We can give to this observable the name  $Z'\to\psi\bar{\psi}X$ where $X=\emptyset, Z$ and call it the ``standard'' inclusive observable. 
This observable corresponds to adding all possible cuts of diagram $S_1$ in fig. \ref{ASJ}. 
As far as leading (double) logs are concerned, this is the end of the story: this observable is free from IR divergences that would occur in the limit of vanishing $Z$ mass. 
However, since we are interested in mass suppressed double logs, there are other diagrams contributing to the observable we just defined and that produce this kind of terms. First, we must add   the two cuts of  $S_2(a)$ diagram (see eq.\,(\ref{eq:Virtual5})) 
and the diagram $A_1(a)$ (see eq.\,(\ref{eq:Virtual7}))  that correspond to virtual corrections to the two body decay into fermion-antifermion. Secondly, we must also add diagram $A_1(c)$   relative to real $Z$ emission. Adding all these diagrams we have therefore:
	\be\label{dfghc}
	Z'\to\psi\bar{\psi}X;X=\emptyset, Z\quad\Rightarrow\quad
	\sum_{{\rm all\,cuts}} S_1+ S_2(a)+A_1(a)+A_1(c)= S_2(a)
	\ee
where use has been made of the ``IR unitarity theorem'' of appendix\,\ref{p+k}, 
implying, in this case, $\sum_{\rm all\,cuts} S_1=A_1(a)+A_1(c)=0$. 
The contribution of diagram $S_2(a)$ can be read off from eq.\,(\ref{eq:Virtual5}). 
We find, at order $\epsilon_i^2$:
\be
\Gamma(
Z'\to\psi\bar{\psi}X;X=\emptyset, Z)
=\Gamma_{tree}-\frac{1}{192\,\pi^3}\epsilon_\psi^2{\cal F}_{h\psi}^2
({\cal F}_{Z'L}^2+{\cal F}_{Z'R}^2) QL^2 
\ee
where ${\cal F}_{h\psi}$ is related to the Yukawa coupling between the light Higgs and the fermion in the mass
eigenstate basis while ${\cal F}_{Z'L}$ and ${\cal F}_{Z'R}$ refer, in the same basis, to the chirality-dependent 
couplings of the heavy 
$Z^{\prime}$ gauge boson with $\psi$.
Therefore, we obtain that   the width for this observable features uncancelled mass suppressed double logs of IR origin ($LL_{PS}$). Since real emissions tend to cancel virtual corrections, by looking at $S_2$ we are   led to enlarge the definition of the observable by including the $h$ in the final state as well: $Z'\to\psi\bar{\psi}X;X=\emptyset, Z,h$. We must then add also diagram $A_1(b)$ that contributes to the same observable, and we obtain:
 
	\be\label{iyg}
	Z'\to\psi\bar{\psi}X;X=\emptyset, Z,h\quad\quad\Longrightarrow\quad\quad
	\sum_{\rm all\,cuts} S_1+\sum_{\rm all\,cuts} S_2+\sum_{i= a,b,c}A_1(i)=- A_1(d)
	\ee
	The contribution from the cut $A_1(d)$ corresponds, in our language, to $V_{Zh}^{(\psi)}$ which can be read from 
	eq.\,(\ref{eq:Virtual7}), and obviously features uncanceled power suppressed double logs.
	The reason for the non-cancellation is simple: as we have seen in eq.\,(\ref{pkth}), 
	the sum over all possible cuts of a given diagram gives 0; this is the ``IR unitarity theorem''. 
	However, on the right-hand side of eq. (\ref{iyg}) the contribution of diagram $A_1$ is unbalanced: summing over cuts $a,b,c$ we obtain minus the contribution of cut $d$, and, therefore, a power suppressed double logarithms. 
\vspace{0.5cm}
\begin{align}
    \boxed{S_1}~~~~ &
    \scalebox{0.8}{
    \raisebox{-13mm}{
	\begin{tikzpicture}
	\draw[thick,style={decorate, decoration={snake}}] (-0.5,0)--(0,0);
	\draw[->-=.6,>=Latex][thick] (0,0)--(1.25,1.25); 
	\draw[->-=.6,>=Latex][thick] (1.25,-1.25)--(0,0);
		\draw[thick,style={decorate, decoration={snake}}] (1.25,1.25)--(1.25,-1.25);
	\draw[->-=.6,>=Latex][thick] (1.25,1.25)--(2.5,0); 
	\draw[->-=.6,>=Latex][thick] (2.5,0)--(1.25,-1.25); 	
	\draw[thick,style={decorate, decoration={snake}}] (2.5,0)--(2.5+0.5,0);
	\draw[black,fill=tangerineyellow,thick] (0,0)circle(2.75pt); 
	\draw[black,fill=tangerineyellow,thick] (2.5,0)circle(2.75pt); 
	\draw[black,fill=tangerineyellow,thick] (1.25,1.25)circle(2.75pt); 
	\draw[black,fill=tangerineyellow,thick] (1.25,-1.25)circle(2.75pt); 
	\end{tikzpicture}}
	}
	\nn \\ & \nn \\
    \boxed{S_2}~~~~ &
    \scalebox{0.8}{
    \raisebox{-19mm}{
	\begin{tikzpicture}
	\draw[thick,style={decorate, decoration={snake}}] (-0.5,0)--(0,0);
	\draw[->-=.8,>=Latex][thick] (0,0)--(1.25/2,1.25/2); 
	\draw[->-=.6,>=Latex][thick] (1.25/2,1.25/2)--(1.25,1.25); 
	\draw[->-=.8,>=Latex][thick] (1.25,-1.25)--(1.25/2,-1.25/2);
	\draw[->-=.6,>=Latex][thick] (1.25/2,-1.25/2)--(0,0);
		\draw[thick,dashed] (1.25,1.25)--(1.25,-1.25);
	\draw[->-=.8,>=Latex][thick] (1.25,1.25)--(1.25+1.25/2,1.25-1.25/2);
    \draw[->-=.6,>=Latex][thick] (1.25+1.25/2,1.25-1.25/2)--(2.5,0); 
	\draw[->-=.8,>=Latex][thick] (2.5,0)--(1.25+1.25/2,-1.25/2);
	\draw[->-=.6,>=Latex][thick] (1.25+1.25/2,-1.25/2)--(1.25,-1.25);	
	\draw[thick,style={decorate, decoration={snake}}] (2.5,0)--(2.5+0.5,0);
	\draw[thick,dash dot,persianblue][thick] (1.25/2,1.25)--(1.25/2,-1.25); 
    \draw[thick,dash dot,persianblue][thick] (1.25+1.25/2,1.25)--(1.25+1.25/2,-1.25); 	
	\draw[black,fill=tangerineyellow,thick] (0,0)circle(2.75pt); 
	\draw[black,fill=tangerineyellow,thick] (2.5,0)circle(2.75pt); 
	\draw[black,fill=tangerineyellow,thick] (1.25,1.25)circle(2.75pt); 
	\draw[black,fill=tangerineyellow,thick] (1.25,-1.25)circle(2.75pt); 
	\node at (1.25,-1.75) {\scalebox{1}{{\it (a)}}};
	\end{tikzpicture}}
	}
	\hspace{0.2cm}
	\scalebox{0.8}{
	\raisebox{-19mm}{
	\begin{tikzpicture}
	\draw[thick,style={decorate, decoration={snake}}] (-0.5,0)--(0,0);
	\draw[->-=.8,>=Latex][thick] (0,0)--(1.25/2,1.25/2); 
	\draw[->-=.6,>=Latex][thick] (1.25/2,1.25/2)--(1.25,1.25); 
	\draw[->-=.8,>=Latex][thick] (1.25,-1.25)--(1.25/2,-1.25/2);
	\draw[->-=.6,>=Latex][thick] (1.25/2,-1.25/2)--(0,0);
		\draw[thick,dashed] (1.25,1.25)--(1.25,-1.25);
	\draw[->-=.8,>=Latex][thick] (1.25,1.25)--(1.25+1.25/2,1.25-1.25/2);
    \draw[->-=.6,>=Latex][thick] (1.25+1.25/2,1.25-1.25/2)--(2.5,0); 
	\draw[->-=.8,>=Latex][thick] (2.5,0)--(1.25+1.25/2,-1.25/2);
	\draw[->-=.6,>=Latex][thick] (1.25+1.25/2,-1.25/2)--(1.25,-1.25);	
	\draw[thick,style={decorate, decoration={snake}}] (2.5,0)--(2.5+0.5,0);
	\draw[thick,dash dot,persianblue][thick] (0,1.25)--(2.5,-1.25); 
	\draw[thick,dash dot,persianblue][thick] (2.5,1.25)--(0,-1.25); 
	\draw[black,fill=tangerineyellow,thick] (0,0)circle(2.75pt); 
	\draw[black,fill=tangerineyellow,thick] (2.5,0)circle(2.75pt); 
	\draw[black,fill=tangerineyellow,thick] (1.25,1.25)circle(2.75pt); 
	\draw[black,fill=tangerineyellow,thick] (1.25,-1.25)circle(2.75pt); 
	\node at (1.25,-1.75) {\scalebox{1}{{\it (b)}}};
	\end{tikzpicture}}
	} \nn  \\ & \nn \\
    \boxed{S_3}~~~~ &
    \scalebox{0.8}{
    \raisebox{-19mm}{
	\begin{tikzpicture}
	\draw[thick,style={decorate, decoration={snake}}] (-0.5,0)--(0,0);
	\draw[thick,style={decorate, decoration={snake}}] (0,0)--(1.25,1.25); 
	\draw[thick,dashed] (1.25,-1.25)--(0,0);
		\draw[thick,style={decorate, decoration={snake}}] (1.25,1.25)--(1.25,-1.25);
	\draw[thick,dashed] (1.25,1.25)--(2.5,0); 
	\draw[thick,style={decorate, decoration={snake}}] (2.5,0)--(1.25,-1.25); 	
	\draw[thick,style={decorate, decoration={snake}}] (2.5,0)--(2.5+0.5,0);
	\draw[thick,dash dot,persianblue][thick] (1.25/2,1.25)--(1.25/2,-1.25); 
    \draw[thick,dash dot,persianblue][thick] (1.25+1.25/2,1.25)--(1.25+1.25/2,-1.25); 
	\draw[black,fill=tangerineyellow,thick] (0,0)circle(2.75pt); 
	\draw[black,fill=tangerineyellow,thick] (2.5,0)circle(2.75pt); 
	\draw[black,fill=tangerineyellow,thick] (1.25,1.25)circle(2.75pt); 
	\draw[black,fill=tangerineyellow,thick] (1.25,-1.25)circle(2.75pt); 
	\node at (1.25,-1.75) {\scalebox{1}{{\it (a)}}};
	\end{tikzpicture}}
	}
	\hspace{0.25cm}
	\scalebox{0.8}{
	\raisebox{-19mm}{
	\begin{tikzpicture}
	\draw[thick,style={decorate, decoration={snake}}] (-0.5,0)--(0,0);
	\draw[thick,style={decorate, decoration={snake}}] (0,0)--(1.25,1.25); 
	\draw[thick,dashed] (1.25,-1.25)--(0,0);
		\draw[thick,style={decorate, decoration={snake}}] (1.25,1.25)--(1.25,-1.25);
	\draw[thick,dashed] (1.25,1.25)--(2.5,0); 
	\draw[thick,style={decorate, decoration={snake}}] (2.5,0)--(1.25,-1.25); 	
	\draw[thick,style={decorate, decoration={snake}}] (2.5,0)--(2.5+0.5,0);
    \draw[thick,dash dot,persianblue][thick] (0,1.25)--(2.5,-1.25); 
	\draw[black,fill=tangerineyellow,thick] (0,0)circle(2.75pt); 
	\draw[black,fill=tangerineyellow,thick] (2.5,0)circle(2.75pt); 
	\draw[black,fill=tangerineyellow,thick] (1.25,1.25)circle(2.75pt); 
	\draw[black,fill=tangerineyellow,thick] (1.25,-1.25)circle(2.75pt); 
	\node at (1.25,-1.75) {\scalebox{1}{{\it (b)}}};
	\end{tikzpicture}}
	}
	\hspace{0.35cm}
	\scalebox{0.8}{
	\raisebox{-19mm}{
	\begin{tikzpicture}
	\draw[thick,style={decorate, decoration={snake}}] (-0.5,0)--(0,0);
	\draw[thick,style={decorate, decoration={snake}}] (0,0)--(1.25,1.25); 
	\draw[thick,dashed] (1.25,-1.25)--(0,0);
		\draw[thick,style={decorate, decoration={snake}}] (1.25,1.25)--(1.25,-1.25);
	\draw[thick,dashed] (1.25,1.25)--(2.5,0); 
	\draw[thick,style={decorate, decoration={snake}}] (2.5,0)--(1.25,-1.25); 	
	\draw[thick,style={decorate, decoration={snake}}] (2.5,0)--(2.5+0.5,0);
	\draw[thick,dash dot,persianblue][thick] (2.5,1.25)--(0,-1.25); 
	\draw[black,fill=tangerineyellow,thick] (0,0)circle(2.75pt); 
	\draw[black,fill=tangerineyellow,thick] (2.5,0)circle(2.75pt); 
	\draw[black,fill=tangerineyellow,thick] (1.25,1.25)circle(2.75pt); 
	\draw[black,fill=tangerineyellow,thick] (1.25,-1.25)circle(2.75pt); 
	\node at (1.25,-1.75) {\scalebox{1}{{\it (c)}}};
	\end{tikzpicture}}
	}	\\ & \nn \\
    \boxed{A_1}~~~~ & 
    \scalebox{0.8}{
    \raisebox{-19mm}{
	\begin{tikzpicture}
	\draw[thick,style={decorate, decoration={snake}}] (-0.5,0)--(0,0);
	\draw[->-=.8,>=Latex][thick] (0,0)--(1.25/2,1.25/2); 
	\draw[->-=.6,>=Latex][thick] (1.25/2,1.25/2)--(1.25,1.25); 
	\draw[->-=.8,>=Latex][thick] (1.25,-1.25)--(1.25/2,-1.25/2);
	\draw[->-=.6,>=Latex][thick] (1.25/2,-1.25/2)--(0,0);
	\draw[->-=.6,>=Latex][thick] (1.25,1.25)--(1.25,-1.25);
	\draw[thick,style={decorate, decoration={snake}}] (1.25,1.25)--(2.5,0); 
	\draw[thick,dashed] (2.5,0)--(1.25,-1.25); 	
	\draw[thick,style={decorate, decoration={snake}}] (2.5,0)--(2.5+0.5,0);
	\draw[thick,dash dot,persianblue][thick] (1.25/2,1.25)--(1.25/2,-1.25); 
	\draw[black,fill=tangerineyellow,thick] (0,0)circle(2.75pt); 
	\draw[black,fill=tangerineyellow,thick] (2.5,0)circle(2.75pt); 
	\draw[black,fill=tangerineyellow,thick] (1.25,1.25)circle(2.75pt); 
	\draw[black,fill=tangerineyellow,thick] (1.25,-1.25)circle(2.75pt); 
	\node at (1.25,-1.75) {\scalebox{1}{{\it (a)}}};
	\end{tikzpicture}}
	}
	\hspace{0.3cm}
	\scalebox{0.8}{
	\raisebox{-19mm}{
	\begin{tikzpicture}
	\draw[thick,style={decorate, decoration={snake}}] (-0.5,0)--(0,0);
	\draw[->-=.8,>=Latex][thick] (0,0)--(1.25/2,1.25/2); 
	\draw[->-=.6,>=Latex][thick] (1.25/2,1.25/2)--(1.25,1.25); 
	\draw[->-=.6,>=Latex][thick] (1.25,-1.25)--(0,0);
	\draw[->-=.6,>=Latex][thick] (1.25,1.25)--(1.25,0);
	\draw[->-=.6,>=Latex][thick] (1.25,0)--(1.25,-1.25);
	\draw[thick,style={decorate, decoration={snake}}] (1.25,1.25)--(2.5,0); 
	\draw[thick,dashed] (2.5,0)--(1.25,-1.25); 	
	\draw[thick,style={decorate, decoration={snake}}] (2.5,0)--(2.5+0.5,0);
	\draw[thick,dash dot,persianblue][thick] (0,1.25)--(2.5,-1.25); 
	\draw[black,fill=tangerineyellow,thick] (0,0)circle(2.75pt); 
	\draw[black,fill=tangerineyellow,thick] (2.5,0)circle(2.75pt); 
	\draw[black,fill=tangerineyellow,thick] (1.25,1.25)circle(2.75pt); 
	\draw[black,fill=tangerineyellow,thick] (1.25,-1.25)circle(2.75pt); 
	\node at (1.25,-1.75) {\scalebox{1}{{\it (b)}}};
	\end{tikzpicture}}
	}
	\hspace{0.3cm}
	\scalebox{0.8}{
	\raisebox{-19mm}{
	\begin{tikzpicture}
	\draw[thick,style={decorate, decoration={snake}}] (-0.5,0)--(0,0);
	\draw[->-=.6,>=Latex][thick] (0,0)--(1.25,1.25); 
	\draw[->-=.8,>=Latex][thick] (1.25,-1.25)--(1.25/2,-1.25/2);
	\draw[->-=.6,>=Latex][thick] (1.25/2,-1.25/2)--(0,0);
	\draw[->-=.6,>=Latex][thick] (1.25,1.25)--(1.25,0);
	\draw[->-=.6,>=Latex][thick] (1.25,0)--(1.25,-1.25);
	\draw[thick,style={decorate, decoration={snake}}] (1.25,1.25)--(2.5,0); 
	\draw[thick,dashed] (2.5,0)--(1.25,-1.25); 	
	\draw[thick,style={decorate, decoration={snake}}] (2.5,0)--(2.5+0.5,0);
	\draw[thick,dash dot,persianblue][thick] (2.5,1.25)--(0,-1.25); 	
	\draw[black,fill=tangerineyellow,thick] (0,0)circle(2.75pt); 
	\draw[black,fill=tangerineyellow,thick] (2.5,0)circle(2.75pt); 
	\draw[black,fill=tangerineyellow,thick] (1.25,1.25)circle(2.75pt); 
	\draw[black,fill=tangerineyellow,thick] (1.25,-1.25)circle(2.75pt); 
	\node at (1.25,-1.75) {\scalebox{1}{{\it (c)}}};
	\end{tikzpicture}}
	}
	\hspace{0.3cm}
	\scalebox{0.8}{
    \raisebox{-19mm}{
    \begin{tikzpicture}
	\draw[thick,style={decorate, decoration={snake}}] (-0.5,0)--(0,0);
	\draw[->-=.6,>=Latex][thick] (0,0)--(1.25,1.25); 
	\draw[->-=.6,>=Latex][thick] (1.25,-1.25)--(0,0);
	\draw[->-=.6,>=Latex][thick] (1.25,1.25)--(1.25,-1.25);
	\draw[thick,style={decorate, decoration={snake}}] (1.25,1.25)--(2.5,0); 
	\draw[thick,dashed] (2.5,0)--(1.25,-1.25); 	
	\draw[thick,style={decorate, decoration={snake}}] (2.5,0)--(2.5+0.5,0);
    \draw[thick,dash dot,persianblue][thick] (1.25+1.25/2,1.25)--(1.25+1.25/2,-1.25); 
	\draw[black,fill=tangerineyellow,thick] (0,0)circle(2.75pt); 
	\draw[black,fill=tangerineyellow,thick] (2.5,0)circle(2.75pt); 
	\draw[black,fill=tangerineyellow,thick] (1.25,1.25)circle(2.75pt); 
	\draw[black,fill=tangerineyellow,thick] (1.25,-1.25)circle(2.75pt); 
	\node at (1.25,-1.75) {\scalebox{1}{{\it (d)}}};
	\end{tikzpicture}}
	}
	\hspace{0.3cm}
	\nn \\ & \nn \\
    \boxed{A_2} ~~~~ &
    \scalebox{0.8}{
    \raisebox{-13mm}{
	\begin{tikzpicture}
	\draw[thick,style={decorate, decoration={snake}}] (-0.5,0)--(0,0);
	\draw[thick,style={decorate, decoration={snake}}] (0,0)--(1.25,1.25); 
	\draw[thick,dashed] (1.25,-1.25)--(0,0);
	\draw[thick,dashed] (1.25,1.25)--(1.25,-1.25);
	\draw[thick,style={decorate, decoration={snake}}] (1.25,1.25)--(2.5,0); 
	\draw[thick,dashed] (2.5,0)--(1.25,-1.25); 	
	\draw[thick,style={decorate, decoration={snake}}] (2.5,0)--(2.5+0.5,0);
	\draw[black,fill=tangerineyellow,thick] (0,0)circle(2.75pt); 
	\draw[black,fill=tangerineyellow,thick] (2.5,0)circle(2.75pt); 
	\draw[black,fill=tangerineyellow,thick] (1.25,1.25)circle(2.75pt); 
	\draw[black,fill=tangerineyellow,thick] (1.25,-1.25)circle(2.75pt); 
	\end{tikzpicture}}	
	}\nn \nn
\end{align}
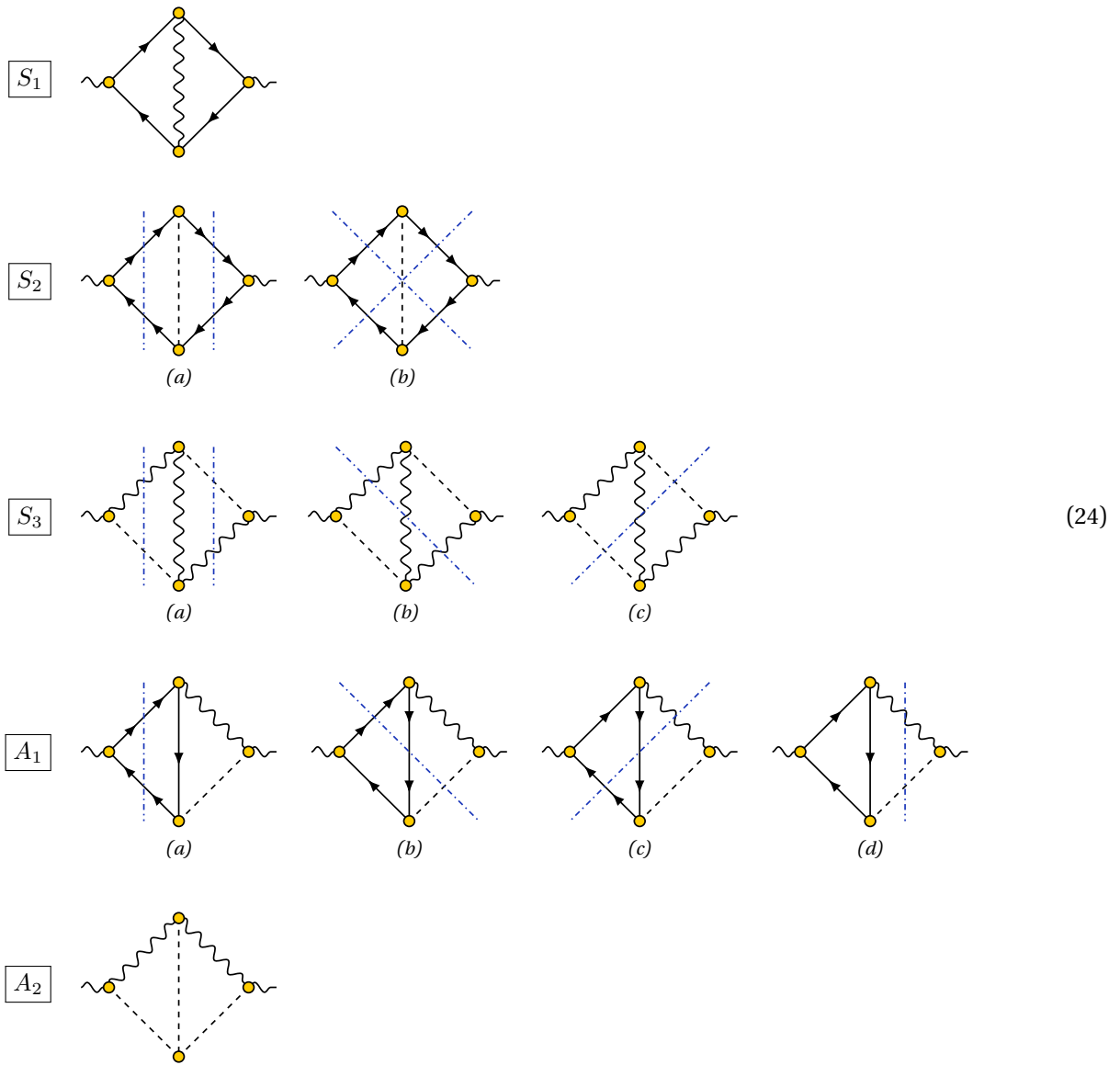
\captionof{figure}{{\em Unitarity cuts contributing to various two-body decay channels (virtual corrections) and three body channels (real emissions).
Dashed lines represent the light Higgs boson $h$ (both real and virtual, depending on the specific diagram considered). 
In virtual corrections, $Z$ internal lines represent the sum over the Feynman $Z$ propagator and the Goldstone $\phi$ propagator. 
In the case of real $Z$ emission, we sum over transverse and longitudinal polarizations. 
We collect all analytical expressions for the various diagrams contributing to each bubble in appendix\,\ref{app:2b}. 
As a side comment, notice that
we named with $S_{i=1,2,3}$ the diagrams $S$ymmetric under reflection over the center of each bubble, while the rest of the diagrams are named $A_{i=1,2}$. 
}}\label{ASJ}\vspace{0.5cm}
	
	We can  obtain a full cancellation by including {\sl all} the cuts present in $A_1$. However, this generates new final states: in this case the $Z h$ final state of $A_1(d)$. In turn, this entails an enlargement in the definition of the observable that forces to include an analogous cut diagram present in $A_2$. Again, in order to have full cancellation one must include all possible $A_2$ cuts and this  forces to include all cut diagrams of $S_3$ in fig.\,\ref{ASJ}. 
	Ultimately, all of the diagrams of fig \ref{ASJ} have to be included and one has to sum over {\it all} possible decay channels in order to cancel power-suppressed double logarithms of IR origin. 
	 {We note that a similar observation was made in ref.\,\cite{Frye:2018xjj} while discussing the possibility to limit the sum implied by the KLN theorem to the initial or final state only. 
	}
	
All the cancellations discussed above can be checked using the expressions collected in appendix\,\ref{app:2b}.
We remark that the explicit expressions of $V_{AB}^{(X)}$ and $R_{AD}^{X}$ corresponding 
to each diagrams in fig.\,\ref{ASJ} were obtained using the Ward identities discussed in 
appendix\,\ref{app:Coupl}. This technical point greatly simplifies the computation, and highlight more clearly  
the pattern of cancellations among virtual and real diagrams.
		
	Let us now start with  another final state, namely $Zh$; in this case there are qualitative differences with respect to the previous case, as we shall see shortly. We define the ``standard'' inclusive observable 
	as $Z'\to Z h X$. The soft particle $X$ can be $\emptyset,Z,h$. By looking at fig.\,\ref{ASJ}, we can write:
	\be
	Z'\to Zh X;X=\emptyset, Z,h\quad\Rightarrow\quad
	 S_3(a)+A_1(d)+\sum_{\rm all\,cuts}A_2=S_3(a)+A_1(d) 
	 \quad\Rightarrow\quad
	 V_{Zh}^{(Z)}+V_{Zh}^{(\psi)}\,.
	 \label{eq:ZhStandard}
	\ee
Notice that in the definition of the ``standard'' inclusive observable we do not include 
the diagram $S_3(c)$ in. fig.\,\ref{ASJ} since the latter features 
two hard Higgs particles in the final state. 	
Coming back to eq.\,(\ref{eq:ZhStandard}), 
we have that the sum over all cuts of $A_2$ diagram vanishes 
but the contributions of $S_3(a)$ and $A_1(d)$ do not. 
In the notation discussed in section\,\ref{sec:Notation}, the contribution  
from the cuts $S_3(a)$ and $A_1(d)$ corresponds, respectively, to 
$V_{Zh}^{(Z)}$ and $V_{Zh}^{(\psi)}$.
What is interesting to notice is that $S_3(a)$ gives rise to an unsuppressed $LL$ contribution:
	\be
	S_3(a)~~~\Longrightarrow~~~ V_{Zh}^{(Z)}=
	-\frac{1}{12} \mathcal{F}_{ZZh}^2 \mathcal{F}_{Z'Zh}^2\left[
	1 + \mathcal{O}(\epsilon^2)
	\right]\,,
	\ee
	which can be read off from eq.\,(\ref{eq:VirSqZh4}). 
	In the above expression $\mathcal{F}_{ZZh}$ and $\mathcal{F}_{Z'Zh}$ refer to the three-linear 
	couplings $Z$-$Z$-$h$ and $Z^{\prime}$-$Z$-$h$ in the mass eigenstate basis, see appendix\,\ref{app:Coupl}.
	It is rather surprising: we have here a ``standard'' inclusive observable in which even IR divergent leading logs are present! In order to understand this, we must recall that, because of the Ward Identities of the theory, a final state longitudinal gauge boson is equivalent to the corresponding Goldstone boson: this is the equivalence theorem \cite{equivalence}. So let's see what happens by looking at fig. \ref{eikonal}, panel (c): A spin 0 Higgs boson transforms into a spin 1 gauge boson along the hard line, emitting a soft gauge boson. Since the spin changes, apparently this is not an eikonal contribution giving rise to double logs, 
	as we have discussed in section\,\ref{sec:MasslessQED}. 
	This is true for transverse polarizations.  However, because of the equivalence theorem, longitudinal $Z$s are equivalent to spin 0 Goldstone bosons, and the spin is unchanged along the hard line, giving riso to eikonal-type double logs. Mechanisms of this type are present both in real and virtual contributions present in the $S_3$ bubble: one has to sum over all possible cuts in order to have an infrared-free result. Then a domino effect similar to the one explained in the $\psi\bar{\psi}$ 
	case starts: because of the presence of diagram $A_1(d)$ one has to include all of the cuts of $A_1$. The observable is then enlarged to include fermions final states and this leads to include all $S_2$ cuts. Ultimately, all available cuts have to be included. 
	
\begin{figure}[!htb!]
\begin{center}
\includegraphics[width=.88\textwidth]{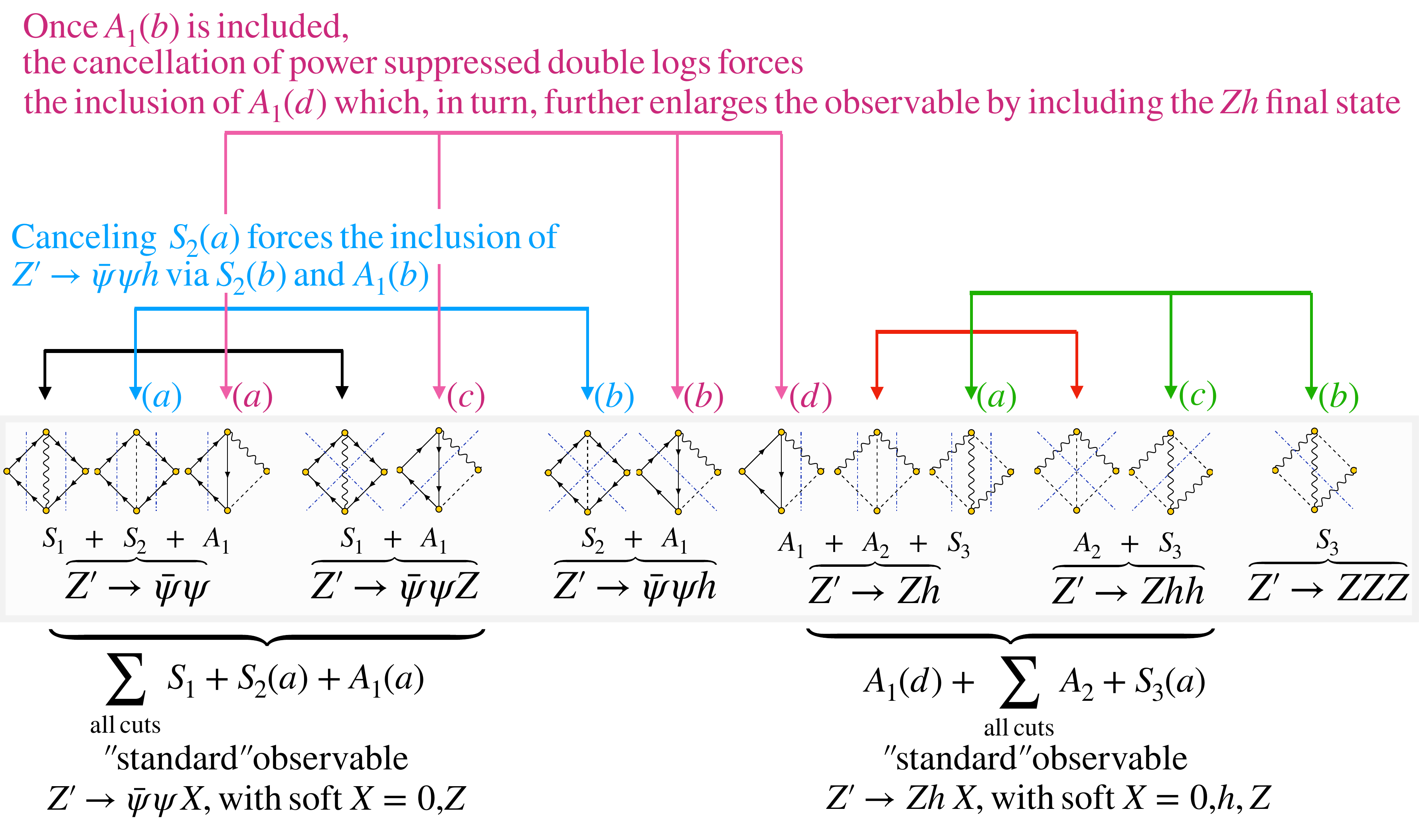}\vspace{-0.15cm}
\caption{\em 
Schematic of the relations between various two- and three-body decay channels describing, respectively, 
virtual and real corrections in the context of the model discussed in section\,\ref{sponte}. 
The colored arrows on the top indicate diagrams that are connected by the IR unitarity theorem (see also 
fig.\,\ref{ASJ}).
}\label{fig:CWInflXi}
\end{center}
\end{figure}

To sum up, we can represent	this situation in the schematic summarized in fig.\,\ref{fig:CWInflXi}.
The central line in this graph is given by the various possible decay channels, and on the top of them there are the various diagrams that contribute (see also fig.\,\ref{ASJ}): for instance $A_1,A_2$ and $S_3$ all contribute to $Z'\to Zh$. The colored lines connect decay channels that share at least one graph:  for instance $Z'\to Z h$ and $Z'\to Zhh$ are connected because $S_3$  contributes to both. This graph makes it  clear that all channels are interconnected and to have full cancellation all of them have to be considered.
We may summarize the situation in this way: 
	
	In order to cancel leading ($LL$) and subleading ($LL_{PS}$) double logs of IR origin, one must define an observable that includes   {\it all} possible final states.
	

\section{Discussion: Degeneracy and cancellation of infrared double logarithms}
\label{kln}

When calculating observables in a given theory, singularities may arise that are not related to the UV behaviour and are instead of IR origin and are related to the  presence of massless particles in the spectrum. This, in turn, entails the existence of degenerate states.
For instance, the states with an electron and an arbitrary number of soft photons are degenerate in energy with the single electron state, in the limit of vanishing photon energies; this singularity is related to the masslessness of the photon. Additionaly, a massless quark accompanied by an arbitrary number of collinear gluons is degenerate with a single quark state even if the gluons energies are not vanishing. In the latter case one speaks about ``collinear'' divergencies; we treat them together with the properly 'infrared' divergencies related with vanishing photon (or gluon) energy and collectively call them infrared divergencies.

The KNL theorem  discussed first in refs.\,\cite{Kinoshita:1962ur,Lee:1964is}  is the keystone of the cancellation of potential infrared singularities mentioned above (see also 
\cite{Muta:1987mz}).
 The essence of this theorem relies on   very general grounds: perturbation theory with degenerate states,  unitarity of the theory.
The theorem states that the transition amplitude squared is finite once we sum over initial and final degenerate states
\be\label{dfsa}
\overline{|S|^2}\equiv
\sum_{\phi_i,\;\phi_f\in {\cal D}(E)} \left|\bra{\phi_f} S\ket{\phi_i}\right |^2<\infty
\ee
where $\phi_i$  and $\phi_f$ are initial and final states (eigenstates of the free Hamiltonian), and the ensemble ${\cal D}(E)$
is given by all the states degenerate in \underline{\it energy} in a $\Delta>0$ range , i.e. $|E_{i/f}-E|<\Delta$. 
The demonstration is done in perturbation theory using the interaction picture and  relies on the use of old fashioned time-ordered perturbation theory  with a generic Hamiltonian $H= H_0+ V$ (with $H_0$ the free Hamiltonian and $ V$ the interaction perturbation) \cite{Muta:1987mz}. 
 At first order  in perturbation theory,   such  divergences manifest themselves through the presence of 
 single and double logs whose argument is a ratio of scales, the heavy hard scale of the process $Q$ and a light cut off scale $m$: $\log^n(Q^2/m^2)$, with $n=1,2$. 
The scale $m$ might be a cutoff scale introduced to give meaning to divergent integrals: this is the case of a fictitious mass for the photon in QED. 
 In other cases, in presence of light massive particles with physical mass $m\ll Q$, the cut off is provided by the mass $m$ itself, so that the truly IR divergences become a ``would be'' 
 IR divergences whose phenomenological importance can be relevant\,\cite{cc1},\cite{BNV}.
  In general,  a  { double} averaging, over the initial states $\phi_i$ and the final state $\phi_f$ is necessary for the complete IR safeness of the S matrix (see the comments in \cite{Contopanagos:1991yb}
   for the initial coherent state structure suggested by the KLN theorem).
%
In this  paper, however,  we   concentrate uniquely on the final state sum, analysing  processes that do not need the initial averaging sum. For this purpose, we consider an initial state given by a neutral heavy particle  $\Phi$  (in this case the hard scale $Q$ is given by the heavy mass of $\Phi$)  that  decays in all possible allowed light states.
On top of  the  leading IR divergences, 
$LL\equiv \log^2(Q^2/m^2)$   there are also similar corrections that give Power Suppressed infrared divergences of the form 
$LL_{PS}\equiv (m^2/Q^2)\,\log^2(Q^2/m^2)$, that also come  from the   presence of almost degenerate particles in the spectrum.
Notice that the leading double log terms come from states that are both infrared and collinear, each type of singularity giving rise to a logarithm. The $LL_{PS}$ need not be  true divergences when   we take the limit $m\to 0$ but 
there are cases where the different mass composition of the light particles allows IR contribution 
of the form $(m_1^2/Q^2)\,\log^2(Q^2/m_2^2)$ where $m_1\neq m_2$ and the  limit $m_2\to 0$ generates a true divergence.

The  most general structure of  the degenerate ensemble  ${\cal D}(E)$, 
  related to a process initiated by a neutral heavy particle  $\Phi$    that  decays in all possible allowed light states, can be summarized as follows:
\be\label{DE}
\sum_{{\rm Final\,states}\,\phi_f\,{\rm with}\,|E_f-E|<\Delta}\left|\bra{\phi_f} S\ket{\Phi}\right |^2\;
=\sum_{{\rm Kinematics}}\quad
 \sum_{{\rm Quantum\,Numbers}} \quad 
\sum_{{\rm Channels}}\quad
\left|\bra{\phi_f} S\ket{\Phi}\right |^2\,.
\ee
We now give a   description of the final states involved in the sum in eq. (\ref{DE}). First, we have two body final states that we denote by
$\bra{\bar{p} p}$ where $p$ is a generic particle. These two body channel define the energy $E$ appearing in  ${\cal D}(E)$ ensemble that contains all the   states degenerate in Energy with the two hard final partons; namely, $E={Q}$ where $Q$ is the $\Phi$ mass. One loop corrections to the two body decay are included. 
Then we have three body states, summed over three conceptually different substates: kinematically degenerate states, quantum number degenerate states and all possible different channels. 
The first sum in eq.\,(\ref{DE}) is related to kinematical degeneracy: 
a hard particle state $A$ with three momentum $\boldsymbol{p}+\boldsymbol{k}$ is degenerate with the state formed by two particles $B,X$ with momenta $\boldsymbol{p}$
and $\boldsymbol{k}$ only when one of the two particles has vanishing energy and/or the two particles are collinear. 

To be more definite,
we denote as  $\ket{A(p+k)B(p),\,X(k)}$ such a state, where the first two states $p_{A,B}$ 
are hard  partons while the third is soft (infrared+collinear) (see also fig.(\ref{eikonal})).  Note that the ordering of the three body states is important:
 for example $\ket{ \bar q\,q,\,Z}$ and $\ket{\bar q\,Z,\,q}$ provide two different configurations where the  hard states are the first two particles while the third one is the IR/collinear one. We think of $k$ as a discrete variable over which we sum; the continuum limit can be performed as a final step. This sum   is related to infrared and collinear singularities (generally called as IR) that show up as double   logs both leading, $LL\sim \log^2(Q^2/m^2)$, and power suppressed, 
 $LL_{PS}\sim (m^2/Q^2)\log^2(Q^2/m^2)$. 
 We give now an example where the kinematical sum is sufficient to eliminate the infrared singularities, by considering in QED the decay into charged particles $p$ that interact with massless photons $\gamma$:
\bea
{\rm Kinematical\; Sum}:\left\{\begin{array}{l}
  \left|\bra{\bar q \,q} S\ket{\Phi}\right|^2\to
	LL+LL_{PS}\\
 \\
 \sum\limits_k \left|\bra{\bar q \,q,\,\gamma(k)} S\ket{\Phi}\right|^2 \to  LL+ \;LL_{PS},\\
  \\
  \left|\bra{\bar q \,q} S\ket{\Phi}\right|^2+\sum_k \left|\bra{\bar q \,q,\,\gamma(k)} S\ket{\Phi}\right|^2\to  NLL,\\
 \end{array}
 \right.\label{eq:KinSum}
\eea
In eq.\,(\ref{eq:KinSum}) $NLL$ 
stands for next to leading log, meaning either a single log or no log at all. Indeed in QED, as we have seen in section \ref{QEDmassive}, including final (soft) photons is necessary and sufficient in order to obtain an observable free from double logs.

  The next sum appearing in eq.\,(\ref{DE}) is over quantum numbers. This is  the case in the SM for color  quantum numbers for instance, or isospin. Collectively indicating with $\alpha$ the hard particles quantum numbers and with $a$ the gauge bosons quantum numbers, we have:
 \bea
\mbox{Quantum Numbers Sum}:\left\{\begin{array}{l}
  \left|\bra{\bar q_\alpha \,q_\beta} S\ket{\Phi}\right|^2\to
	LL+LL_{PS}\\
 \\
 \left|\bra{\bar q_\alpha \,q_\beta} S\ket{\Phi}\right|^2+\sum\limits_{k,a} \left|\bra{\bar q_\alpha \,q_\beta,\, g_a(k)} S\ket{\Phi}\right|^2 \to  LL+\;LL_{PS},\\
  \\
  \sum\limits_{\alpha,\beta}\left|\bra{\bar q_\alpha \,q_\beta} S\ket{\Phi}\right|^2+\sum\limits_{k,a,\alpha,\beta} \left|\bra{\bar q_\alpha \,q_\beta,\, g_a(k)} S\ket{\Phi}\right|^2\to  NLL,\\
 \end{array}
 \right.
\eea 
 Here, to cancel all the $LL$ logs, we need to sum over all the colours of the quarks and over the emission of IR gluons of all possible colours (note that, at  single IR log level, there are non cancelling power suppressed IR divergences in QCD \cite{Doria:1980ak}
 ).
  Without this sum the decay process is IR divergent\,\cite{Ito:1981}.
  The necessity of summing over final particels' quantum numbers is related to the fact that states in the same kinematical configuration but with different quantum numbers are of course degenerate, and therefore belong to the same set ${\cal D}(E)$ appearing in 
  eq.\,(\ref{dfsa}).
A similar effects is present also for $SU(2)_L$ multiples with the  \underline{Isospin sum}. The difference is that the gauge group is spontaneously broken and the gauge mediators are massive  
(see\,\cite{cc1, BNV}). 
Notice that  summing over isospin quantum numbers entails summing  over particles of different nature connected by isospin, like electron and neutrino for instance (see \cite{Ciafaloni:2010ti} 
for the relevance of the above effects in a cosmological setting).

  Finally the    \underline{Channels sum} is the   new feature  that we want to discuss in this paper.
The example that we give corresponds to the description of the toy model of chapter \ref{sponte}, a spontaneously broken 
$U(1)$ model.  At tree level, the heavy $Z^{\prime}$ 
can   decay  into two light states: a fermion-antifermion state ($Z'\to
\ket{\bar q\,q}
$) and a light Z gauge boson-light higgs state ($Z'\to \ket{ Z\,h} $). We assume also that the fermion $q$ interact with the previous gauge boson 
 Then we have the following set of possible, degenerate in energy, final states.
 \begin{align}
{\rm Channel\,sum}:&\left\{
\begin{array}{l}
 \left.
 \begin{array}{l} 
  \left.\begin{array}{l} 
 \left|\bra{ \bar{q} q}S\ket{\Phi}\right|^2 \to LL\, +\, LL_{PS}\\
 \\
  \sum\limits_{X = {Z,h}}\,\sum\limits_k\left|\bra{ q\,\bar{q},\,X(k) }S\ket{Z'}\right|^2 \to  LL\, +\, LL_{PS}\end{array}
 \right\} \hspace{0.2cm} \to LL_{PS}
   \\
 \\
  \sum\limits_{X = {Z,h}}\,\sum\limits_{k }\left( 	 \left|\bra{ X\,\bar{q},\,q(k) }S\ket{\Phi}\right|^2
	+
	 \left|\bra{ X\,{q},\,\bar{q}(k) }S\ket{\Phi}\right|^2\right)
	\to \, LL_{PS}
  \end{array}
 \right\} \hspace{0.2cm} \to LL_{PS}
 \\
 \\
 \left.
 \begin{array}{l} \left.
 \begin{array}{l}  \left|\bra{Zh}S\ket{Z'}\right|^2 \to LL\, +\, LL_{PS}\\
 \\
     \sum\limits_k\left|\bra{Zh,\,h(k) }S\ket{\Phi}\right|^2 \to   LL_{PS}
	\end{array}
 \right\} \hspace{0.2cm} \to LL\, +\,LL_{PS}
	\\
 \\
	 \sum\limits_{k }\left(\left|\bra{ Z\,Z,\,Z(k) }S\ket{\Phi}\right|^2
	+
	 \left|\bra{ h\,h,\,Z(k) }S\ket{\Phi}\right|^2\right)
	\to LL\, +\,\, LL_{PS}
 \end{array}
 \right\} \hspace{0.2cm} \to LL_{PS}
 \end{array}
 \right\} \hspace{0.3cm}\to\;NLL\nn\\
 &
\end{align}
Note that the $\ket{\bar q\,q}$ state contains the virtual ($t$-channel) $Z$-$h$ corrections.

The cancellation of the leading $LL$ IR divergences for ``anomalous'' channel $Z'\to Zh$ requires the introduction of the three body decays $Z'\to hh,\,Z({\rm soft})$ and $Z'\to ZZ,\,Z({\rm soft})$ where we see a change for the hard state from $Zh $ to $hh$ and $ZZ$. This can be understood in terms of what we have described in section\,\ref{sponte}: 
the longitudinal component of a $Z$ is in reality a Goldstone boson $\phi$.
This phenomenon  was already discovered in \,\cite{g/h} where the resummation of the initial state emission from boson gauge fusion were performed. As we have noticed in section\,\ref{sponte},
when we look carefully to the $Z'\to Zh$ decay process we discover that the leading logs come from the process $Z'\to \phi h$ where   the longitudinal part of the $Z$ gauge boson   is playing the leading role.
The interaction in between the light scalar sector is ``non diagonal'' in the sense that  the eikonal interaction induced by a $Z$-soft emission mixes different hypercharge states  in the scalar  $\phi$-$h$ system:
$(\phi\;\partial_\mu h-h\;\partial_\mu \phi)\;Z^\mu$.
 The emission of a soft $Z$ is flipping a goldstone $\phi$ into an higgs $h$ or viceversa.
The KLN cancellation for this channel is then explicable with the ``Quantum Number Sum'' in the sense that substituting $Zh\to \phi h $ and applying the sum over the $\phi$-$h$ quantum numbers, we automatically get
\bea
 |\bra{\phi  h}|S \ket{\Phi}|^2+\sum_k (|\bra{h h,\,Z(k)}|S \ket{\Phi}|^2+
|\bra{\phi\phi,\,Z(k)}|S \ket{\Phi}|^2 )\to LL_{PS}
\eea
that translated into the physical massive gauge boson ($\phi\to Z$) becomes
\bea
 |\bra{Z  h}|S \ket{\Phi}|^2+\sum_k \left(|\bra{h h,\,Z(k)}|S \ket{\Phi}|^2+
|\bra{ZZ,\,Z(k)}|S \ket{\Phi}|^2 \right)\to LL_{PS}\,.
\eea
With the emission of a soft transverse Z gauge boson  we generate the $LL$ corrections and, as well known, the only change in the final state induced by such a process is a quantum number change (eikonal emission is blind for what concern the spin of the hard emitting state). Once we average over the appropriate quantum numbers, LL corrections are cancelled. Then it remains
   the sum over all the other possible decay channels that cancel the higher twist double log corrections.
The above sum include hard states of different spin (here fermion and bosons) and it is because of this feature that we named the sum as "channel sum".
The two particle hard state that contains the double logs (here $\ket{\bar q\,q}$ and $\ket{Z\,h}$) define the Leading IR subsystem of  ${\cal D}(E)$ that we call   ${\cal D}(E)_{LL}$ 
composed by the hard state $\ket{\bar q\,q}$ plus  $\ket{\bar q\,q,\;Z(h)}$ containing a soft $Z/h$  boson
(or $\ket{Z\,h }$ plus  $\ket{h\,h,\;Z}$ and $\ket{Z\,Z,\;Z}$ containing a soft $Z$  boson). 
Note that in this case the hard states retain their intrinsic spin structure 
(the real amplitudes for  $\ket{Z\,h }$ are $\ket{\phi\,h }$plus  $\ket{h\,h,\;Z}$ and $\ket{\phi\,\phi\phi,\;Z}$ containing a soft $Z$  boson) (for non abelian interactions the hard states can change colour/isospin).
The rest of the degenerate states  allowed by the interactions will belong to the remaining subset of  ${\cal D}(E)$ that we call   ${\cal D}(E)_{LL_{PS}}$. 
Such a subset
 ${\cal D}(E)_{LL_{PS}}$
(in our simplified case consist of the states $ \ket{Z\,q,\;\bar q},\; \ket{Z\,\bar q,\; q}$ and $ \ket{Z\,h,\;h} $   ) contains as  hard particles both bosons and also   mixed spin states, boson plus fermions, and it's the   
ensemble needed  to cancel the higher twist double logs.
Very schematically we can write the following decomposition
\be
{\cal D}(E)={\cal D}(E)_{LL_{PS}}\cup{\cal D}(E)_{LL}
\;\;{\rm where}\;\; \left\{
\begin{array}{l}
{\cal D}(E)_{LL}=\{\ket{\bar q\,q}+\ket{\bar q\,q,\;Z}+\ket{\bar q\,q,\;h}\}+\{\ket{Z\,h }+\ket{h\,h,\;Z}+\ket{Z\,Z,\;Z}\}\\
\\
{\cal D}(E)_{LL_{PS}}=\{\ket{Z\,q,\;\bar q}+ \ket{Z\,\bar q,\; q}+ \ket{Z\,h,\; h}    \}
\end{array}\right.
\ee
where the sum over each single ensemble at the level of amplitude squared   generate the following pattern of double logs cancellations
\bea
{\rm No\; average:}&&  ~~~~\, \hspace{1.cm} |S|^2\to\alpha\;\log^2\!Q^2 +\alpha\;\frac{m^2}{Q^2}\;\log^2\!Q^2 \label{eq:Sum1}\\\nonumber
\\
{\rm Average \,over\,{\cal D}(E)_{LL }:}&& ~~~~~\, \sum_{ {\cal D}(E)_{LL }} \hspace{-0.2cm} |S|^2\to \alpha+\alpha\;\frac{m^2}{Q^2}\;\log^2\!Q^2 \label{eq:Sum2}\\\nonumber
 \\
{\rm Average \,over\,{\cal D}(E)_{LL }+{\cal D}(E)_{LL_{PS}}:} && \sum_{\tiny\begin{array}{l} {\cal D}(E)_{LL }+\\
  {\cal D}(E)_{LL_{PS}}
  \end{array}
  } \hspace{-0.2cm} |S|^2\to \alpha\label{eq:Sum3}
\eea
{To be crystal-clear, let us stress that eqs.\,(\ref{eq:Sum1}-\ref{eq:Sum3}) are valid up to $NLL$ terms which are set to zero here for ease of reading. 
Furthermore, $\alpha$ is a shorthand notation 
that replaces a one-loop suppression factor of the kind $\mathrm{g}^2/4\pi^2$ with the appropriate coupling 
$\mathrm{g} = g,g^{\prime},\dots$ that depends on the specific interaction involved.}

\section{Conclusions}\label{sec:Conclusions}
In this paper we have considered  corrections of infrared origin in a spontaneously broken $U'(1) \otimes U(1)$ gauge theory. The model we have chosen  includes a heavy $Z'$ with mass $Q$ and a spectrum of light particles $i$ with mass $m_i=\epsilon_i Q$, $\epsilon_i\ll 1$. We have calculated one loop virtual corrections that scale like
\begin{align} 
\frac{1}{(Q^2)^k}\log^2 Q^2\,,~~~~~~ k=0,1,\dots\,,
\end{align}
 along with the relevant real emission counterparts. 
  The terms with $k=0$ are the leading contributions, while   $k\ge 1$ corresponds to the power suppressed terms we focus our interest in; our calculations are exact,
 meaning that  they include contributions for all values of $k$. The main result of our work is the following. Consider the decay of the heavy particle in a given channel, for instance $Z'\to f\bar{f}$; this observable features corrections of infrared origin of the type discussed above. In the standard treatment, one deals with such infrared corrections by defining an inclusive observable obtained by adding the emission of a light particle  $X$. However this standard treatment fails in our case, since $\Gamma(Z' \to f \bar{f})$ +
$\sum_X\Gamma(Z' \to f \bar{f}X)$ still features uncancelled double logs of infrared origin. Our detailed analysis then shows that the only way of cancelling all leading and power suppressed terms is to include {\it all} possible   decay channels, and this must include channels very different from the starting one like $Z'\to Z h$, $Z$ being the light
 $U(1)$ gauge boson and $h$ the light Higgs boson.

The above  IR phenomenon is at work only for  specific features  of the fermionic sector. 
Specifically, all the following   requirements are necessary, at the same time (see appendix\,\ref{app:Coupl} and table\,\ref{Limit}):
\begin{itemize}
\item $\epsilon_{\psi}\neq0$, massive fermions.
\item $y_{\phi}=y_R-y_L\neq0$, chiral  fermions under $ U(1)$ gauge group.
\item $f_{\phi}=f_R-f_L\neq0$, chiral  fermions under $ U'(1)$ gauge group.
\end{itemize}  
In the absence of only one of the above features the ``standard'' KLN cancellation mechanism
  works also for $LL_{PS}$ corrections. 
In the SM we can get a sort of parallelism with our toy model as soon as:
the physical $Z$ gauge boson can simulate our hypothetical  $Z’$ ( and in this case $U'(1)$ is generated by 
the combination $T_3-s_W^2Q$ of weak isospin and the charge operators), the photon $\gamma$ 
can simulate our hypothetical $Z$ ($U(1)$ in this case is generated by the charge operator $Q$) and the fermion sector is given by the massive electron field.
As you see, in SM we miss only the last requirement necessary to have the full exploitation of the KNL theorem: 
$U(1)$ electric gauge group is  vector-like.
So, the only possible phenomenological models  where the above three requirements can be present is in a new physical scenario where an extra heavy 
$Z^{\prime}$ gauge boson can be produced at future colliders.
\\
Several technical results have been achieved as a byproduct of the calculations performed here. In first place, we give an expression for the scalar function $C_0$,
 which is relevant for one loop calculations in any theory. This expression holds at the double-logarithmic level and is valid for  power suppressed terms at all orders. Secondly, we give a recipe for calculating power suppressed terms in {\it any} theory,  by defining suitable substitutions that isolate such terms. Thirdly, we demonstrate a theorem, that we call the ``{\it IR unitarity theorem}'' that states that by summing over all cuts of a given self-energy diagram we obtain exactly zero. The surprising feature is that this is true not only for terms that are IR divergent when a mass $\lambda$ goes to zero and a mass $m$ is kept finite,
like $(m^2/Q^2)\log^2(Q^2/\lambda^2)$, but also for IR finite terms like $(\lambda^2/Q^2)\log^2(Q^2/\lambda^2)$. Finally, we derive the Ward identities for the theory after mixing is taken into account, i.e. in the mass eigenstate basis.

\begin{acknowledgments}
The research of A.U. was supported in part by the
MIUR under contract 2017FMJFMW  (``{New Avenues in Strong Dynamics},'' PRIN\,2017) 
while the research of D.C.  
was supported in part by Grant No. 2017X7X85K
``The dark universe: A synergic multimessenger approach'' under the program PRIN 2017 funded by Ministero dell’Istruzione, Universit\`a e della Ricerca (MIUR).
\end{acknowledgments}

\appendix

\section{Mass spectrum, Ward Identities and Feynman rules.}
\label{app:Coupl}

The Lagrangian (\ref{scalar}) in the scalar sector can be written in a matricial formalism which is useful to derive the Ward Identities of the theory. For this purpose, we collect the two gauge fields $Z_\mu,Z'_\mu$ in one doublet 
$\boldsymbol{Z}^T \equiv(Z,\,Z')$ and we do the same for the scalars
 $\boldsymbol{\varphi}^T = (\varphi,\,\varphi')$. 
 The hypercharge operators become the $2\times2$ diagonal matrices 
$\boldsymbol{Y}={\rm diag}(y_\phi,\,y_{\phi'}=0)$ and  
$\boldsymbol{F}={\rm diag}(f_\phi,\,f_{\phi'})$. 
Additionaly we have
$\boldsymbol{\varphi}=(\braket{\boldsymbol{\varphi}}+\boldsymbol{h}+i\,\boldsymbol{\phi})/\sqrt{2}$. Then:
\be\label{xcvb}
{\cal L}_s=(\boldsymbol{D}_\mu\boldsymbol{\varphi})^\dagger
\,(\boldsymbol{D}_\mu\boldsymbol{\varphi})\,\quad\quad
\boldsymbol{D}_\mu \equiv \partial_\mu+i\, g \,Z_\mu \boldsymbol{Y}
+i\,g'\,Z'_\mu \boldsymbol{F}\,.
\ee
The gauge fixing Lagrangian is tailored to cancel the $\phi$-$Z$ mixing terms, which are given by
\be
|(\partial_\mu+i\,g\, Z_\mu \boldsymbol{Y}
+i\,g'\,Z'_\mu \boldsymbol{F})\,\boldsymbol{\varphi}|^2~~~\rightarrow~~~ 
\frac{1}{2}\,
\braket{\boldsymbol{\varphi}}^\dagger(i\,g \,Z_\mu \,\boldsymbol{Y}
+i\,g'\,Z'_\mu\, \boldsymbol{F})\partial_\mu\boldsymbol{\phi}+h.c.
\ee
and are canceled by 
\be\label{cessacciu}
{\cal L}_{gf}=\frac{1}{2}\,|\de^\mu Z_\mu-g\braket{\boldsymbol{\varphi}}^\dagger \boldsymbol{Y} \boldsymbol{\phi}|^2+\frac{1}{2}\,
|\de^\mu Z'_\mu-g'\braket{\boldsymbol{\varphi}}^\dagger \boldsymbol{F}
\boldsymbol{\phi}|^2=\frac{1}{2}\,|\de^\mu \boldsymbol{Z}_\mu- \boldsymbol{W}\,\boldsymbol{\phi}|^2
\ee
having introduced the matrix
\be \boldsymbol{W}=
\begin{pmatrix}
g\,\braket{\boldsymbol{\varphi}}^\dagger \,\boldsymbol{Y}
\\
g'\,\braket{\boldsymbol{\varphi}}^\dagger \,\boldsymbol{F}
\end{pmatrix}
=
\begin{pmatrix}
g\,v\,y_\phi  &0
\\
g' \,v\, f_{\phi} & g'\,v'\,f_{\phi'}
\end{pmatrix}
\ee

From eq. (\ref{xcvb}) it is straightforward to obtain that the mass terms for gauge bosons are given by 
$\boldsymbol{Z}^T\boldsymbol{W}
\boldsymbol{W}^T \boldsymbol{Z}$ while eq. (\ref{cessacciu}) generates a term $\boldsymbol{\varphi}^T\boldsymbol{W}^T
\boldsymbol{W} \boldsymbol{\varphi}$. Since the mass matrix in the $Z$-$Z'$ sector $\boldsymbol{W}\boldsymbol{W}^T$ and the mass matrix in the 
$\varphi$-$\varphi'$ sector $\boldsymbol{W}^T\boldsymbol{W}$ 
have same trace and same determinant, the eigenvalues i.e. the physical masses are the same, and they are given by
\bea\label{massesMQ}
M^2/Q^2=\frac{1}{2} \left\{ g'^2 \left(v^2\, f_{\phi }^2+v'^2 \,f_{\phi '}^2\right)+g^2 \,v^2\,
   y_{\phi }^2\mp\sqrt{\left[ g'^2 \left(v^2 \,f_{\phi }^2+v'^2\, f_{\phi
   '}^2\right)+g^2\, v^2 \,y_{\phi }^2\right]^2-4\, g^2\, v^2\, v'^2 \, g'^2\, y_{\phi }^2
  \, f_{\phi '}^2}\right\}\,.
\eea
Remember that we are working in the  hierarchical regime $Q\gg M$ obtained in the limit $g'\,f_{\phi'}\,v'\gg g\,y_{\phi'}\,v$.
Let us now come to the Ward Identities, which can be read off directly from the gauge fixing term   (\ref{cessacciu}). When the operator 
$\de^\mu \boldsymbol{Z}_\mu- \boldsymbol{W}\boldsymbol{\phi}$ is applied to a given Green function, the result is 0; this relates Green functions with an external gauge boson and an arbitrary number of remaining external legs to Green functions with Goldston bosons and the same remaining external legs. This holds in the gauge basis, what happens in the mass eigenstate basis? Let us first notice that any (real) matrix can be asymmetrically  diagonalized with two unitary matrices $\boldsymbol{U}_Z,\boldsymbol{U}_\varphi$ in the following way:
\be\label{best}
\boldsymbol{D}=\boldsymbol{U}_Z^T \boldsymbol{W}\boldsymbol{U}_\varphi
\ee
where $\boldsymbol{D}$ is a diagonal matrix. Let us now show that $\boldsymbol{U}_Z$ and $\boldsymbol{U}_\phi$ are precisely the unitary matrices that diagonalize the mass matrices
through the transformations $\boldsymbol{Z}=
\boldsymbol{U}_Z\boldsymbol{Z}_m$, $\boldsymbol{\varphi}=
\boldsymbol{U}_\varphi\boldsymbol{\varphi}_m$. Indeed, using eq. (\ref{best}) we have:
\be
\boldsymbol{Z}^T\boldsymbol{W}\boldsymbol{W}^T \boldsymbol{Z}=
\boldsymbol{Z}_m^T\boldsymbol{U}_Z^T \boldsymbol{W}\boldsymbol{U}_\varphi
(\boldsymbol{U}_Z^T \boldsymbol{W}\boldsymbol{U}_\varphi)^T\boldsymbol{Z}_m
=\boldsymbol{Z}_m^T \boldsymbol{D}^2\boldsymbol{Z}_m;\quad
\boldsymbol{\varphi}^T\boldsymbol{W}^T\boldsymbol{W} \boldsymbol{\varphi}=
\boldsymbol{\varphi}_m^T \boldsymbol{D}^2\boldsymbol{\varphi}_m
\ee
The diagonalisation procedure allows to write the unitary matrices in the following form:
\be
\boldsymbol{U}_Z=\exp[i \, \theta \, \boldsymbol{\sigma}_2]\,,\quad
\tan(2\theta )=\frac{2\,g\,g'\, f_\phi \,y_\phi \,v^2}{g^{\prime\,2}(v^2\,f_{\phi}^2 + v^{\prime\,2}\,f_{\phi^{\prime}}^2)-g^2 \,v^2\,y_{\phi}^2}
\ee
\be
\boldsymbol{U}_\varphi=\exp[i\,\beta\,\boldsymbol{\sigma}_2]\,,\quad
\tan(2\beta)=\frac{2\,g^{\prime\,2}\,f_{\phi}\,f_{\phi^{\prime}}\,v\,v^{\prime}}{
g^{\prime\,2}\,f_{\phi^{\prime}}^2 \,v^{\prime\,2}-(g^2\,y_{\phi}^2 + g^{\prime\,2}\,f_{\phi}^2)v^2}
\ee
The Ward Identities (WI) in the mass eigenstates base are now easily written
by using eq.\,(\ref{best})
\be
\de_\mu \boldsymbol{Z}^\mu=\boldsymbol{W}\boldsymbol{\varphi}~~~~~~\Longrightarrow~~~~~~
\de_\mu \boldsymbol{Z}^\mu_m=\boldsymbol{D}\boldsymbol{\varphi}_m\,,\qquad
\boldsymbol{D}={\rm diag}(M,Q)\,.
\ee
Therefore WI acquire, in the mass eigenstates base, a very simple form, as follows. Take an amplitude with an external light gauge boson ($Z_m$) and make the substitution $\epsilon_\mu(k)\to ik_\mu/M$; this amplitude is equal to the one obtained by substituting  the light gauge boson with the corresponding light Goldstone $\phi_m$. The same holds for heavy states with $M\to Q$.
\\
{\bf Note that from now on we drop the index $m$ in $Z_m,\varphi_m$ so in our graphs and Feynman rules by $Z$ we mean the light gauge boson mass eigenstate, and so on.}
\\
We define the dimensionless parameter $\epsilon \equiv M/Q \ll 1$, and we use it as an expansion parameter. 
For instance, we find (notice that such expansion holds only for $y_\phi\neq 0$)
\begin{align}
\cos\theta & = 1 - \frac{g^{\prime\,2}\, f_{\phi}^2 \,\epsilon^4}{2\,g^2\,y_{\phi}^2} + O(\epsilon^4)\,,\quad
\sin\theta  = -\frac{g^{\prime}\,f_{\phi}\,\epsilon^2}{g \,y_{\phi}} + O(\epsilon^4)\,,\\
\cos\beta & = 1-\frac{g^{\prime\,2}\,f_{\phi}^2\,\epsilon^2}{2\,g^2\,y_{\phi}^2} + O(\epsilon^4)\,,\quad
\sin\beta   = -
\frac{g^{\prime}\,f_{\phi}\,\epsilon}{g\, y_{\phi}} + 
+ O(\epsilon^3)\,.\label{eq:IdeMarch}
\end{align}
As far as the masses of the goldstones are concerned, we find the exact expressions $m_{\phi}^2 = M^2$ and $m_{\phi^{\prime}}^2 = Q^2$.
We now move to consider cubic interactions in the mass eigenstates basis. 
We will write both the exact expressions for the interaction vertices and their expansion in $\epsilon$. The vertices listed below are only those necessary for
calculating  the quantities of relevance in this paper.
\\
Gauge boson/fermion interactions.
  \begin{align}\label{eq:Fey1}
  	\raisebox{-14.2mm}{
	\begin{tikzpicture}
	\draw[->-=.65,>=Latex][thick] (0,0)--(0.75,+1.5);
	\draw[-<-=.65,>=Latex][thick] (0,0)--(0.75,-1.5); 
	\draw [thick,style={decorate, decoration={snake}}] (-1.5,0)--(0,0); 
	\draw[black,fill=tangerineyellow,thick] (0,0)circle(6pt);
    \node at (-1,0.5) {\scalebox{1}{$a^{\mu}$}};
    \node at (1,-0.85) {\scalebox{1}{$\psi_{b}$}};
    \node at (1,+0.85) {\scalebox{1}{$\bar{\psi_{b}}$}};	   
	\end{tikzpicture}} \hspace{0.5cm} =\hspace{0.5cm} \gamma^{\mu} \left\{
\begin{array}{ccc}
 -i\,(g\,y_L\,c_{\theta} + g^{\prime}\,f_L\,s_{\theta}) \simeq 
 -i\,g\,y_L\,\left(1 - \frac{g^{\prime\,2}\,f_L\,f_{\phi}\,\epsilon^2}{g^2\,y_L\, y_{\phi}}\right) \equiv -i\,\mathcal{F}_{ZL} 
     & & \hspace{0.5cm}\{a,b\} = \{Z,L\}   \\
    & &  \\
  -i\,(g\,y_R\,c_{\theta} + g^{\prime}\,f_R\,s_{\theta}) \simeq -i\,g\,y_R\,\left(1 - \frac{g^{\prime\,2}\,f_R\,f_{\phi}\,\epsilon^2}{g^2\,y_R\, y_{\phi}}\right) \equiv -i\,\mathcal{F}_{ZR} & & \hspace{0.5cm}\{a,b\} = \{Z,R\} \\
    & &  \\
  -i\,(g^{\prime}\,f_L\,c_{\theta} - g\,y_L\,s_{\theta}) \simeq -i\,g^{\prime}\,f_L\,\left(1 + \frac{y_L \,f_{\phi}\,\epsilon^2}{f_L\, y_{\phi}}\right) \equiv -i\,\mathcal{F}_{Z'L} & & \hspace{0.5cm}\{a,b\} = \{Z^{\prime},L\}\\
    & &  \\
  -i\,(g^{\prime} \,f_R\,c_{\theta} - g\,y_R\,s_{\theta}) \simeq
  -i\,g^{\prime}\,f_R\,\left(1 + \frac{y_R\, f_{\phi}\,\epsilon^2}{f_R\, y_{\phi}}\right) \equiv -i\mathcal{F}_{Z'R} & & \hspace{0.5cm}\{a,b\} = \{Z^{\prime},R\}
\end{array}
	\right.
\end{align}
Scalar/fermion interactions.
  \begin{align}\label{eq:Fey2}
  	\raisebox{-14.2mm}{
	\begin{tikzpicture}
	\draw[->-=.65,>=Latex][thick] (0,0)--(0.75,+1.5);
	\draw[-<-=.65,>=Latex][thick] (0,0)--(0.75,-1.5); 
	\draw [thick,dashed] (-1.5,0)--(0,0); 
	\draw[black,fill=tangerineyellow,thick] (0,0)circle(6pt);
    \node at (-1.2,0.35) {\scalebox{1}{$s$}};
    \node at (1,-0.85) {\scalebox{1}{$\psi$}};
    \node at (1,+0.85) {\scalebox{1}{$\bar{\psi}$}};	   
	\end{tikzpicture}}
	\hspace{0.5cm} =\hspace{0.5cm}\left\{
\begin{array}{ccc}
   -i\,h_f/\sqrt{2} \equiv -i\,\mathcal{F}_{h\psi} & & \hspace{0.5cm} s = h  \\
    & &  \\
   -h_f\,c_{\beta} \,\gamma^5/\sqrt{2} \simeq -\frac{h_f}{\sqrt{2}}\bigg(1-\frac{g^{\prime\,2}f_{\phi}^2\,\epsilon^2}{2\,g^2\,y_{\phi}^2}\bigg)\,\gamma^5 \equiv \mathcal{F}_{\phi\psi}\gamma^5 & & \hspace{0.5cm} s = \phi 
\end{array}
	\right.
\end{align}
\\
Gauge boson/scalar interactions.
  \begin{align}\label{eq:Fey3}	
  \small
  	\raisebox{-14.2mm}{
	\begin{tikzpicture}
	\draw[->-=.65,>=Latex][thick,dashed] (0,0)--(0.75,+1.5);
	\draw[-<-=.65,>=Latex][thick,dashed] (0,0)--(0.75,-1.5); 
	\draw [thick,style={decorate, decoration={snake}}] (-1.5,0)--(0,0); 
	\draw[black,fill=tangerineyellow,thick] (0,0)circle(6pt);
    \node at (-1,0.5) {\scalebox{1}{$a^{\mu}$}};
    \node at (1,-0.85) {\scalebox{1}{$h$}};
    \node at (0.05,-0.85) {\scalebox{1}{$p_1$}};
    \node at (1,+0.85) {\scalebox{1}{$\phi$}};	
    \node at (0.05,+0.85) {\scalebox{1}{$p_2$}};   
	\end{tikzpicture}} \hspace{-0.6cm} \hspace{0.2cm} =\hspace{0.2cm} (p_h + p_\phi)^{\mu}
	\left\{	
\begin{array}{ccc}
 -(g^{\prime}\,f_{\phi}\,s_{\theta}\,c_{\beta} + g\,y_{\phi}\,c_{\theta}\,c_{\beta}) = 
 -g\,y_{\phi}\,\left(1 - \frac{3\,g^{\prime\,2}\,f_{\phi}^2\,\epsilon^2}{2\,g^2\,y_{\phi}^2}\right)  \equiv \mathcal{F}_{Zh\phi} & & \hspace{-0.2cm}a =
  Z\\
    & &  \\
 -(g^{\prime}\,f_{\phi}\,c_{\theta}\,c_{\beta} - g\,y_{\phi}\,s_{\theta}\,c_{\beta}) = 
 -f_{\phi}\,g^{\prime}\,\left[
 1 + \left(1 - \frac{g^{\prime\,2}\,f_{\phi}^2}{2\,g^2\,y_{\phi}^2}\right)\,\epsilon^2
 \right] \equiv \mathcal{F}_{Z'h\phi}
 & & \hspace{-0.2cm}a =
  Z^\prime\\
    & &  \\
\end{array}
		\right. 
\end{align}
  \begin{align}
\small
  	\raisebox{-14.2mm}{
	\begin{tikzpicture}
	\draw[thick,style={decorate, decoration={snake}}] (0,0)--(0.75,+1.5);
	\draw[thick,style={decorate, decoration={snake}}] (0,0)--(0.75,-1.5); 
	\draw [thick,dashed] (-1.5,0)--(0,0); 
	\draw[black,fill=tangerineyellow,thick] (0,0)circle(6pt);
    \node at (-1.2,0.35) {\scalebox{1}{$h$}};
    \node at (1,-0.85) {\scalebox{1}{$a^{\mu}$}};
    \node at (1,+0.85) {\scalebox{1}{$Z^{\nu}$}};	   
	\end{tikzpicture}} 
	\hspace{-0.2cm} =  
	\hspace{0.02cm} 
		g^{\mu\nu}  
		\left\{
\begin{array}{ccc}
2\,i\,v\,(g^{\prime}\,f_{\phi}\,s_{\theta} + g\,y_{\phi}\,c_{\theta})^2 \simeq
2\,i\,g\,y_{\phi}\,M\left(1 - \frac{3\,g^{\prime\,2}\,f_{\phi}^2\,\epsilon^2}{2\,g^2\,y_{\phi}^2}\right) \equiv i\,M\mathcal{F}_{ZZh}
  & & \hspace{0.25cm}  a = Z \\
    & &  \\
2\,i\,v\,[(g^{\prime\,2}\,f_{\phi}^2 - g^2\,y_{\phi}^2)\,s_{\theta}\,c_{\theta} 
+g\,g^{\prime}\,f_{\phi}\,y_{\phi}\,(c^2_{\theta} - s^2_{\theta})] \simeq \vspace{0.25cm} \\
\hspace{3.5cm} 2\,i\,g^{\prime}\,f_{\phi}\,M\left[1 + 
\left(1 - \frac{g^{\prime\,2}\,f_{\phi}^2}{2\,g^2\,y_{\phi}^2}\right)\,\epsilon^2\right] \equiv i\,M\mathcal{F}_{Z'Zh}
  & & \hspace{0.25cm}   a  =  Z^{\prime} 
\end{array}
\right.         \label{eq:Fey4} 
\end{align}

Scalar interactions.

  \begin{align}\label{eq:Fey5}
  	\raisebox{-14.2mm}{
	\begin{tikzpicture}
	\draw[thick,dashed] (0,0)--(0.75,+1.5);
	\draw[thick,dashed] (0,0)--(0.75,-1.5); 
	\draw [thick,dashed] (-1.5,0)--(0,0); 
	\draw[black,fill=tangerineyellow,thick] (0,0)circle(6pt);
    \node at (-1.2,0.35) {\scalebox{1}{$h$}};
    \node at (1,-0.85) {\scalebox{1}{$s_2$}};
    \node at (1,+0.85) {\scalebox{1}{$s_3$}};	   
	\end{tikzpicture}}
	\hspace{0.5cm} =\hspace{0.5cm}\left\{
\begin{array}{ccc}
  -6\,i\,\lambda \,v \simeq
 -\frac{3\,i\,g\, m_h^2\, y_{\phi}}{M}\left(1- \frac{g^{\prime\,2}\,f_{\phi}^2\, \epsilon^2}{2\,g^2\,y_{\phi}^2}\right)  \equiv -\frac{i\,m_h^2}{M}\,\mathcal{F}_{hhh}
    & & \hspace{0.5cm} \{s_2,s_3\} = \{h,h\}   \\
    & &  \\
  -2\,i\,\lambda \,v\,c^2_{\beta} \,\simeq
-\frac{i\,g\, m_h^2\, y_{\phi}}{M}\left( 
1 - \frac{3\,g^{\prime\,2}\,f_{\phi}^2 \,\epsilon^2}{2\,g^2\, y_{\phi}^2}
\right)  \equiv -\frac{i\, m_h^2}{M}\mathcal{F}_{h\phi\phi}
    & & \hspace{0.5cm} \{s_2,s_3\} = \{\phi,\phi\}  \\
    & &  \\
\end{array}
	\right.
\end{align}
Because of the WI, not all the vertices listed above are independent. Indeed,   we find the following relationships:
\be\label{fqe}
\frac{m_{\psi}}{M}({\cal F}_{ZL}-{\cal F}_{ZR})={\cal F}_{\phi\psi},
\qquad
{\cal F}_{Z'Zh}=2 \,{\cal F}_{Z'\phi h},\qquad
{\cal F}_{ZZh}=2 \,{\cal F}_{Z\phi h}=2\,{\cal F}_{h\phi \phi} 
\ee
so that one can express all observables in term of vertices involving gauge bosons: all vertices involving Goldstones in place of gauge bosons can be derived from the former. One can check the validity of identities (\ref{fqe}) by using the following equations obtained using (\ref{best}):
\be
g \,y_\phi\, s_\theta- g'\,f_\phi \,c_\theta=\frac{Q}{v}\,s_\beta,
\qquad
g \,y_\phi \,c_\theta+ g'\,f_\phi \,s_\theta=\frac{M}{v}\,c_\beta\;\;\to\;\; tg\beta=\epsilon\,
\frac{g\,y_\phi\,tg\theta-g'\,f_\phi}
{g\,y_\phi+g'\,f_\phi\,tg \theta}
\ee
Some limits deserve some care (in Table \ref{Limit} we report our findings in a compact form) :
	 \begin{itemize}
\item   $y_\phi=0$: The $Z$ gauge boson is massless $M=0$, 
 	 fermions are  $U(1)$ vector like $y_L=y_R$, and the mixing angles for the gauge and goldstone bosons are: 
 	 $c_\theta=1$, 
 	 $\,c_\beta=\frac{f_{\phi'}\,v'}
 	 {\sqrt{f_{\phi}^2\,v^2+f_{\phi'}^2\,v'^2 }}$. 
	 In this limit there are no couplings  with transverse gauge bosons 
	 ( $Z' \,Z\, h$ and $Z\;Z h$ are null) while the coupling with goldstone mode 
	 $Z' \,\phi\, h$, $\phi' \,\phi\, h$, $\phi \,\phi\, h$ are non zero. We also have that the ${\cal F}_{ZR}={\cal F}_{ZL}$ and this implies that the $A_1$ diagram is  zero.
	 
\item  $f_\phi=0$: both gauge and goldstone bosons mass matrices  are diagonals ($c_\theta=1$ and $c_\beta=1$) with eigenstates $M=g \,y_\phi\, v$ and $ Q=g' \,f_\phi' \,v'$. 
   Fermions are  $U'(1)$ vector like $f_L=f_R$.
   In this case the decay channel $Z'\to Z \;h$ is zero so the heavy $Z'$ can decay only in light  fermions. The $A_1$ diagram is zero.

\item Massless fermions $\epsilon_{\psi}=0$:  Higgs and goldstone fields do not interact anymore with fermions. The $S_2$ and $A_1$ diagrams are clearly zero.

\item Massless higgs $\epsilon_{h}=0$: Only the diagram $A_2$ is zero being proportional to the 
coupling $\lambda \propto m_h\to 0$.

\item Zero vev $v=0$: This limits implies that all the light spectrum is massless ($m_h=m_\psi=M=0$).
All power suppressed corrections are going to zero and only the leading log corrections, proportional to the Sudakov double logs, remains. In $S_1$ we have the Sudakov of the fermion decay channel 
while in $S_3$ we have the Sudakov related to the decay $Z'\to \phi\;h$ channel.
   
  \end{itemize}
  As a final comment, let us remember that the  presence of the $A_1$ bubble
   is the element that relate the decay rate of the $Z'$ into fermionic and purely bosonic channels (see the 
   schematic in fig.\,\ref{fig:CWInflXi}).
   The fact that the $A_1$ bubble is zero in some cases, it means that the cancellation of the $LL$ and $LL_{PS}$ corrections proceed in the {\it usual way}.
   From the above discussion we realize that  the $A_1$ bubble is non zero only for \underline{massive fermions} and 
  both \underline{gauge groups being chirals} (for example in the SM, we missed the chirality of the photon gauge couplings).
  The   configuration with non trivial $A_1$ could be present in the presence of new physics related to an extra heavy $Z'$ gauge boson to be produced in the future colliders.
  
  {\Large  \begin{table}[htp]
\begin{center}
\resizebox{.6\textwidth}{!}{
\begin{tabular}{|c||c|c|c|c|c|}
\hline
& $\quad S_1\quad$ & $\quad S_2\quad $  & $\quad S_3\quad $ & $\quad A_1\quad $  & $\quad A_2\quad $ 
\\
\hline \hline 
$y_\phi=0,\,(\epsilon=0)$ & {{\color{indiagreen}{\ding{51}}}} & {{\color{indiagreen}{\ding{51}}}} & {{\color{indiagreen}{\ding{51}}}} & 
{{\color{cornellred}{\ding{55}}}} & {{\color{indiagreen}{\ding{51}}}}
\\
\hline 
 $f_\phi=0$ & {{\color{indiagreen}{\ding{51}}}} & {{\color{indiagreen}{\ding{51}}}} & {{\color{cornellred}{\ding{55}}}} & {{\color{cornellred}{\ding{55}}}} & {{\color{cornellred}{\ding{55}}}} \\
 \hline 
 $\epsilon_\psi=0$ & {{\color{indiagreen}{\ding{51}}}} & {{\color{cornellred}{\ding{55}}}} & {{\color{indiagreen}{\ding{51}}}} & {{\color{cornellred}{\ding{55}}}} & {{\color{indiagreen}{\ding{51}}}} \\
 \hline 
 $\epsilon_h=0$ & {{\color{indiagreen}{\ding{51}}}} & {{\color{indiagreen}{\ding{51}}}} & {{\color{indiagreen}{\ding{51}}}} & {{\color{indiagreen}{\ding{51}}}} & {{\color{cornellred}{\ding{55}}}} \\
 \hline
 $v=0,\,(\epsilon=\epsilon_{\psi,h}=0)$& {{\color{indiagreen}{\ding{51}}}} & {{\color{cornellred}{\ding{55}}}} & {{\color{indiagreen}{\ding{51}}}} & {{\color{cornellred}{\ding{55}}}} & {{\color{cornellred}{\ding{55}}}} \\
 \hline
\end{tabular}
}
\end{center}\vspace{-0.4cm}
\caption{{\it Presence ({{\color{indiagreen}{\ding{51}}}}) or absence ({{\color{cornellred}{\ding{55}}}}) of 
 bubble diagrams for specific limits. See fig.\,\ref{ASJ} for the diagrammatic definition of $S_{1,2,3}$ and $A_{1,2}$.}}
\label{Limit}
\end{table}
}


\section{Numerator substitutions and the ``$p+k$ theorem''}
\label{p+k}
In order to calculate the double log power suppressed terms we are interested in, we make a certain number of simplifications that we illustrate here. Consider diagram $V_{AB}^{(X)}$ in fig.\,\ref{subs}.

\vspace{0.5cm}
\begin{adjustbox}{max width=0.94\textwidth} 
\hspace{2.75cm}
    \scalebox{1.2}{
    \raisebox{-19mm}{
	\begin{tikzpicture}
	\draw[thick,style={decorate, decoration={snake}}] (-0.5,0)--(0,0);
	\draw[->-=.8,>=Latex][thick] (0,0)--(1.25/2,1.25/2); 
	\draw[->-=.6,>=Latex][thick] (1.25/2,1.25/2)--(1.25,1.25); 
	\draw[-<-=.7,>=Latex][thick] (1.25,-1.25)--(1.25/2,-1.25/2);
	\draw[-<-=.6,>=Latex][thick] (1.25/2,-1.25/2)--(0,0);
	\draw[-<-=.5,>=Latex][thick] (1.25,1.25)--(1.25,-1.25);
	\draw[->-=.6,>=Latex][thick] (1.25,1.25)--(2.5,0); 
	\draw[-<-=.5,>=Latex][thick] (2.5,0)--(1.25,-1.25);	
	\draw[thick,style={decorate, decoration={snake}}] (2.5,0)--(2.5+0.5,0);
	\draw[thick,dash dot,persianblue][thick] (1.25/2,1.25)--(1.25/2,-1.25); 	
	\draw[black,fill=tangerineyellow,thick] (0,0)circle(2.75pt); 
	\draw[black,fill=tangerineyellow,thick] (2.5,0)circle(2.75pt); 
	\draw[black,fill=tangerineyellow,thick] (1.25,1.25)circle(2.75pt); 
	\draw[black,fill=tangerineyellow,thick] (1.25,-1.25)circle(2.75pt); 
    \node at (0.2,0.9) {\scalebox{1}{{\color{oucrimsonred}{$A$}}}};
    \node at (0.1,0.5) {\scalebox{0.9}{$p_A$}}; 
    \node at (2.75,0.5) {\scalebox{0.9}{$p_A + k$}};
    \node at (2.75,-0.5) {\scalebox{0.9}{$p_B - k$}}; 
    \node at (0.1,-0.5) {\scalebox{0.9}{$p_B$}};
    \node at (1.5,-0.2) {\scalebox{0.9}{$k$}}; 
    \node at (0.2,-0.9) {\scalebox{1}{{\color{oucrimsonred}{$B$}}}};
    \node at (2.4,0.9) {\scalebox{1}{{\color{oucrimsonred}{$C$}}}}; 
    \node at (2.1,-0.9) {\scalebox{1}{{\color{oucrimsonred}{$D$}}}}; 
	\node at (1.5,0.45) {\scalebox{1}{{\color{oucrimsonred}{$X$}}}};
		\node at (1.25,-1.85) {\scalebox{1}{{$V_{AB}^{(X)}$}}};
	\end{tikzpicture}}
	} \hspace{2cm}
	\scalebox{1.2}{
    \raisebox{-19mm}{
	\begin{tikzpicture}
	\draw[thick,style={decorate, decoration={snake}}] (-0.5,0)--(0,0);
	\draw[->-=.8,>=Latex][thick] (0,0)--(1.25/2,1.25/2); 
	\draw[->-=.6,>=Latex][thick] (1.25/2,1.25/2)--(1.25,1.25); 
	\draw[-<-=.6,>=Latex][thick] (1.25,-1.25)--(0,0);
	\draw[-<-=.5,>=Latex][thick] (1.25,1.25)--(1.25,0);
	\draw[-<-=.5,>=Latex][thick] (1.25,0)--(1.25,-1.25);
	\draw[->-=.6,>=Latex][thick] (1.25,1.25)--(2.5,0); 
	\draw[-<-=.7,>=Latex][thick] (2.5,0)--(1.25+1.25/2,-1.25/2);
	\draw[-<-=.5,>=Latex][thick] (1.25+1.25/2,-1.25/2)--(1.25,-1.25); 	
	\draw[thick,style={decorate, decoration={snake}}] (2.5,0)--(2.5+0.5,0);
	\draw[thick,dash dot,persianblue][thick] (0,1.25)--(2.5,-1.25); 	
	\draw[black,fill=tangerineyellow,thick] (0,0)circle(2.75pt); 
	\draw[black,fill=tangerineyellow,thick] (2.5,0)circle(2.75pt); 
	\draw[black,fill=tangerineyellow,thick] (1.25,1.25)circle(2.75pt); 
	\draw[black,fill=tangerineyellow,thick] (1.25,-1.25)circle(2.75pt); 
    \node at (0.15,0.7) {\scalebox{1}{{\color{oucrimsonred}{$A$}}}}; 
    \node at (0,0.4) {\scalebox{0.9}{$p_A$}};
    \node at (.65,-1.) {\scalebox{0.9}{$p_B$}};
    \node at (2.75,0.5) {\scalebox{0.9}{$p_A + k$}};
    \node at (1.95,-1.3) {\scalebox{0.9}{$p_B - k$}};
    \node at (1.05,-0.4) {\scalebox{0.9}{$k$}}; 
    \node at (0.2,-0.7) {\scalebox{1}{{\color{oucrimsonred}{$B$}}}};
    \node at (2.1,0.9)  {\scalebox{1}{{\color{oucrimsonred}{$C$}}}}; 
    \node at (2.4,-0.53) {\scalebox{1}{{\color{oucrimsonred}{$D$}}}}; 
    \node at (1.5,0.45) {\scalebox{1}{{\color{oucrimsonred}{$X$}}}};	
    	\node at (1.25,-1.85) {\scalebox{1}{{$R_{AD}^{(X)}$}}};
	\end{tikzpicture}}	
	}
	\end{adjustbox}	
	
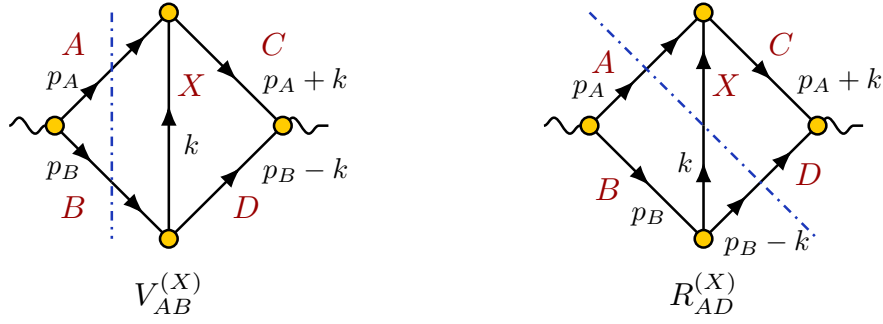
\captionof{figure}{{\em 
Diagrammatic comparison between a one-loop two-body decay graph (on the left-side, when the dot-dashed blue cut encounters two particle lines) and a three-body decay process (on the right-side, when the dot-dashed blue  cut encounters three particle lines).
The indices $X$, $A$, $B$, $C$, $D$ label the particle in the corresponding leg. 
The flow of momenta $p_{A}$, $p_B$ and $k$ follows the black arrows. 
The virtual diagram $V^{(X)}_{AB}$ is related to the $Z'\to AB$ decay with particles $X$, $C$ and $D$ entering as internal particles 
in the one-loop correction.
The real diagram $R^{(X)}_{AD}$ is related to the three-body decay $Z'\to ADX$ with particles $C$ and $B$ entering as 
internal particles.
	}}\label{subs}\vspace{0.5cm}
 The calculation features a numerator constructed from a calculation involving the vertices (fermion loops and so on), and a denominator which is the product of three propagators
\be\label{tewjn}
D_K=k^2-m_X^2\,,\qquad
D_C=(p_A+k)^2-m_C^2\,,\qquad
D_D=(p_B-k)^2-m_D^2\,.
\ee
Now, in the Feynman gauge (Lorentz invariant gauge) we work in, the numerator is composed by Lorentz invariant terms that do not depend on the integration variable $k$ (like $Q^2,p_A^2,...$)  and by terms that do depend on $k$ (like $p_A\cdot k,k^2,...$). The former are directly proportional to the $\mathcal{C}_0$ function defined in eq.\,(\ref{eq:ScalarC0}). One should then perform a separate integral for the   $k$-dependent terms; however such integration is not necessary. Let us consider for instance a $(p_A\cdot k)$ term. 
Using eq.\,(\ref{tewjn}) we can write:
\be
2\,(p_A\cdot k)=D_C-D_k+m_C^2-p_A^2-m_k^2
\ee
Now, the terms $D_C,D_k$ in the numerator cannot produce double logs by simple power counting: these terms cancel a propagator in the denominator and lower the degree of IR divergence so that no maximal (double log) terms can be generated.\footnote{Another way of seeing it is that these terms become proportional to a scalar two point integral, also known as $B_0$ function, that is free from double logs, see \cite{tHooft:1978jhc}} Ultimately, we can calculate the $k\cdot p_A$ term simply by making the substitution $2\,(p_A\cdot k)\to m_C^2-p_A^2-m_k^2$, so that also this term becomes proportional to the $C_0$ function. Similar substitutions in the numerator hold for all $k$-dependent terms.
\\
The ``$p+k$ theorem'' states  that the algebraic manipulations done for the cut of a given bubble lead to the same result (apart from a sign) for virtual and real contributions. Using the definitions given in eqs.\,(\ref{gvrw},\,\ref{oiuh}) and referring to fig.\,\ref{fig:Bubble}, this statement becomes
\be\label{mannaggia}
R^{(X)}_{AD}=R^{(X)}_{BC}=-V^{(X)}_{CD}=-V^{(X)}_{AB}\,.
\ee
A corollary of this theorem is the IR unitarity theorem quoted in eq. (\ref{pkth}) that we use throughout this paper.
 The theorem is connected to the fact that 
 \begin{itemize}
 \item[{\it i)}] The role of $p_B$ in the virtual cut is played by $p_B+k$ in the real cut, as is apparent from fig.\,\ref{subs}, and 
 \item[{\it ii)}] the same happens for four-momentum conservation.
 \end{itemize}
Let us come to the demonstration. 
When evaluating the cuts, in both cases we have a numerator constructed involving the vertices and spinorial indeces (fermion loops and so on) and a denominator given by the propagators. Since we work in a covariant gauge, the numerator can only depend on the six independent invariants constructed with the momenta $p_A,\,p_B,\,k$. However $p_A^2=m_A^2$ and $k^2=M^2$ are the same in both graphs, so we are left with 4 relevant invariants. Moreover, $(p_A\cdot k)$ is handled in the same way in both graphs, so we are left with three invariants that we can choose to be 
$p_B^2,\,p_A\cdot p_B,\,p_B\cdot k$.
 Momentum conservation allows to write a linear condition between these three invariants, so only two are left: we can write the virtual numerator as $N[p_B^2,\,p_A\cdot p_B]$. Point {\it i)} ensures that the very same function $N$ appears in the real contribution; the numerator, once the substitution $p_B\to p_B+k$ is made, can be  written
as $N[(p_B+k)^2,p_A\cdot (p_B+k)]$. Here comes the only non-trivial step. For virtual contributions $p_B^2=m_B^2$ holds; for real contributions $(p_B+k)^2$ appears instead. However we can write $(p_B+k)^2=D_B+m_B^2$, $D_B$ being the relevant propagator dominator. Since the term in  $D_B$ at the numerator cannot produce double logs, we can make the substitution $(p_B+k)^2\to m_B^2$. We can thus write 
$N[m_B^2,\,p_A\cdot p_B]$ for virtuals and $N[m_B^2,\,p_A\cdot (p_B+k)]$ for reals. A similar mechanism holds for the variable $p_A\cdot p_B$. Indeed, for virtual terms momentum conservation $Q=p_A+p_B$ implies $2\,p_A\cdot p_B=Q^2-p_A^2-p_B^2\to
Q^2-m_A^2-m_B^2$ while for real terms $Q=p_A+p_B+k$ implies
$2\,p_A\cdot (p_B+k)=Q^2-p_A^2-(p_B+k)^2\to Q^2-m_A^2-m_B^2$. The final result after doing the substitutions, is $N[m_B^2,Q^2-m_B^2-m_A^2]$ both for real and virtual contributions.
\\
Some comments are in order.
\begin{itemize}
\item[$\circ$] The demonstration holds for any combination of internal/external masses and for any kind of interactions since it relies solely on Lorentz invariance and momentum conservation.
\item[$\circ$] Clearly, it holds only for double-log terms, both leading and power suppressed. For instance we have $p_B^2=m_B^2$ for virtuals which is equivalent to  
$(p_A+p_B)^2\to m_B^2$ for reals only if we  neglect subleading terms. 
\item[$\circ$] We have shown that $V^{(X)}_{AB}=-R^{(X)}_{AD}$ but, by swapping $p_A$ with $p_B$ in the demonstration, we can show that 
$V^{(X)}_{AB}=-R^{(X)}_{BC}$. And with the same line of reasoning we can also demonstrate that $V^{(X)}_{CD}=-R^{(X)}_{BC}$. Therefore, 
eq.\,(\ref{mannaggia}) holds.
\end{itemize}

\section{The ``Integral theorem''}\label{app:Vir}

In this appendix we show that the equality reported in eq.\,(\ref{final}) holds, 
${\cal I}_V$ and ${\cal I}_R$ being defined in eq.\,(\ref{gvrw}) and
eq.\,(\ref{oiuh}). Let us begin with the integral associated with virtual corrections.

\subsection{Virtual corrections: the double-log structure of the scalar three-point integral $\mathcal{C}_0$\label{app:c0}}
In the case of virtual corrections, we start from the Lorentz-invariant two-body phase space. 
Consider a particle with four-momentum $q$ and mass $Q$ decaying into two (distinct) particles with four-momenta $p_a$ and $p_b$ and masses, respectively, $m_a$ and $m_b$. 
The conservation of energy and momentum is $q=p_a + p_b$.
In the rest frame of the decaying particle, the total decay rate is (an additional factor $1/2$ is present if the final-state particles are identical)
\begin{align}\label{eq:totalDR}
\Gamma^V = |\mathcal{M}_{q\to p_a p_b}|^2\underbrace{\frac{1}{16\pi Q}\sqrt{1 - \frac{2(m_a^2 + m_b^2)}{Q^2} + \frac{(m_a^2 - m_b^2)^2}{Q^4}}}_{\Phi_{2}(m_a^2,m_b^2)}\,.
\end{align}
Consider the modulus squared of the amplitude $|\mathcal{M}_{q\to p_a p_b}|^2$. 
On dimensional ground, the factor $|\mathcal{M}_{q\to p_a p_b}|^2$ has mass-dimension $2$.
Diagrammatically, we have
\begin{align}
 |\mathcal{M}_{q\to p_a p_b}|^2 =	& 
    \raisebox{-12mm}{
	\begin{tikzpicture}
	\draw (-1.1,-1.25)--(-1.1,1.25); 
	\draw[->-=.65,>=Latex][thick] (-0.75,0)--(0,0);
	\draw[->-=.6,>=Latex][thick] (0,0)--(1.25,1.25); 
	\draw[->-=.6,>=Latex][thick] (0,0)--(1.25,-1.25);
	\draw[black,fill=violachiaro,thick] (0,0)circle(2.75pt);  
	\node at (-0.6,0.35) {\scalebox{1}{$q$}};
	\node at (1.2,0.8) {\scalebox{1}{$p_a$}};
	\node at (1.2,-0.8) {\scalebox{1}{$p_b$}};
	\end{tikzpicture}} ~~~~~~+~~~~~~ 
  	\raisebox{-12mm}{
	\begin{tikzpicture}
	\draw (1.75,-1.25)--(1.75,1.25); 
	\draw[->-=.65,>=Latex][thick] (-0.75,0)--(0,0);
	\draw[->-=.7,>=Latex][thick] (0,0)--(1.25/2,1.25/2); 
	\draw[->-=.7,>=Latex][thick] (1.25/2,1.25/2)--(1.25,1.25); 
	\draw[->-=.7,>=Latex][thick] (1.25/2,1.25/2)--(1.25/2,-1.25/2);
	\draw[->-=.7,>=Latex][thick] (0,0)--(1.25/2,-1.25/2);
	\draw[->-=.7,>=Latex][thick] (1.25/2,-1.25/2)--(1.25,-1.25);
	\draw[black,fill=violachiaro,thick] (0,0)circle(2.75pt);
	\draw[black,fill=violachiaro,thick] (1.25/2,-1.25/2)circle(2.75pt);
	\draw[black,fill=violachiaro,thick] (1.25/2,+1.25/2)circle(2.75pt);
	\node at (-0.6,0.35) {\scalebox{1}{$q$}};
	\node at (1.2,0.8) {\scalebox{1}{$p_a$}};
	\node at (0.2,0.65) {\scalebox{1}{${\color{oucrimsonred}{A}}$}};
	\node at (0.2,-0.65) {\scalebox{1}{${\color{oucrimsonred}{B}}$}};
	\node at (0.9,0.) {\scalebox{1}{${\color{oucrimsonred}{X}}$}};
	\node at (1.2,-0.8) {\scalebox{1}{$p_b$}};
	\node at (2.,1.25) {\scalebox{0.85}{2}};   
	\end{tikzpicture}}~~~~~~=~~~~~~ \nn \\
	&
    \raisebox{-12mm}{
	\begin{tikzpicture} 
	\draw[->-=.65,>=Latex][thick] (-0.75,0)--(0,0);
	\draw[->-=.6,>=Latex][thick] (0,0)--(1.25,1.25); 
	\draw[->-=.6,>=Latex][thick] (0,0)--(1.25,-1.25);
	\draw[->-=.6,>=Latex][thick] (1.25,1.25)--(2.5,0);
	\draw[->-=.6,>=Latex][thick] (1.25,-1.25)--(2.5,0);
	\draw[->-=.7,>=Latex][thick] (2.5,0)--(2.5+0.75,0);
	\draw[thick,dash dot,persianblue][thick] (1.25,1.25)--(1.25,-1.25);
	\draw[black,fill=violachiaro,thick] (0,0)circle(2.75pt);
	\draw[black,fill=violachiaro,thick] (2.5,0)circle(2.75pt);  
	\end{tikzpicture}} ~~~~~~+~~~~~~
	  	\raisebox{-12mm}{
	\begin{tikzpicture}
	\draw[->-=.65,>=Latex][thick] (-0.75,0)--(0,0);
	\draw[->-=.7,>=Latex][thick] (0,0)--(1.25/2,1.25/2); 
	\draw[->-=.7,>=Latex][thick] (1.25/2,1.25/2)--(1.25,1.25); 
	\draw[->-=.7,>=Latex][thick] (1.25/2,1.25/2)--(1.25/2,-1.25/2);
	\draw[->-=.7,>=Latex][thick] (0,0)--(1.25/2,-1.25/2);
	\draw[->-=.7,>=Latex][thick] (1.25/2,-1.25/2)--(1.25,-1.25);
	\draw[->-=.6,>=Latex][thick] (1.25,1.25)--(2.5,0);
	\draw[->-=.6,>=Latex][thick] (1.25,-1.25)--(2.5,0);
	\draw[->-=.7,>=Latex][thick] (2.5,0)--(2.5+0.75,0);
	\draw[thick,dash dot,persianblue][thick] (1.25,1.25)--(1.25,-1.25);
	\draw[black,fill=violachiaro,thick] (0,0)circle(2.75pt);
	\draw[black,fill=violachiaro,thick] (1.25/2,-1.25/2)circle(2.75pt);
	\draw[black,fill=violachiaro,thick] (1.25/2,+1.25/2)circle(2.75pt);   
	\draw[black,fill=violachiaro,thick] (2.5,0)circle(2.75pt); 
	\end{tikzpicture}} ~~~~~~+~~~~~~
	\raisebox{-12mm}{
	\begin{tikzpicture}
	\draw[->-=.65,>=Latex][thick] (-0.75,0)--(0,0);
	\draw[->-=.6,>=Latex][thick] (0,0)--(1.25,1.25); 
	\draw[->-=.6,>=Latex][thick] (0,0)--(1.25,-1.25);
	\draw[->-=.7,>=Latex][thick] (1.25,1.25)--(1.25+1.25/2,1.25/2);
	\draw[->-=.7,>=Latex][thick] (1.25+1.25/2,1.25/2)--(2.5,0);
	\draw[->-=.7,>=Latex][thick] (1.25,-1.25)--(1.25+1.25/2,-1.25/2);
	\draw[->-=.7,>=Latex][thick] (1.25+1.25/2,-1.25/2)--(2.5,0);
	\draw[->-=.7,>=Latex][thick] (1.25+1.25/2,-1.25/2)--(2.5,0);
	\draw[->-=.7,>=Latex][thick] (1.25+1.25/2,1.25/2)--(1.25+1.25/2,-1.25/2);	
	\draw[->-=.7,>=Latex][thick] (2.5,0)--(2.5+0.75,0);	
	\draw[thick,dash dot,persianblue][thick] (1.25,1.25)--(1.25,-1.25);
	\draw[black,fill=violachiaro,thick] (0,0)circle(2.75pt);
	\draw[black,fill=violachiaro,thick] (1.25+1.25/2,1.25/2)circle(2.75pt);
	\draw[black,fill=violachiaro,thick] (1.25+1.25/2,-1.25/2)circle(2.75pt); 
	\draw[black,fill=violachiaro,thick] (2.5,0)circle(2.75pt);  
	\end{tikzpicture}}\label{eq:VirtualCorrections2}
\end{align}
The first diagram in eq.\,(\ref{eq:VirtualCorrections2}) gives the tree-level contribution to the decay rate. 
The second and third diagram are one-loop virtual corrections. 
We are interested in the double-log terms generated by the last two diagrams. 
To this end, it is necessary to investigate the structure of the scalar three-point integral.
\\
We consider the scalar three-point integral defined by\footnote{We stick to the case of $d=4$ space-time dimensions since 
our attention is focused on the IR structure of the integral rather than on its ultraviolet behavior. 
Notice also that in some cases an overall constant $i/16\pi^2$ is factored out of the integral
measure to simplify the output. For instance, our $\mathcal{C}_0$ is related to the 
scalar three-point integral $\mathcal{I}_3$ defined in ref.\,\cite{Ellis:2007qk} by 
$\mathcal{C}_0 = (i/16\pi^2)\,\mathcal{I}_3$.
}
\begin{align}\label{eq:ScalarC0}
\mathcal{C}_0(q^2,p_a^2,p_b^2,m_A^2,m_B^2,m_k^2) & = 
  	\raisebox{-13.8mm}{
	\begin{tikzpicture}
	\draw[->-=.65,>=Latex][thick] (-1.,0)--(0,0);
	\draw[->-=.6,>=Latex][thick] (0,0)--(1.25,1.25); 
	\draw[->-=.6,>=Latex][thick] (1.25,-1.25)--(0,0);
	\draw[->-=.55,>=Latex][thick] (1.25,1.25)--(1.25,-1.25); 
	\draw[->-=.65,>=Latex][thick] (1.25,1.25)--(2.25,1.25);
	\draw[->-=.65,>=Latex][thick] (1.25,-1.25)--(2.25,-1.25);
	\draw[black,fill=violachiaro,thick] (0,0)circle(2.75pt); 
	\draw[black,fill=violachiaro,thick] (1.25,1.25)circle(2.75pt); 
	\draw[black,fill=violachiaro,thick] (1.25,-1.25)circle(2.75pt);   
	\node at (0.,0.75) {\scalebox{1}{$p_a + k$}};
	\node at (0.,-0.75) {\scalebox{1}{$k - p_b$}};
	\node at (1.55,0.2) {\scalebox{1}{$k$}}; 
	\node at (2.6,1.25) {\scalebox{1}{$p_a$}}; 
	\node at (2.6,-1.25) {\scalebox{1}{$p_b$}}; 
	\node at (-1.2,0.1) {\scalebox{1}{$q$}};
	\node at (1.55,-0.2) {\scalebox{0.9}{{\color{oucrimsonred}{$m_k$}}}};
	\node at (0.9,0.42) {\scalebox{0.9}{{\color{oucrimsonred}{$m_A$}}}}; 
	\node at (0.9,-0.42) {\scalebox{0.9}{{\color{oucrimsonred}{$m_B$}}}}; 
	\end{tikzpicture}} \\
& 	= \int\frac{d^4 k}{(2\pi)^4}\frac{1}{(k^2-m_k^2+i\epsilon)[(p_a+k)^2-m_A^2+i\epsilon][(p_b-k)^2-m_B^2+i\epsilon]}\,,\nn
\end{align}
with $q = p_a + p_b$.
First, we introduce the Feynman-parameter representation and integrate over the four-momentum $k$. We find
\begin{align}\label{eq:ScalarC0Fey}
\mathcal{C}_0(q^2,p_a^2,p_b^2,m_A^2,m_B^2,m_k^2) & = \frac{(-i)}{(4\pi)^2}\int_0^{1}
dx_1 dx_2 dx_3 \delta(1-x_1-x_2-x_3) \frac{1}{\left[g(x_1,x_2,x_3) - q^2 x_2 x_3\right]}\,,\\
g(x_1,x_2,x_3) & \equiv m_k^2 x_1 +m_A^2 x_2 +m_B^2 x_3 - p_a^2 x_1x_2 -p_b^2 x_1x_3 - i\epsilon\,.
\end{align}
We are interested in the high-energy limit $q^2 \gg p_a^2,p_b^2,m_A^2,m_B^2,m_k^2$
 of eq.\,(\ref{eq:ScalarC0Fey}) and, in particular, our aim is to isolate the double-log terms. 
 To this end, we introduce the Mellin transform of the scalar three-point integral.  
 We follow the method outlined in ref.\,\cite{Roth:1996pd} that we generalize to include power-suppressed double-log corrections (at all orders). 
 \\
Before proceeding, let us give few standard definitions. 
If the function $f(r)$ fulfils  certain integrability conditions, the Mellin transform of $f(r)$ is defined by  
 \begin{align}\label{eq:MellinTra}
  {\rm M}[f(r),\xi] = \int_0^\infty r^{\xi-1}f(r)dr\,,
  \end{align}
and it converges absolutely and is holomorphic in a vertical strip $\alpha < {\rm Re}(\xi) < \beta$. 
The exact values of $\alpha$ and  $\beta$ depend on the function $f(r)$. 
The Mellin transform can be continued analytically on the entire complex $\xi$-plane;  its analytical continuation contains poles in the two complex half-planes with ${\rm Re}(\xi) \geqslant \beta$ and 
${\rm Re}(\xi) \leqslant \alpha$.  The inverse transform is 
\begin{align}\label{eq:Mellin}
 f(r)  = \frac{1}{2\pi i}\int_{c-i\infty}^{c+i\infty}r^{-\xi}\,{\rm M}[f(r),\xi]\,d\xi\,,\hspace{1cm}    
  	\raisebox{-29mm}{
	\begin{tikzpicture}
	\draw[->,>=Latex][thick] (-1,0)--(4.5,0);
	\draw[->,>=Latex][thick] (0,-2.5)--(0,2.5);
	\fill[fill=green!15] (0.5,-2.5)rectangle(1.75,2.5);
	\draw[thick,dashed] (0.5,-2.5)--(0.5,2.5);
	\draw[thick,dashed] (1.75,-2.5)--(1.75,2.5);
	\draw[->-=.8,>=Latex][thick,firebrick] (1.25,-2.5)--(1.25,2.5);
    \draw[->-=.8,>=Latex][thick,firebrick] (1.25,2.5) arc (90:-90:2.5);	
    \node at (5.,0) {\scalebox{1}{${\rm Re}(\xi)$}};
    \node at (-0.5,2.75) {\scalebox{1}{${\rm Im}(\xi)$}};
    \node at (0.4,-2.75) {\scalebox{0.75}{${\rm Re}(\xi)=\alpha$}};
    \node at (1.95,-2.75) {\scalebox{0.75}{${\rm Re}(\xi)=\beta$}};
    \node at (1,0) [rotate=90,firebrick] {\scalebox{0.75}{${\rm Re}(\xi)=c$}};
    \node at (1,1.75) [rotate=90,firebrick] {\scalebox{0.75}{$\xi \to c+i\infty$}};
    \node at (1.,-1.7) [rotate=-90,firebrick] {\scalebox{0.75}{$\xi \to c-i\infty$}};
	\end{tikzpicture}}
\end{align}
where the integral is a line integral taken over a vertical line in the complex plane whose real part $c$ is arbitrary as long as $\alpha < c < \beta$.
To compute the integral,  as illustrated in the picture above, we close the integration contour in the right-hand complex half-pane and apply the theorem of residues so that the poles with 
${\rm Re}(\xi) \geqslant \beta$ will reconstruct the function $f(r)$ (assuming we drop the contribution from the arch at infinity). 
The utility of the Mellin transform can be illustrated with the help of the following example. 
Suppose that the Mellin transform of a certain function has a pole of order $(n+1)$ at $\xi = \xi_0$. 
The contribution of this pole to the inverse Mellin transform is 
\begin{align}
 \frac{1}{2\pi i}\int_{c-i\infty}^{c+i\infty}d\xi r^{-\xi}\,\frac{n!}{(\xi - \xi_0)^{n+1}} = \frac{(-1)^n}{r^{\xi_0}}\log^n r\,.
\end{align}
From this simple computation it follows that we can obtain the double-logs of the function $f(r)$ from the third-order poles of its Mellin transform. 
Furthermore, we see that the leading term corresponds to the pole that sits on the right edge of the convergence domain, that is the pole with ${\rm Re}(\xi_0) = \beta$; 
sub-leading double logs (which are important for us) can be computed including the subsequent third-order poles that one encounters moving rightwards from ${\rm Re}(\xi_0) = \beta$.

In order to exploit this result, the rationale will be the following.
As a first step, we need to compute the Mellin transform of eq.\,(\ref{eq:ScalarC0Fey}), 
${\rm M}[\mathcal{C}_0(q^2,p_a^2,p_b^2,m_A^2,m_B^2,m_k^2),\xi]$, that we shall indicate in the following with the shorthand notation 
${\rm M}[\mathcal{C}_0(q^2),\xi]$.
Second, by looking at the third-order poles of ${\rm M}[\mathcal{C}_0(q^2),\xi]$ it will be possible to extract, by means of the inverse Mellin transform, 
the double-log terms of the scalar three-point integral.

The Mellin transform of the scalar three-point integral can be computed starting from the following identity\,\cite{Scharf:1993ds}
\begin{align}\label{eq:Id}
\int_0^{\infty}dr_0\dots dr_n\frac{\prod_{i=0}^{n}r_i^{\xi_i -1} \delta(1-\sum_{i=0}^n\alpha_i r_i)}{(\sum_{i=0}^n A_i r_i)^{\eta}} = 
\frac{\prod_{i=0}^{n}\Gamma(\xi_i)}{\Gamma(\eta)\prod_{i=0}^n A_i^{\xi_i}}\,,
\end{align}
where $\eta = \sum_{i=0}^n \xi_i$ and ${\rm Re}(\xi_i) > 0$;  $\alpha_i \geqslant 0$ are arbitrary real and non-negative constants, which do not all vanish simultaneously. 
We consider the special case with $\alpha_0 = 1$ and $\alpha_{i>0} = 0$; the delta function in eq.\,(\ref{eq:Id}) reduces to $\delta(1-r_0)$, and we can use it to integrate over $r_0$.  We find
\begin{align}\label{eq:Id2}
\int_0^{\infty}dr_1\dots dr_n \frac{r_1^{\xi_1 - 1}\dots r_n^{\xi_n - 1}}{(A_0 + A_1 r_1 + \dots + A_n r_n)^{\eta}} 
= \frac{\prod_{i=0}^{n}\Gamma(\xi_i)}{\Gamma(\eta)\prod_{i=0}^n A_i^{\xi_i}}\,.
\end{align}
Consider now the simple case with $n=1$. We have 
\begin{align}\label{eq:MellinTraSim}
\int_0^{\infty}dr_1 \frac{r_1^{\xi_1 -1}}{(A_0 + r_1 A_1)^{\eta}} = \frac{\Gamma(\eta - \xi_1)\Gamma(\xi_1)}{\Gamma(\eta)A_0^{\eta - \xi_1}A_1^{\xi_1}}\,,
\end{align}
where we used $\eta = \xi_0 + \xi_1$ to rewrite $\xi_0 = \eta - \xi_1$. 
Eq.\,(\ref{eq:MellinTraSim}) is the Mellin transform of $(A_0 + r_1 A_1)^{-\eta}$, 
and can be used to compute the Mellin transform of eq.\,(\ref{eq:ScalarC0Fey}) if we set $\eta = 1$ and identify 
$A_0 = g(x_1,x_2,x_3)$, $r_1 = |q^2|$ and $A_1 = (-\sigma - i\epsilon)x_2 x_3$ with $\sigma \equiv q^2/|q^2| = \pm 1$.
All in all, we find the Mellin transform of the scalar three-point function
\begin{align}\label{eq:C0Mellin}
{\rm M}[\mathcal{C}_0(q^2),\xi] = \frac{(-i)\Gamma(1-\xi)\Gamma(\xi)}{(4\pi)^2(-\sigma - i\epsilon)^{\xi}}
\int_0^1dx_2\int_0^{1-x_2}dx_3\,\frac{1}{g(1-x_2-x_3,x_2,x_3)^{1-\xi} x_2^{\xi}x_3^{\xi}}\,,
\end{align}
where we used the delta function in eq.\,(\ref{eq:ScalarC0Fey}) to integrate over $x_1$. Notice that we introduced an infinitesimal imaginary part both in $A_0$ (which was actually already present 
in the definition of $g(x_1,x_2,x_3)$) and $A_1$ to ensure the applicability
 of eq.\,(\ref{eq:Id2})\,\cite{Scharf:1993ds}. 
 In eq.\,(\ref{eq:C0Mellin}), $\Gamma(\xi)$ is holomorphic in the complex half-plane $\{\xi \in \mathbb{C}\,|\,{\rm Re}(\xi) >0\}$ 
 and $\Gamma(1-\xi)$ is holomorphic in the complex half-plane $\{\xi \in \mathbb{C}\,|\,{\rm Re}(\xi) < 1\}$. 
 The strip of convergence of eq.\,(\ref{eq:C0Mellin}), therefore, is given by $0<{\rm Re}(\xi)<1$. 
 The Gamma function $\Gamma(1-\xi)$ in eq.\,(\ref{eq:C0Mellin}) has simple poles at $\xi = 1,2,3,\dots$. 
 According to our general argument, the leading term in the high-energy expansion of the three-point 
 function is given by the residue of the pole at $\xi = 1$. 
To understand the order of this pole, we need to expand the integrand in eq.\,(\ref{eq:C0Mellin}).
First, we define $h(x_2,x_3) \equiv g(1-x_2-x_3,x_2,x_3)$ with $h(0,0) = m_k^2$.
We Taylor-expand around $\{x_2,x_3\} = \{0,0\}$, and, up to terms with two powers of $x_{i=2,3}$,
we find
\begin{align}
\frac{1}{h(x_2,x_3)^{1-\xi}} = & ~~~~~\,\frac{1}{(m_k^2-i\epsilon)^{1-\xi}} + \frac{x_2(1-\xi)(m_k^2 -m_A^2 + p_a^2)}{(m_k^2-i\epsilon)^{2-\xi}} 
+  \frac{x_3(1-\xi)(m_k^2 -m_B^2 + p_b^2)}{(m_k^2-i\epsilon)^{2-\xi}} \nn\\
& + \frac{x_2x_3(1-\xi)}{(m_k^2-i\epsilon)^{3-\xi}}\left\{
[m_k^4 - m_k^2(m_A^2 + m_B^2) + (m_A^2 - p_a^2)(m_B^2 - p_b^2)](2-\xi) + m_k^2(p_a^2 + p_b^2)(1-\xi)
\right\}\nn \\
& +  \frac{x_2^2(1-\xi)}{4(m_k^2-i\epsilon)^{5-\xi}}\left\{
2m_k^4[m_k^4 - 2m_k^2 m_A^2 +(m_A^2 - p_a^2)^2](2-\xi) + 4m_k^6p_a^2(1-\xi)
\right\}\nn\\
& +  \frac{x_3^2(1-\xi)}{4(m_k^2-i\epsilon)^{5-\xi}}\left\{
2m_k^4[m_k^4 - 2m_k^2 m_B^2 +(m_B^2 - p_b^2)^2](2-\xi) + 4m_k^6p_b^2(1-\xi)
\right\}\,.\label{eq:Expansion}
\end{align}
We now integrate over $x_{2,3}$. We find  
\begin{align}\label{eq:FundamentalIntegral}
\int_0^1 dx_2 \int_0^{1-x_2}dx_3\frac{x_2^n x_3^m}{x_2^{\xi}x_3^{\xi}} = 
\frac{\Gamma(2+m-\xi)\Gamma(1+n-\xi)}{(1+m-\xi)\Gamma(3+m+n-2\xi)}\,,~~~~~{\rm with}~~~{\rm Re}(\xi) < 1+m \land {\rm Re}(\xi) < 1+n\,.
\end{align}
From this integral, combined with the $\Gamma(1-\xi)$ factor already present in eq.\,(\ref{eq:C0Mellin}), it is clear that only the first term in the expansion of 
eq.\,(\ref{eq:Expansion}), that is the one with $n=m=0$ in eq.\,(\ref{eq:FundamentalIntegral}), gives a third-order pole at $\xi = 1$ for the Mellin transform ${\rm M}[\mathcal{C}_0(q^2),\xi]$. 
If we compute its residue we obtain, via the inverse Mellin transform, the corresponding term in the high-energy expansion of the three-point function. We find (setting $q^2 = Q^2$) the 
well-known result
\begin{align}
\mathcal{C}_0(Q^2,p_a^2,p_b^2,m_A^2,m_B^2,m_k^2) = \frac{i}{32\pi^2 Q^2}\log^2 Q^2\,.
\end{align}
In our analysis we are interested in sub-leading corrections to the above expression. 
This means that we need to move to the next pole at $\xi = 2$, and check which term in the  expansion in eq.\,(\ref{eq:Expansion}) gives, once integrated by means of 
eq.\,(\ref{eq:FundamentalIntegral}) and combined with 
the $\Gamma(1-\xi)$ factor in eq.\,(\ref{eq:C0Mellin}), a third-order pole. 
It is easy to see, in this respect, that the only relevant term in eq.\,(\ref{eq:Expansion})  is the one with $n=m=1$. 
If we compute its residue, we find
\begin{align}\label{eq:C0Asy}
\mathcal{C}_0(Q^2,p_a^2,p_b^2,m_A^2,m_B^2,m_k^2) = \frac{i}{32\pi^2 Q^2}\left[
1 + \frac{(p_a^2 + p_b^2)}{Q^2}
\right] \log^2 Q^2\,.
\end{align}
In the computation of the decay rate, the amplitude squared in eq.\,(\ref{eq:VirtualCorrections2}) 
is multiplied by the two-body phase space factor $\Phi^{(2)}_{\rm LIPS}$ in eq.\,(\ref{eq:totalDR}). 
If we set $p_a^2 = m_a^2$ and $p_b^2 = m_b^2$ in eq.\,(\ref{eq:C0Asy}) and expand at the 
quadratic order in the parameters $\epsilon_a \equiv m_a/Q$ and $\epsilon_b \equiv m_b/Q$, we find
\begin{align}\label{eq:ClaimC0}
\mathcal{C}_0(Q^2,p_a^2,p_b^2,m_A^2,m_B^2,m_k^2)\Phi_{2}(m_a^2,m_b^2) = 
\frac{i}{32\pi^2 Q^2}\log^2 Q^2\,,~~~~~~~~~~~{\rm at\,order\,}O(\epsilon_{a,b}^2)\,.
\end{align}
It is natural to ask what happens including power-suppressed terms. 
To answer this question, we need to generalize the expansion in eq.\,(\ref{eq:C0Asy}). 
The claim is that eq.\,(\ref{eq:ClaimC0}) is valid at all orders as far as power-suppressed terms are concerned. 

From our previous discussion, 
we know that the power-suppressed correction of order $O(\epsilon_{a,b}^{2(p-1)})$ is given by the inverse Mellin transform, computed 
by means of the residue theorem, of the third-order poles of ${\rm M}[\mathcal{C}_0(q^2),\xi]$ at 
$\xi = p$ for $p =1,2,3,\dots$.
We rewrite the Mellin transform ${\rm M}[\mathcal{C}_0(q^2),\xi]$
 as
\begin{align}
{\rm M}[\mathcal{C}_0(q^2),\xi] & = 
\frac{(-i)\Gamma(1-\xi)\Gamma(\xi)}{(4\pi)^2(-\sigma - i\epsilon)^{\xi}}
\int_0^1dx_2\int_0^{1-x_2}dx_3\,\frac{1}{g(1-x_2-x_3,x_2,x_3)^{1-\xi} x_2^{\xi}x_3^{\xi}}\\
& = \frac{(-i)\Gamma(1-\xi)\Gamma(\xi)}{(4\pi)^2(-\sigma - i\epsilon)^{\xi}}\left[
\sum_{n,m}
\left.\frac{1}{h(x_2,x_3)^{1-\xi}}\right|_{\{n,m\}}\frac{\Gamma(2+m-\xi)\Gamma(1+n-\xi)}{(1+m-\xi)\Gamma(3+m+n-2\xi)}\right]\,,
\label{eq:MellinDeco}
\end{align}
where $\left.1/h(x_2,x_3)^{1-\xi}\right|_{\{n,m\}}$ represents 
a generalization of eq.\,(\ref{eq:Expansion}), that is
the  
 term of order $O(x_2^nx_3^m)$ in the Taylor expansion of $1/h(x_2,x_3)^{1-\xi}$ around $\{x_2,x_3\} = \{0,0\}$. 
 For fixed $\{n,m\}$ and positive $\xi$, therefore, the singular structure of ${\rm M}[\mathcal{C}_0(q^2),\xi]$ is entirely given by the factor 
\begin{align}\label{eq:Singular}
\frac{\Gamma(1-\xi)\Gamma(2+m-\xi)\Gamma(1+n-\xi)}{(1+m-\xi)\Gamma(3+m+n-2\xi)}\,.
\end{align} 
The strategy of the computation is the following. 
First, we need to identify for which values of $\{n,m\}$ eq.\,(\ref{eq:Singular}) has third-order poles at 
$\xi = p$. Second, for these values of $\{n,m\}$ we compute the inverse Mellin transform of eq.\,(\ref{eq:MellinDeco}) by means 
of the residue theorem and sum over all possible $\{n,m\}$ pairs; actually, as we shall see next, this last sum is not needed since 
for $\xi = p$ we only have one third-order pole. 
\\
The key point is to realize that eq.\,(\ref{eq:Singular}) admits only one third-order pole at $\xi = p$, given by 
$n=m=p-1$. The proof goes as follows.
\begin{itemize}
\item [$\ast$] The third-order poles in eq.\,(\ref{eq:Singular}) are obtained as a product of three simple poles.
\item [$\ast$] The factor $\Gamma(1-\xi)$ always gives a simple pole at $\xi = p$ for $p \in \mathbb{Z}^+$ 
since $\Gamma(1-\xi)$ has simple poles at $\xi = 1,2,3,\dots$. 
Another simple pole at $\xi = p$ comes from the factor $1/(1+m-\xi)$ if we take $m = p-1$. This fixes the value of $m$. 
Consider now the two $\Gamma$ functions at the numerator of eq.\,(\ref{eq:Singular}). Since we just fixed $m = p-1$, 
we have $\Gamma(2+m-\xi) = \Gamma(1+p-\xi)$ which has poles at $\xi = p+1,p+2,\dots \neq p$. 
This means that the only possibility to get another simple pole at $\xi = p$ is to extract it from the other $\Gamma$ function $\Gamma(1+n-\xi)$. 
The latter has simple poles at $\xi = n+1,n+2,\dots = p$ from which we have $n = p-1,p-2,\dots$. 
Apparently, therefore, we have third-order poles at 
$\{n,m\} = \{p-1,p-1\},\,\{p-2,p-1\},\,\{p-3,p-1\},\dots$.

However, not all the above values of $n$ are allowed. The only values of $n$ that are allowed are those for which the 
$\Gamma$ function $\Gamma(3+m+n-2\xi)$ in the denominator of eq.\,(\ref{eq:Singular}) is non-singular (otherwise 
it would decrease the order of the pole from three to two).  
This means that we need to avoid values of $\{n,m\}$ such that $n+m= 2p-3,2p-4,\dots$ which are the poles of 
$\Gamma(3+m+n-2\xi)$ for $\xi = p$. 
For instance, $\{n,m\} = \{p-2,p-1\}$ gives $n+m=2p-3$ which is invalidated by the above argument. 
It follows that the only valid third-order pole comes from $\{n,m\} = \{p-1,p-1\}$.  

\item [$\ast$] Consider now the following situation. 
As before, the factor $\Gamma(1-\xi)$ always gives a simple pole at $\xi = p$. 
However, this time we do not fix $m$ by requiring the factor $1/(1+m-\xi)$ to be singular at $\xi = p$. 
On the contrary, we try to get the remaining two simple poles out of the two 
$\Gamma$ functions $\Gamma(2+m-\xi)$ and $\Gamma(1+n-\xi)$ in the numerator of eq.\,(\ref{eq:Singular}).
This means that, apparently, we have all possible combinations of 
$m= p-2,p-3,p-4,\dots$ and $n=p-1,p-2,p-3,\dots$ at our disposal.
However, as before, we need to avoid poles of the 
$\Gamma$ function $\Gamma(3+m+n-2\xi)$ in the denominator of eq.\,(\ref{eq:Singular}).
This condition (which, as before, translates into $n+m \neq 2p-3,2p-4,\dots$) automatically excludes all above possibilities.
\end{itemize}
In conclusion, at a given order $O(\epsilon_{a,b}^{2(p-1)})$ we only need to compute, by means of the residue theorem, 
the inverse Mellin transform of eq.\,(\ref{eq:MellinDeco}) for fixed $\{n,m\} = \{p-1,p-1\}$. 
We are now in the position of generalizing eq.\,(\ref{eq:C0Asy}).
\\
Consider first the case $p_a^2 = p_b^2 = m^2$. The residue of the third-order pole at $\xi = p$ gives 
\begin{align}\label{eq:C0AsyP}
\mathcal{C}_0^{(p)}(Q^2, m^2, m^2,m_A^2,m_B^2,m_k^2) = \frac{i}{32\pi^2 Q^2}\left[
4^{p-1}\left(\frac{m^2}{Q^2}\right)^{p-1}\frac{(1/2)_{p-1}}{(1)_{p-1}}
\right] \log^2 Q^2\,,
\end{align}
where $(a)_n$ is the Pochhammer symbol. If we now sum over the whole tower of third-order poles, we find
\begin{align}
\mathcal{C}_0(Q^2, m^2, m^2,m_A^2,m_B^2,m_k^2) = 
\sum_{p=1}^{\infty}\mathcal{C}_0^{(p)}(Q^2, m^2, m^2,m_A^2,m_B^2,m_k^2) = 
\frac{1}{\sqrt{1- \frac{4m^2}{Q^2}}}\,\frac{i}{32\pi^2 Q^2}\log^2 Q^2\,,
\end{align}
where the factor $1/\sqrt{1- \frac{4m^2}{Q^2}}$ reproduces precisely the 
inverse of the Lorentz-invariant two-body phase space. 
The case with $m_a^2 \neq m_b^2$ can be solved with the same technique.  We find
\begin{align}\label{eq:C0AsyP}
\mathcal{C}_0^{(p)}(Q^2, m_a^2, m_b^2,m_A^2,m_B^2,m_k^2) = \frac{i}{32\pi^2 Q^2}\left[
\left(\frac{m_a^2 - m_b^2}{Q^2}\right)^{p-1}C_{p-1}^{(1/2)}\left(\frac{m_a^2 + m_b^2}{m_a^2 - m_b^2}\right)
\right] \log^2 Q^2\,,
\end{align}
where $C_n^{(m)}(x)$ is the Gegenbauer polynomial. 
If we now sum over the whole tower of third-order poles, we find
\begin{align}\label{,,,,}
\mathcal{C}_0(Q^2, m_a^2, m_b^2,m_A^2,m_B^2,m_k^2) = 
\sum_{p=1}^{\infty}\mathcal{C}_0^{(p)}(Q^2, m_a^2, m_b^2,m_A^2,m_B^2,m_k^2) = 
\frac{1}{
\sqrt{1 - \frac{2(m_a^2 + m_b^2)}{Q^2} + \frac{(m_a^2 - m_b^2)^2}{Q^4}}
}\,\frac{i}{32\pi^2 Q^2}\log^2 Q^2\,.
\end{align}
Consequently, given the expression for the two body phase space $\Phi_{2}(m_a^2,m_b^2)$ in eq.\,(\ref{eq:totalDR}), 
at all orders in the expansion in power-suppressed terms we find
\begin{align} \label{cofi}
\mathcal{C}_0(Q^2, m_a^2, m_b^2,m_A^2,m_B^2,m_k^2)\Phi_{2}(m_a^2,m_b^2) = 
\frac{i}{512\pi^3 Q^3}\log^2 Q^2\,.
\end{align} 
As a consequence of the ``$p+k$ theorem'' discussed in appendix\,\ref{p+k}, we can now write
the contribution of the second diagram in eq.\,(\ref{eq:VirtualCorrections2}) to the total decay width as
\begin{align}\label{eq:ExpressionIV}
\Gamma^V \ni Q\,V_{AB}^{(X)}\,\mathcal{I}^V\,,~~~~~~~~~{\rm with}~~~~
\mathcal{I}^V = -iQ^3\mathcal{C}_0\Phi_{2}\,.
\end{align}
In the above expression $\mathcal{I}^V$ is the a-dimensional phase-space integral that includes the 
scalar $\mathcal{C}_0$ function, and that we introduced 
in eq.\,(\ref{gvrw}); on the contrary, $V_{AB}^{(X)}$ is the a-dimensional combination of masses and couplings 
that follows from the application of the ``$p+k$ theorem'' to the one-loop 
diagram shown in the top row of eq.\,(\ref{eq:VirtualCorrections2}). 
Notice that in $V_{AB}^{(X)}$ we are using the same notation introduced in eq.\,(\ref{pkth}). 
In the double-log approximation of eq.\,(\ref{cofi}), we find
\begin{align}\label{eq:AsiIV}
\mathcal{I}^V = \frac{1}{512\,\pi^3}\log^2 Q^2\,,
\end{align}
which is the asymptotic behavior quoted in eq.\,(\ref{final}).

\subsection{Real corrections}\label{app:Re}

In the case of real emission, we need the Lorentz-invariant three-body phase space. 
Consider a particle with four-momentum $q$ and mass $Q$ decaying into three particles with four-momenta $p_a$, $p_b$ and $k$ and masses, respectively, $m_a$, $m_b$ and $m_k$. 
The conservation of energy and momentum is $q=p_a + p_b + k$.
The partial decay rate in the rest frame of the decaying particle is 
\begin{align}\label{eq:Main3Body}
d\Gamma^R = \frac{(2\pi)^4}{2Q}|\mathcal{M}_{q\to p_a p_b k}|^2 d\Phi_3(q;p_a,p_b,k)\,,~~~~~~
d\Phi_3(q;p_a,p_b,k) \equiv \frac{d^3\vec{p_a}}{(2\pi)^3 2E_a}\frac{d^3\vec{p_b}}{(2\pi)^3 2E_b}\frac{d^3\vec{k}}{(2\pi)^32E_k}\delta(q - p_a - p_b - k)\,.
\end{align}
where $d\Phi_3(q;p_a,p_b,k)$ is the Lorentz invariant phase space. 
We indicate with $|\mathcal{M}_{q\to p_a p_b p_k}|^2$ the (dimensionless) matrix element squared that includes the sum over the final-state spins and the average over the initial state polarizations. 
In our case, the decay process depends only on two independent kinematic variables. This is because the three final state three-vectors (nine variables) are constrained by four energy-momentum 
conservation equations; furthermore, the decay is isotropic in the rest frame of $q$ (this is true if the decaying particle is a scalar or, in the case of a vector, if we average over its polarization states, the latter being our case) 
so that the final state cannot depend on the three angles that describe its orientation. 
From this argument, it follows that only two independent variables are left.
We rewrite the partial decay rate in the form
\begin{align}\label{eq:GammaC}
d\Gamma^R = \frac{|\mathcal{M}_{q\to p_a p_b k}|^2}{16\pi Q}\frac{dE_a dE_k}{4\pi^2}\,,
\end{align}
where we use as independent variables the two energies $E_a$ and $E_k$. 
\\
To compare with the total decay rate in eq.\,(\ref{oiuh}), we need to integrate over the energies.
 The boundary of the physical region is determined by the configurations in which the three-momentum $\vec{p}_a$ is parallel or antiparallel to $\vec{k}$; these two conditions read
 \begin{align}
 Q^2 - 2Q(E_a + E_k) +2 E_a E_k + m_a^2 + m_k^2 - m_b^2 = \pm 2\sqrt{(E_a^2 - m_a^2)(E_k^2 - m_k^2)}\,.
 \end{align}
From these two boundaries, 
one extracts the minimal and maximal energies $E_{k,{\rm min}}(E_a)$ and $E_{k,{\rm max}}(E_a)$ solving 
for $E_k$ as function of $E_a$ (the opposite slicing is also possible).
The analytical expression for $E_{k,{\rm min}}(E_a)$ and $E_{k,{\rm max}}(E_a)$ contain a square root, and the positivity of its argument corresponds to the boundaries $E_{a,{\rm min}}$ and $E_{a,{\rm max}}$. 
For completeness, we find 
\begin{align}
\frac{E_{k,
{\rm max}
}}{Q} & =  \frac{(1-x_a)(1 - 2x_a + \epsilon_k^2 + \epsilon_a^2 - \epsilon_b^2) + 
\sqrt{(x_a^2 - \epsilon_a^2)[1 - 2x_a + \epsilon_a^2 - (\epsilon_b-\epsilon_k)^2][1-2x_a + \epsilon_a^2 - (\epsilon_b+\epsilon_k)^2]}}
{2[\epsilon_a^2 + (1-2x_a)]}\,,\label{eq:BC1}\\
\frac{E_{k,{\rm min}}}{Q} & =  \frac{(1-x_a)(1 - 2x_a + \epsilon_k^2 + \epsilon_a^2 - \epsilon_b^2) - 
\sqrt{(x_a^2 - \epsilon_a^2)[1 - 2x_a + \epsilon_a^2 - (\epsilon_b-\epsilon_k)^2][1-2x_a + \epsilon_a^2 - (\epsilon_b+\epsilon_k)^2]}}
{2[\epsilon_a^2 + (1-2x_a)]}\,,\label{eq:BC2}\\
\frac{E_{a,{\rm max}}}{Q} & = \frac{1}{2}[1 + \epsilon_a^2 - (\epsilon_b + \epsilon_k)^2]\,,\label{eq:BC3}\\
\frac{E_{a,{\rm min}}}{Q} & = \epsilon_a\,,\label{eq:BC4}
\end{align}
with $\epsilon_i \equiv m_i/Q$ and $x_a \equiv E_a/Q$.
The total decay rate, therefore, is given by
\begin{align}\label{eq:TotalGammaC}
\Gamma^R = \frac{1}{64\pi^3 Q}\int_{E_{a,{\rm min}}}^{E_{a,{\rm max}}}dE_a
\int_{E_{k,{\rm min}}(E_a)}^{E_{k,{\rm max}}(E_a)}dE_k\,
|\mathcal{M}_{q\to p_a p_b k}|^2
\equiv 
\frac{1}{64\pi^3 Q}\int_{\rm BC}dE_a dE_k|\mathcal{M}_{q\to p_a p_b k}|^2
\,.
\end{align}
We now move to discuss the structure of the amplitude squared $|\mathcal{M}_{q\to p_a p_b k}|^2$.
\\
Diagrammatically, we have
\begin{align}\label{eq:RealEmission}
  |\mathcal{M}_{q\to p_a p_b k}|^2 =	& 
    \raisebox{-12mm}{
	\begin{tikzpicture}
	\draw (-1.1,-1.25)--(-1.1,1.25); 
	\end{tikzpicture}}
        \raisebox{-12mm}{
	\begin{tikzpicture}
	\draw[->-=.65,>=Latex][thick] (-0.75,0)--(0,0);
	\draw[->-=.7,>=Latex][thick] (0,0)--(1.25/2,1.25/2); 
	\draw[->-=.7,>=Latex][thick] (1.25/2,1.25/2)--(1.25,1.25); 
	\draw[->-=.7,>=Latex][thick] (1.25/2,1.25/2)--(1.25,0); 
	\draw[->-=.6,>=Latex][thick] (0,0)--(1.25,-1.25);
	\draw[black,fill=violachiaro,thick] (0,0)circle(2.75pt);
	\draw[black,fill=violachiaro,thick] (1.25/2,1.25/2)circle(2.75pt); 
	\node at (-0.6,0.35) {\scalebox{1}{$q$}};  
	\node at (0.15,0.7) {\scalebox{1}{{\color{oucrimsonred}{$A$}}}}; 
	\node at (0.7,0.1) {\scalebox{1}{{\color{oucrimsonred}{$X$}}}};  
	\node at (1.2,0.8) {\scalebox{1}{$p_a$}};
	\node at (1.2,-0.8) {\scalebox{1}{$p_b$}};
	\node at (1.2,-0.3) {\scalebox{1}{$k$}};
	\end{tikzpicture}} ~~~~~~+~~~~~~ 
  	\raisebox{-12mm}{
	\begin{tikzpicture}
	\draw[->-=.65,>=Latex][thick] (-0.75,0)--(0,0);
	\draw[->-=.6,>=Latex][thick] (0,0)--(1.25,1.25); 
	\draw[->-=.7,>=Latex][thick] (0,0)--(1.25/2,-1.25/2);
	\draw[->-=.7,>=Latex][thick] (1.25/2,-1.25/2)--(1.25,-1.25);
	\draw[->-=.7,>=Latex][thick] (1.25/2,-1.25/2)--(1.25,0);
	\node at (0.15,-0.7) {\scalebox{1}{{\color{oucrimsonred}{$B$}}}};
	\node at (0.7,-0.1) {\scalebox{1}{{\color{oucrimsonred}{$X$}}}}; 
	\draw[black,fill=violachiaro,thick] (0,0)circle(2.75pt);
	\draw[black,fill=violachiaro,thick] (1.25/2,-1.25/2)circle(2.75pt);
	\node at (-0.6,0.35) {\scalebox{1}{$q$}}; 
	\node at (1.2,0.8) {\scalebox{1}{$p_a$}};
	\node at (1.2,-0.8) {\scalebox{1}{$p_b$}};	
	\node at (1.2,+0.3) {\scalebox{1}{$k$}};
	\end{tikzpicture}}
	\raisebox{-12mm}{
	\begin{tikzpicture}
	\draw (-0.5,-1.25)--(-0.5,1.25); 
	\node at (-.2,1.25) {\scalebox{0.85}{2}}; 
	\end{tikzpicture}}
	~~~~~~=~~~~~~ 
\raisebox{-12mm}{
	\begin{tikzpicture}
	\draw[->-=.65,>=Latex][thick] (-0.75,0)--(0,0);
	\draw[->-=.7,>=Latex][thick] (0,0)--(1.25/2,1.25/2); 
	\draw[->-=.7,>=Latex][thick] (1.25/2,1.25/2)--(1.25,1.25); 
	\draw[->-=.7,>=Latex][thick] (1.25/2,1.25/2)--(1.25,0); 
	\draw[->-=.6,>=Latex][thick] (0,0)--(1.25,-1.25);
	\draw[->-=.7,>=Latex][thick] (1.25,1.25)--(1.25+1.25/2,1.25/2);
	\draw[->-=.7,>=Latex][thick] (1.25+1.25/2,1.25/2)--(2.5,0);
	\draw[->-=.7,>=Latex][thick] (1.25,0)--(1.25+1.25/2,1.25/2); 	
	\draw[->-=.7,>=Latex][thick] (1.25,-1.25)--(2.5,0);
	\draw[->-=.7,>=Latex][thick] (2.5,0)--(2.5+0.75,0);
	\draw[thick,dash dot,persianblue][thick] (1.25,1.25)--(1.25,-1.25);
	\draw[black,fill=violachiaro,thick] (0,0)circle(2.75pt);
	\draw[black,fill=violachiaro,thick] (1.25/2,1.25/2)circle(2.75pt);
	\draw[black,fill=violachiaro,thick] (1.25+1.25/2,1.25/2)circle(2.75pt);
	\draw[black,fill=violachiaro,thick] (2.5,0)circle(2.75pt);   
	\end{tikzpicture}}	~~~~~~+ \nn \\ &
	\raisebox{-12mm}{
	\begin{tikzpicture}
	\draw[->-=.65,>=Latex][thick] (-0.75,0)--(0,0);
	\draw[->-=.6,>=Latex][thick] (0,0)--(1.25,1.25); 
	\draw[->-=.7,>=Latex][thick] (0,0)--(1.25/2,-1.25/2);
	\draw[->-=.7,>=Latex][thick] (1.25/2,-1.25/2)--(1.25,-1.25);
	\draw[->-=.7,>=Latex][thick] (1.25/2,-1.25/2)--(1.25,0);
	\draw[->-=.7,>=Latex][thick] (1.25,1.25)--(2.5,0);
	\draw[->-=.7,>=Latex][thick] (1.25,-1.25)--(1.25+1.25/2,-1.25/2);
	\draw[->-=.7,>=Latex][thick] (1.25+1.25/2,-1.25/2)--(2.5,0);	
	\draw[->-=.7,>=Latex][thick] (1.25,0)--(1.25+1.25/2,-1.25/2);
	\draw[->-=.7,>=Latex][thick] (2.5,0)--(2.5+0.75,0);
	\draw[thick,dash dot,persianblue][thick] (1.25,1.25)--(1.25,-1.25);
	\draw[black,fill=violachiaro,thick] (0,0)circle(2.75pt);
	\draw[black,fill=violachiaro,thick] (2.5,0)circle(2.75pt); 
	\draw[black,fill=violachiaro,thick] (1.25/2,-1.25/2)circle(2.75pt);
	\draw[black,fill=violachiaro,thick] (1.25+1.25/2,-1.25/2)circle(2.75pt);   
	\end{tikzpicture}}
	~~~~~~+~~~~~~
	\raisebox{-12mm}{
	\begin{tikzpicture}
	\draw[->-=.65,>=Latex][thick] (-0.75,0)--(0,0);
	\draw[->-=.7,>=Latex][thick] (0,0)--(1.25/2,1.25/2); 
	\draw[->-=.7,>=Latex][thick] (1.25/2,1.25/2)--(1.25,1.25); 
	\draw[->-=.7,>=Latex][thick] (1.25/2,1.25/2)--(1.25,0); 
	\draw[->-=.6,>=Latex][thick] (0,0)--(1.25,-1.25);
	\draw[->-=.7,>=Latex][thick] (1.25,1.25)--(2.5,0);
	\draw[->-=.7,>=Latex][thick] (1.25,-1.25)--(1.25+1.25/2,-1.25/2);
	\draw[->-=.7,>=Latex][thick] (1.25+1.25/2,-1.25/2)--(2.5,0);	
	\draw[->-=.7,>=Latex][thick] (1.25,0)--(1.25+1.25/2,-1.25/2);	
	\draw[->-=.7,>=Latex][thick] (2.5,0)--(2.5+0.75,0);
	\draw[thick,dash dot,persianblue][thick] (1.25,1.25)--(1.25,-1.25);
	\draw[black,fill=violachiaro,thick] (0,0)circle(2.75pt);
	\draw[black,fill=violachiaro,thick] (1.25+1.25/2,-1.25/2)circle(2.75pt); 
	\draw[black,fill=violachiaro,thick] (1.25/2,1.25/2)circle(2.75pt);  
	\draw[black,fill=violachiaro,thick] (2.5,0)circle(2.75pt);
	\end{tikzpicture}}~~~~~~+~~~~~~
	\raisebox{-12mm}{
	\begin{tikzpicture}
	\draw[->-=.65,>=Latex][thick] (-0.75,0)--(0,0);
	\draw[->-=.6,>=Latex][thick] (0,0)--(1.25,1.25); 
	\draw[->-=.7,>=Latex][thick] (0,0)--(1.25/2,-1.25/2);
	\draw[->-=.7,>=Latex][thick] (1.25/2,-1.25/2)--(1.25,-1.25);
	\draw[->-=.7,>=Latex][thick] (1.25/2,-1.25/2)--(1.25,0);
	\draw[->-=.7,>=Latex][thick] (1.25,1.25)--(1.25+1.25/2,1.25/2);
	\draw[->-=.7,>=Latex][thick] (1.25+1.25/2,1.25/2)--(2.5,0);
	\draw[->-=.7,>=Latex][thick] (1.25,0)--(1.25+1.25/2,1.25/2); 	
	\draw[->-=.7,>=Latex][thick] (1.25,-1.25)--(2.5,0);
	\draw[->-=.7,>=Latex][thick] (2.5,0)--(2.5+0.75,0);
	\draw[thick,dash dot,persianblue][thick] (1.25,1.25)--(1.25,-1.25);
	\draw[black,fill=violachiaro,thick] (0,0)circle(2.75pt);
	\draw[black,fill=violachiaro,thick] (2.5,0)circle(2.75pt);
	\draw[black,fill=violachiaro,thick] (1.25+1.25/2,1.25/2)circle(2.75pt);
	\draw[black,fill=violachiaro,thick] (1.25/2,-1.25/2)circle(2.75pt);   
	\end{tikzpicture}}
\end{align}
where $A$ and $B$ indicate intermediate propagators 
with masses given, in full generality, by $m_A$ and $m_B$, respectively.  
Furthermore, we indicate with $X$ the soft particle with four-momentum $k$ that is radiated in the final state.
In the explicit evaluation of eq.\,(\ref{eq:RealEmission}) we can make use of the same technical substitutions   
discussed in appendix\,\ref{p+k}.
Schematically, the four terms on the right-hand side of eq.\,(\ref{eq:RealEmission}) will have, after these substitutions, the following form
\begin{align}\label{eq:SquaredEx}
	\raisebox{-12mm}{
	\begin{tikzpicture}
	\draw[->-=.65,>=Latex][thick] (-0.75,0)--(0,0);
	\draw[->-=.7,>=Latex][thick] (0,0)--(1.25/2,1.25/2); 
	\draw[->-=.7,>=Latex][thick] (1.25/2,1.25/2)--(1.25,1.25); 
	\draw[->-=.7,>=Latex][thick] (1.25/2,1.25/2)--(1.25,0); 
	\draw[->-=.6,>=Latex][thick] (0,0)--(1.25,-1.25);
	\draw[->-=.7,>=Latex][thick] (1.25,1.25)--(2.5,0);
	\draw[->-=.7,>=Latex][thick] (1.25,-1.25)--(1.25+1.25/2,-1.25/2);
	\draw[->-=.7,>=Latex][thick] (1.25+1.25/2,-1.25/2)--(2.5,0);	
	\draw[->-=.7,>=Latex][thick] (1.25,0)--(1.25+1.25/2,-1.25/2);	
	\draw[->-=.7,>=Latex][thick] (2.5,0)--(2.5+0.75,0);
	\draw[thick,dash dot,persianblue][thick] (1.25,1.25)--(1.25,-1.25);
	\draw[black,fill=violachiaro,thick] (0,0)circle(2.75pt);
	\draw[black,fill=violachiaro,thick] (1.25+1.25/2,-1.25/2)circle(2.75pt); 
	\draw[black,fill=violachiaro,thick] (1.25/2,1.25/2)circle(2.75pt);  
	\draw[black,fill=violachiaro,thick] (2.5,0)circle(2.75pt);
	\node at (0.1,0.6) {\scalebox{1}{{\color{oucrimsonred}{$A$}}}};
	\node at (0.7,0.1) {\scalebox{1}{{\color{oucrimsonred}{$X$}}}}; 
	\node at (2.45,-0.5) {\scalebox{1}{{\color{oucrimsonred}{$B$}}}}; 	 	
	\end{tikzpicture}} 
 \equiv 
 \frac{1
 }{D_A D_B}\times 
 \underbrace{{\rm function\,of\,couplings\,and\,masses\,with\,dim\,4}}_{=\,Q^4 R_{AB}^{(X)}}
\end{align}
with analogue expressions for the remaining three diagrams; in eq.\,(\ref{eq:SquaredEx}) we factored out the two propagator denominators, and the remaining amplitude squared is nothing but a combination of couplings and masses (with mass-dimension 4) that factors out from the integration over $\{E_a,E_k\}$ required in eq.\,(\ref{eq:GammaC}). 
The propagator denominators, on the contrary, depend on energy and, therefore, must be included in the integration that gives the total decay rate.  
In terms of the integration variables $\{E_a,E_k\}$, the explicit form of the two propagators is 
\begin{align}
D_A = 2Q(E_a + E_k) - Q^2 +m_b^2 -m_A^2\,,~~~~~~~~~~~~~~D_B = Q(Q - 2E_a) + m_a^2 - m_B^2\,.
\end{align}
In eq.\,(\ref{eq:SquaredEx}) we introduced the a-dimensional function $R_{AB}^{(X)}$ which we already
defined in eq.\,(\ref{oiuh}); the subindices in $R_{AB}^{(X)}$ refer to the internal particles
 while the subscript refers to the  real emission (see also eq.\,(\ref{pkth}) where 
 the same notation was used).   
\\
Notice that the first two diagrams in eq.\,(\ref{eq:RealEmission}) vanish in our approximation since they do not give double-logarithmic infrared corrections. 
 \\
Coming back to eq.\,(\ref{eq:TotalGammaC}), we find that the diagram in eq.\,(\ref{eq:SquaredEx}) contributes to the total 
three-body decay width as follows
\begin{align}\label{eq:ExpressionIR}
\Gamma^R  \ni
Q\,R_{AB}^{(X)}\,\mathcal{I}^R\,,~~~~~~~~~{\rm with}~~~~ 
 \mathcal{I}^R \equiv \frac{Q^2}{64\pi^3}\int_{\rm BC}\frac{dE_a dE_k}{D_A D_B}\,.
\end{align}
In the above expression $\mathcal{I}^R$ is the a-dimensional phase-space integral we introduced 
in eq.\,(\ref{oiuh}). We rewrite it in the form
\begin{align}\label{eq:IntegraleBastardo}
\mathcal{I}^R = \frac{1}{64\pi^3}
\int_{\rm BC}\frac{dx_a dx_k}{
[2(x_a + x_k) - 1 + \epsilon_b^2 - \epsilon_A^2](1-2x_a + \epsilon_a^2 - \epsilon_B^2)}\,,
\end{align}
with boundary conditions in eqs.\,(\ref{eq:BC1}-\ref{eq:BC4}).
We consider the following limits.
\begin{itemize}
\item[$\circ$] {\underline{Massless QED}}.
This case corresponds to $m_a = m_A = m_b = m_B = 0$ and $m_k = \lambda$ with $\lambda$ a fictitious photon mass. 
In this case the integrations can be carried out exactly. 
We find
\begin{align}
\mathcal{I}^R = 
\frac{1}{256\,\pi^3}\,\left[-{\rm Li}_2\left(\frac{1}{1+\epsilon^2}\right) + {\rm Li}_2\left(\frac{\epsilon^2}{1+\epsilon^2}\right)+
2\;\log(\epsilon)\;\log\left(\frac{\epsilon}{1+\epsilon^2}\right)\right]
\end{align}
In the double-log limit, we find
\begin{align}\label{eq:DueIndizi}
\mathcal{I}^R = 
\frac{1}{512\,\pi^3}\log^2 Q^2\,,
\end{align}
which is precisely our result for $\mathcal{I}^V$ in  eq.\,(\ref{eq:AsiIV}).
\end{itemize}
We would like to prove the relation $\mathcal{I}^R = \mathcal{I}^V$ in more general situations beyond the 
simple case of the massless QED. 
{Unfortunately, 
we were not able to perform analytically the integration in eq.\,(\ref{eq:IntegraleBastardo}) in the most generic mass 
configuration. 
Nevertheless, we are in the position to make few interesting comments.
\begin{itemize}
\item[$\circ$] {\underline{Massive QED}}. 
This case corresponds to $m_a = m_A = m_b = m_B = m_{\psi}$ and $m_k = \lambda$ with $\lambda$ a fictitious photon mass.
In this case we can invoke the KLN theorem. As previously discussed, 
the virtual integral $\mathcal{I}^V$ does not contain power-suppressed double logarithms.  
This must be true also for $\mathcal{I}^R$ otherwise terms of the kind $(m_{\psi}^2/Q^2)\log(Q/m_{\psi})\log(Q/\lambda)$ 
would give rise to uncanceled IR divergences in the physical limit $\lambda \to 0$, in contradiction with the KLN theorem.
\item[$\circ$] {\underline{Generic mass configuration}}. 
Another case which can be handled analytically is the mass configuration 
$m_a = m_b = 0$, $m_A = m_B = m_{\psi}$, $m_k = \lambda$. 
We do not report here the full result of the integration in eq.\,(\ref{eq:IntegraleBastardo}) but, at the double-logarithmic level, we find again 
eq.\,(\ref{eq:DueIndizi}) with no power-suppressed terms.
\end{itemize} 
 }

\section{Results \label{app:2b}}

In this appendix we collect our results about the explicit computation of the bubble diagrams classified as in fig.\,\ref{ASJ}. 
We make use of the Feynman rules and WI derived in appendix\,\ref{app:Coupl}.
Remember that whenever a virtual $Z$ appears, we automatically include the $\phi$ Goldstone boson contribution. 

\subsection{$S_1$ diagram}

We start from the bubble dubbed $S_1$ in the classification illustrated in fig.\,\ref{ASJ}. 
The virtual diagrams are given by
\begin{align}
 V_{\psi\bar{\psi}}^{(Z)} = & 
 \raisebox{-12mm}{
	\begin{tikzpicture}
	\draw[thick,style={decorate, decoration={snake}}] (-0.75,0)--(0,0);
	\draw[->-=.6,>=Latex][thick] (0,0)--(1.25/2,1.25/2); 
	\draw[->-=.6,>=Latex][thick] (1.25/2,1.25/2)--(1.25,1.25); 
	\draw[->-=.8,>=Latex][thick] (1.25,-1.25)--(1.25/2,-1.25/2);
	\draw[->-=.8,>=Latex][thick] (1.25/2,-1.25/2)--(0,0);
	\draw[thick,style={decorate, decoration={snake}}] (1.25/2,1.25/2)--(1.25/2,-1.25/2);
	\draw[->-=.6,>=Latex][thick] (1.25,1.25)--(2.5,0); 
	\draw[->-=.6,>=Latex][thick] (2.5,0)--(1.25,-1.25); 	
	\draw[thick,dash dot,persianblue][thick] (1.25,1.25)--(1.25,-1.25); 
	\draw[thick,style={decorate, decoration={snake}}] (2.5,0)--(2.5+0.75,0);
	\draw[black,fill=tangerineyellow,thick] (0,0)circle(2.75pt); 
	\draw[black,fill=tangerineyellow,thick] (2.5,0)circle(2.75pt); 
	\draw[black,fill=tangerineyellow,thick] (1.25/2,1.25/2)circle(2.75pt); 
	\draw[black,fill=tangerineyellow,thick] (1.25/2,-1.25/2)circle(2.75pt); 
	\node at (0.85,0) {\scalebox{0.85}{$Z$}};
	\node at (-0.2,-0.4) {\scalebox{0.85}{$\mathcal{F}_{Z'i}$}};
	\node at (2.7,-0.4) {\scalebox{0.85}{$\mathcal{F}_{Z'i}^*$}};
	\node at (0.35,-0.85) {\scalebox{0.85}{$\mathcal{F}_{Zi}$}};
	\node at (0.35,+0.95) {\scalebox{0.85}{$\mathcal{F}_{Zi}$}};
	\end{tikzpicture}} ~~ + ~~
 \underbrace{\raisebox{-12mm}{
	\begin{tikzpicture}
	\draw[thick,style={decorate, decoration={snake}}] (-0.75,0)--(0,0);
	\draw[->-=.6,>=Latex][thick] (0,0)--(1.25,1.25); 
	\draw[->-=.6,>=Latex][thick] (1.25,-1.25)--(0,0); 
	\draw[->-=.6,>=Latex][thick] (1.25,1.25)--(2.5-1.25/2,1.25/2);
	\draw[->-=.6,>=Latex][thick] (2.5-1.25/2,1.25/2)--(2.5,0); 
	\draw[->-=.8,>=Latex][thick] (2.5,0)--(2.5-1.25/2,-1.25/2); 
	\draw[->-=.6,>=Latex][thick] (2.5-1.25/2,-1.25/2)--(1.25,-1.25); 
	\draw[thick,style={decorate, decoration={snake}}] (2.5-1.25/2,-1.25/2)--(2.5-1.25/2,1.25/2);
	\draw[thick,dash dot,persianblue][thick] (1.25,1.25)--(1.25,-1.25); 
	\draw[thick,style={decorate, decoration={snake}}] (2.5,0)--(2.5+0.75,0);
	\draw[black,fill=tangerineyellow,thick] (2.5-1.25/2,-1.25/2)circle(2.75pt); 
	\draw[black,fill=tangerineyellow,thick] (2.5-1.25/2,1.25/2)circle(2.75pt); 
	\draw[black,fill=tangerineyellow,thick] (0,0)circle(2.75pt); 
	\draw[black,fill=tangerineyellow,thick] (2.5,0)circle(2.75pt); 
	\node at (1.62,0) {\scalebox{0.85}{$Z$}};
	\node at (-0.2,-0.4) {\scalebox{0.85}{$\mathcal{F}_{Z'i}$}};
	\node at (2.7,-0.4) {\scalebox{0.85}{$\mathcal{F}_{Z'i}^*$}};
	\node at (2.25,-0.85) {\scalebox{0.85}{$\mathcal{F}_{Zi}^*$}};
	\node at (2.25,+0.85) {\scalebox{0.85}{$\mathcal{F}_{Zi}^*$}};
	\end{tikzpicture}}}_{h.c.}
 \label{eq:Vir1} \\ = & -\frac{8}{3}\Bigg\{\mathcal{F}_{Z'L}\mathcal{F}_{Z'R}
	\epsilon_{\psi }^2
	\left[2\mathcal{F}_{ZR}
   \mathcal{F}_{ZL}\left(1+ \epsilon ^2 -3\epsilon_{\psi }^2\right) + 
   \left(\mathcal{F}_{ZR}^2+\mathcal{F}_{ZL}^2\right)\left(2 + \epsilon ^2 -3\epsilon_{\psi }^2\right)\right]+
  \nn  \\ & \left. 
   \mathcal{F}_{Z'L}^2 \left[\mathcal{F}_{ZR} \mathcal{F}_{ZL} \left(\left(2 \epsilon
   ^2+1\right) \epsilon _{\psi }^2-3 \epsilon _{\psi }^4\right)+\mathcal{F}_{ZL}^2 \left(-4
   \left(\epsilon ^2+1\right) \epsilon _{\psi }^2+\left(\epsilon ^2+1\right)^2+\frac{5
   \epsilon _{\psi }^4}{2}\right) +\frac{5}{2}
   \mathcal{F}_{ZR}^2 \epsilon _{\psi }^4\right]+
    \right. \nn \\ & 
  \mathcal{F}_{Z'R}^2
   \left[\mathcal{F}_{ZR} \mathcal{F}_{ZL} \left(\left(2 \epsilon
   ^2+1\right) \epsilon _{\psi }^2-3 \epsilon _{\psi }^4\right)+\mathcal{F}_{ZR}^2
   \left(-4 \left(\epsilon ^2+1\right) \epsilon _{\psi }^2+\left(\epsilon
   ^2+1\right)^2+\frac{5 \epsilon _{\psi }^4}{2}\right)+\frac{5}{2}
   \mathcal{F}_{ZL}^2 \epsilon _{\psi }^4\right]\Bigg\} \nn
  \\  \simeq & 
   -\frac{8}{3}  g^2  g^{\prime 2}  \left(f_L^2  y_L^2+f_R^2  y_R^2\right)+
   \frac{2\epsilon ^2}{3  y_{\phi}^2}  \bigg\{-8  g^2  g^{\prime 2}  y_{\phi} \left[f_L^2  y_L^2  \left(y_R-2  y_L\right)+f_R^2 
   y_R^2  \left(2  y_R-y_L\right)+f_L  f_R  \left(y_L^3-y_R^3\right)\right]+
      \nn \\ & \nonumber
   2  g^{\prime 2}  h_{\mathit{f}}^2  \left[-y_L  y_R  \left(f_L+f_R\right)^2+2 
   f_L  y_L^2  \left(2  f_L-f_R\right)-2  f_R  y_R^2  \left(f_L-2  f_R\right)\right]
   - 8  g_1^4  y_{\phi} \left(f_L-f_R\right)  \left(f_L^3  y_L+f_R^3  y_R\right) \bigg\}. \nn 
\end{align}
As stated before, we include the contribution of the virtual $\phi$ exchange. 
The corresponding coupling with fermions $\mathcal{F}_{\phi\psi}$, given in eq.\,(\ref{eq:Fey2}), is rewritten by means of the Ward 
identities in eq.\,(\ref{fqe}). Furthermore, we use the identity $\epsilon_{\phi} = \epsilon$ (see comment 
below eq.\,(\ref{eq:IdeMarch})). 
In eq.\,(\ref{eq:Vir1}), as well as in all following diagrams, {\it h.c.} stands for the hermitian conjugate diagram (from now 
on, we shall not draw them explicitly).

The real diagrams contributing to the same bubble are given by
\begin{align}
R_{\psi\bar{\psi}}^{(Z)} = &
 \raisebox{-12mm}{
	\begin{tikzpicture}
	\draw[thick,style={decorate, decoration={snake}}] (-0.75,0)--(0,0);
	\draw[->-=.7,>=Latex][thick] (0,0)--(1.25/2,1.25/2); 
	\draw[->-=.7,>=Latex][thick] (1.25/2,1.25/2)--(1.25,1.25); 
	\draw[->-=.6,>=Latex][thick] (1.25,-1.25)--(0,0);
	\draw[->-=.7,>=Latex][thick] (1.25,1.25)--(2.5,0);
	\draw[->-=.7,>=Latex][thick] (1.25+1.25/2,-1.25/2)--(1.25,-1.25);
	\draw[->-=.7,>=Latex][thick] (2.5,0)--(1.25+1.25/2,-1.25/2);		
	\draw[thick,style={decorate, decoration={snake}}] (2.5,0)--(2.5+0.75,0);
	\draw[thick,style={decorate, decoration={snake}}] (1.25/2,1.25/2)--(1.25+1.25/2,-1.25/2);
	\draw[thick,dash dot,persianblue][thick] (1.25,1.25)--(1.25,-1.25);
	\draw[black,fill=tangerineyellow,thick] (0,0)circle(2.75pt);
	\draw[black,fill=tangerineyellow,thick] (1.25+1.25/2,-1.25/2)circle(2.75pt); 
	\draw[black,fill=tangerineyellow,thick] (1.25/2,1.25/2)circle(2.75pt); 
	\draw[black,fill=tangerineyellow,thick] (2.5,0)circle(2.75pt);  
	\node at (0.85,-0.13) {\scalebox{0.85}{$Z$}};
	\node at (-0.2,-0.4) {\scalebox{0.85}{$\mathcal{F}_{Z'i}$}};
	\node at (2.7,-0.4) {\scalebox{0.85}{$\mathcal{F}_{Z'i}^*$}};
	\node at (0.35,+0.95) {\scalebox{0.85}{$\mathcal{F}_{Zi}$}};
	\node at (2.1,-0.95) {\scalebox{0.85}{$\mathcal{F}^*_{Zi}$}};
	\end{tikzpicture}} ~~ + ~~ h.c.
\end{align}	
We sum over the three polarizations of the emitted $Z$ boson. 
By explicit computation, we find precisely the relation $R_{\psi\bar{\psi}}^{(Z)} = -V_{\psi\bar{\psi}}^{(Z)}$ 
which is indeed guaranteed by the IR unitarity theorem in eq.\,(\ref{pkth}).

\subsection{$S_2$ diagram}

We now move to consider the bubble $S_2$. 
As far as the virtual diagrams are concerned, we find
\begin{align}
V_{\psi\bar{\psi}}^{(h)} & = 
  	\raisebox{-12mm}{
	\begin{tikzpicture}
	\draw[thick,style={decorate, decoration={snake}}] (-0.75,0)--(0,0);
	\draw[->-=.6,>=Latex][thick] (0,0)--(1.25/2,1.25/2); 
	\draw[->-=.6,>=Latex][thick] (1.25/2,1.25/2)--(1.25,1.25); 
	\draw[->-=.8,>=Latex][thick] (1.25,-1.25)--(1.25/2,-1.25/2);
	\draw[->-=.8,>=Latex][thick] (1.25/2,-1.25/2)--(0,0);
	\draw[thick,dashed] (1.25/2,1.25/2)--(1.25/2,-1.25/2);
	\draw[->-=.6,>=Latex][thick] (1.25,1.25)--(2.5,0); 
	\draw[->-=.6,>=Latex][thick] (2.5,0)--(1.25,-1.25); 	
	\draw[thick,dash dot,persianblue][thick] (1.25,1.25)--(1.25,-1.25); 
	\draw[thick,style={decorate, decoration={snake}}] (2.5,0)--(2.5+0.75,0);
	\draw[black,fill=tangerineyellow,thick] (0,0)circle(2.75pt); 
	\draw[black,fill=tangerineyellow,thick] (2.5,0)circle(2.75pt); 
	\draw[black,fill=tangerineyellow,thick] (1.25/2,1.25/2)circle(2.75pt); 
	\draw[black,fill=tangerineyellow,thick] (1.25/2,-1.25/2)circle(2.75pt); 
	\node at (0.85,0) {\scalebox{0.85}{$h$}};
	\node at (0.35,-0.85) {\scalebox{0.85}{$\mathcal{F}_{h \psi}$}};
	\node at (0.35,+0.95) {\scalebox{0.85}{$\mathcal{F}_{h \psi}$}};	
	\node at (-0.2,-0.4) {\scalebox{0.85}{$\mathcal{F}_{Z'i}$}};
	\node at (2.7,-0.4) {\scalebox{0.85}{$\mathcal{F}_{Z'i}^*$}};
	\end{tikzpicture}} \label{eq:Virtual5} ~~ + ~~ h.c.\\ & =
	-\frac{4}{3}\;\mathcal{F}_{h\psi}^2\bigg[
	(\mathcal{F}_{Z'L}^2+\mathcal{F}_{Z'R}^2)\,\epsilon_{\psi}^2\,(4-4\, \epsilon_{\psi}^2 + \epsilon_h^2) +
	2\,\mathcal{F}_{Z'L}\,\mathcal{F}_{Z'R}\,(\epsilon_h^2-4\,\epsilon_{\psi}^2)\,
	(\epsilon_h^2-3\,\epsilon_{\psi}^2)
	\bigg]\nn
	   \\ & 
	 \simeq
	-\frac{4 \, g^{\prime 2} \, \epsilon ^2 \, h_{\mathit{f}}^4 \, \left(f_L^2+f_R^2\right)}{3  \,g^2 \,
   y_{\phi}^2} \nn
\end{align}
As before, we find $V_{\psi\bar{\psi}}^{(h)} = -R_{\psi\bar{\psi}}^{(h)}$ with, diagrammatically
\begin{align}\label{eq:Re3}
R_{\psi\bar{\psi}}^{(h)} & =
    \raisebox{-12mm}{
	\begin{tikzpicture}
	\draw[thick,style={decorate, decoration={snake}}] (-0.75,0)--(0,0);
	\draw[->-=.7,>=Latex][thick] (0,0)--(1.25/2,1.25/2); 
	\draw[->-=.7,>=Latex][thick] (1.25/2,1.25/2)--(1.25,1.25); 
	\draw[->-=.6,>=Latex][thick] (1.25,-1.25)--(0,0);
	\draw[->-=.7,>=Latex][thick] (1.25,1.25)--(2.5,0);
	\draw[->-=.7,>=Latex][thick] (1.25+1.25/2,-1.25/2)--(1.25,-1.25);
	\draw[->-=.7,>=Latex][thick] (2.5,0)--(1.25+1.25/2,-1.25/2);		
	\draw[thick,style={decorate, decoration={snake}}] (2.5,0)--(2.5+0.75,0);
	\draw[thick,dashed] (1.25/2,1.25/2)--(1.25+1.25/2,-1.25/2);
	\draw[thick,dash dot,persianblue][thick] (1.25,1.25)--(1.25,-1.25);
	\draw[black,fill=tangerineyellow,thick] (0,0)circle(2.75pt);
	\draw[black,fill=tangerineyellow,thick] (1.25+1.25/2,-1.25/2)circle(2.75pt); 
	\draw[black,fill=tangerineyellow,thick] (1.25/2,1.25/2)circle(2.75pt);   
	\draw[black,fill=tangerineyellow,thick] (2.5,0)circle(2.75pt);  
	\node at (0.85,-0.13) {\scalebox{0.85}{$h$}};
	\node at (-0.2,-0.4) {\scalebox{0.85}{$\mathcal{F}_{Z'i}$}};
	\node at (2.7,-0.4) {\scalebox{0.85}{$\mathcal{F}_{Z'i}^*$}};
	\node at (0.35,+0.95) {\scalebox{0.85}{$\mathcal{F}_{h\psi}$}};
	\node at (2.1,-0.95) {\scalebox{0.85}{$\mathcal{F}_{h\psi}$}};
	\end{tikzpicture}}  ~~ + ~~ h.c.
\end{align}
 
 \subsection{$S_3$ diagram}

For the virtual diagram, we find
\begin{align}
V_{Zh}^{(Z)} & =         \raisebox{-12mm}{
	\begin{tikzpicture}
	\draw[thick,style={decorate, decoration={snake}}] (-0.75,0)--(0,0);
	\draw[thick,dashed] (0,0)--(1.25/2,1.25/2); 
	\draw[thick,style={decorate, decoration={snake}}] (1.25/2,-1.25/2)--(0,0);
	\draw[thick,style={decorate, decoration={snake}}](1.25/2,1.25/2)--(1.25/2,-1.25/2);
	\draw[thick,style={decorate, decoration={snake}}] (1.25/2,1.25/2)--(1.25,1.25); 
        \draw[thick,dashed] (1.25,-1.25)--(1.25/2,-1.25/2);
	\draw[thick,dash dot,persianblue][thick] (1.25,1.25)--(1.25,-1.25);
	\draw[thick,style={decorate, decoration={snake}}] (1.25,1.25)--(2.5,0); 
	\draw[thick,dashed] (2.5,0)--(1.25,-1.25); 	
	\draw[thick,dash dot,persianblue][thick] (1.25,1.25)--(1.25,-1.25); 
	\draw[thick,style={decorate, decoration={snake}}] (2.5,0)--(2.5+0.75,0);
	\draw[black,fill=tangerineyellow,thick] (0,0)circle(2.75pt);
	\draw[black,fill=tangerineyellow,thick] (1.25/2,1.25/2)circle(2.75pt);
	\draw[black,fill=tangerineyellow,thick] (1.25/2,-1.25/2)circle(2.75pt);
	\draw[black,fill=tangerineyellow,thick] (2.5,0)circle(2.75pt); 
	\node at (0.85,0) {\scalebox{0.85}{$Z$}};
	\node at (0.15,0.5) {\scalebox{0.85}{$h$}};
	\node at (0.43,-0.2) {\scalebox{0.85}{$Z$}};
	\node at (-0.2,-0.4) {\scalebox{0.85}{$\mathcal{F}_{Z'Zh}$}};
	\node at (2.7,-0.4) {\scalebox{0.85}{$\mathcal{F}_{Z'Zh}$}};
	\node at (0.35,0.95) {\scalebox{0.85}{$\mathcal{F}_{Zh\phi}$}};
	\node at (0.35,-0.85) {\scalebox{0.85}{$\mathcal{F}_{Zh\phi}$}};
	\node at (1,-0.7) {\scalebox{0.85}{$h$}};
	\node at (1,+0.62) {\scalebox{0.85}{$Z$}};	
	\end{tikzpicture}}	~~ + ~~ h.c.  \label{eq:VirSqZh4} \\
&= -\frac{1}{24} \mathcal{F}_{ZZh}^2 \mathcal{F}_{Z'Zh}^2\left[\epsilon^4 \left(30-12 \epsilon_{h}^2\right)+\epsilon^2
   \left(-5 \epsilon_{h}^4-16 \epsilon _{h}^2+6\right)+\frac{\epsilon
   _{h}^4 \left(\epsilon _{h}^4+2 \epsilon
   _{h}^2-1\right)}{\epsilon ^2}+6 \epsilon _{h}^6-2 \epsilon
   _{h}^4-4 \epsilon _{h}^2+10 \epsilon ^6+2\right] \nn
   \\ & 
	 \simeq
	 -\frac{4}{3} g^2 g^{\prime 2} f_{\phi }^2 y_{\phi }^2+\frac{4 g^{\prime 2} \epsilon ^2 f_{\phi }^2 }{3 g^2 y_{\phi }^2}\left[4 g^2 y_{\phi }^2 \left(g^{\prime 2} f_{\phi
   }^2+\lambda \right)-5 g^4 y_{\phi }^4+2 \lambda ^2\right]\,.\nn
\end{align}
Notice the presence of a non-vanishing term in the limit $\epsilon\to 0$ which corresponds to a leading double-logarithm. 
As far as the three-body decay diagrams are concerned, the IR unitarity theorem is verified by the inclusion of
\begin{align}\label{eq:ReZh4}
R_{ZZ}^{(Z)} = 
\raisebox{-12mm}{
	\begin{tikzpicture}
	\draw[thick,style={decorate, decoration={snake}}] (-0.75,0)--(0,0);
	\draw[thick,dashed] (0,0)--(1.25/2,1.25/2); 
	\draw[thick,style={decorate, decoration={snake}}] (1.25/2,1.25/2)--(1.25,1.25); 
	\draw[thick,style={decorate, decoration={snake}}] (1.25,-1.25)--(0,0);
	\draw[thick,style={decorate, decoration={snake}}] (1.25,1.25)--(2.5,0);
	\draw[thick,style={decorate, decoration={snake}}] (1.25+1.25/2,-1.25/2)--(1.25,-1.25);
	\draw[thick,dashed] (2.5,0)--(1.25+1.25/2,-1.25/2);
	\draw[thick,style={decorate, decoration={snake}}] (2.5,0)--(2.5+0.75,0);
	\draw[thick,style={decorate, decoration={snake}}] (1.25/2,1.25/2)--(1.25+1.25/2,-1.25/2);
	\draw[thick,dash dot,persianblue][thick] (1.25,1.25)--(1.25,-1.25);
	\draw[black,fill=tangerineyellow,thick] (0,0)circle(2.75pt);
	\draw[black,fill=tangerineyellow,thick] (1.25+1.25/2,-1.25/2)circle(2.75pt); 
	\draw[black,fill=tangerineyellow,thick] (1.25/2,1.25/2)circle(2.75pt);   
	\draw[black,fill=tangerineyellow,thick] (2.5,0)circle(2.75pt);  
	\node at (0.85,-0.13) {\scalebox{0.85}{$Z$}};
	\node at (0.12,0.5) {\scalebox{0.85}{$h$}};
	\node at (-0.2,-0.4) {\scalebox{0.85}{$\mathcal{F}_{Z'Zh}$}};
	\node at (2.7,-0.4) {\scalebox{0.85}{$\mathcal{F}_{Z'Zh}$}};
	\node at (0.35,+0.95) {\scalebox{0.85}{$\mathcal{F}_{ZZh}$}};
	\node at (2.1,-0.95) {\scalebox{0.85}{$\mathcal{F}_{ZZh}$}};
	\node at (2,-0.15) {\scalebox{0.85}{$h$}};
	\node at (1,+0.65) {\scalebox{0.85}{$Z$}};
	\node at (1.075,-0.62) {\scalebox{0.85}{$Z$}};
	\end{tikzpicture}} ~~ + ~~ h.c.\,,~~~~~~~~~~~~~~
R_{hh}^{(Z)} = 
\raisebox{-12mm}{
	\begin{tikzpicture}
	\draw[thick,style={decorate, decoration={snake}}] (-0.75,0)--(0,0);
	\draw[thick,style={decorate, decoration={snake}}] (0,0)--(1.25/2,1.25/2); 
	\draw[thick,dashed] (1.25/2,1.25/2)--(1.25,1.25); 
	\draw[thick,dashed] (1.25,-1.25)--(0,0);
	\draw[thick,dashed] (1.25,1.25)--(2.5,0);
	\draw[thick,dashed] (1.25+1.25/2,-1.25/2)--(1.25,-1.25);
	\draw[thick,style={decorate, decoration={snake}}] (2.5,0)--(1.25+1.25/2,-1.25/2);
	\draw[thick,style={decorate, decoration={snake}}] (2.5,0)--(2.5+0.75,0);
	\draw[thick,style={decorate, decoration={snake}}] (1.25/2,1.25/2)--(1.25+1.25/2,-1.25/2);
	\draw[thick,dash dot,persianblue][thick] (1.25,1.25)--(1.25,-1.25);
	\draw[black,fill=tangerineyellow,thick] (0,0)circle(2.75pt);
	\draw[black,fill=tangerineyellow,thick] (1.25+1.25/2,-1.25/2)circle(2.75pt); 
	\draw[black,fill=tangerineyellow,thick] (1.25/2,1.25/2)circle(2.75pt);   
	\draw[black,fill=tangerineyellow,thick] (2.5,0)circle(2.75pt);  
	\node at (0.85,-0.13) {\scalebox{0.85}{$Z$}};
	\node at (0.12,0.5) {\scalebox{0.85}{$Z$}};
	\node at (-0.2,-0.4) {\scalebox{0.85}{$\mathcal{F}_{Z'Zh}$}};
	\node at (2.7,-0.4) {\scalebox{0.85}{$\mathcal{F}_{Z'Zh}$}};
	\node at (0.35,+0.95) {\scalebox{0.85}{$\mathcal{F}_{ZZh}$}};
	\node at (2.1,-0.95) {\scalebox{0.85}{$\mathcal{F}_{ZZh}$}};
	\node at (2,-0.15) {\scalebox{0.85}{$Z$}};
	\node at (1,+0.65) {\scalebox{0.85}{$h$}};
	\node at (1,-0.65) {\scalebox{0.85}{$h$}};
	\end{tikzpicture}} ~~ + ~~ h.c.\,.
\end{align}

\subsection{$A_1$ diagram}

Consider now the diagrams that belong to the bubble $A_1$. 
We focus first on diagrams of the kind $A_1(a)$. Explicitly, we have 
(the factor 2 in front of the first diagram accounts for the exchange of the internal particles in the loop)
\begin{align}
V_{\psi\bar{\psi}}^{(\psi)} &= 
  	2\times\raisebox{-12mm}{
	\begin{tikzpicture}
	\draw[thick,style={decorate, decoration={snake}}] (-0.75,0)--(0,0);
	\draw[thick,dashed] (0,0)--(1.25/2,1.25/2); 
	\draw[->-=.6,>=Latex][thick] (1.25/2,1.25/2)--(1.25,1.25); 
	\draw[->-=.8,>=Latex][thick] (1.25,-1.25)--(1.25/2,-1.25/2);
	\draw[thick,style={decorate, decoration={snake}}] (1.25/2,-1.25/2)--(0,0);
	\draw[->-=.6,>=Latex][thick] (1.25/2,-1.25/2)--(1.25/2,1.25/2);
	\draw[->-=.6,>=Latex][thick] (1.25,1.25)--(2.5,0); 
	\draw[->-=.6,>=Latex][thick] (2.5,0)--(1.25,-1.25); 	
	\draw[thick,dash dot,persianblue][thick] (1.25,1.25)--(1.25,-1.25); 
	\draw[thick,style={decorate, decoration={snake}}] (2.5,0)--(2.5+0.75,0);
	\draw[black,fill=tangerineyellow,thick] (0,0)circle(2.75pt); 
	\draw[black,fill=tangerineyellow,thick] (2.5,0)circle(2.75pt); 
	\draw[black,fill=tangerineyellow,thick] (1.25/2,1.25/2)circle(2.75pt); 
	\draw[black,fill=tangerineyellow,thick] (1.25/2,-1.25/2)circle(2.75pt); 
	\node at (0.35,-0.85) {\scalebox{0.85}{$\mathcal{F}_{Z i}$}};
	\node at (0.35,+0.95) {\scalebox{0.85}{$\mathcal{F}_{h \psi}$}};	
	\node at (-0.2,-0.4) {\scalebox{0.85}{$\mathcal{F}_{Z'Zh}$}};
	\node at (2.7,-0.4) {\scalebox{0.85}{$\mathcal{F}_{Z'i}^*$}};
	\end{tikzpicture}} 
	 ~~ + ~~ h.c.	\label{eq:Virtual7}
	   \\ & = -\frac{4}{3}\frac{\epsilon_{\psi}}{\epsilon}\mathcal{F}_{h\psi }
   \mathcal{F}_{Z'Zh} \bigg\{\left(-\epsilon_{h}^2+\epsilon
   ^2+1\right) \left[\left(\epsilon_{h}^2-\epsilon^2-1\right) \epsilon_{\psi
   }^2+2\epsilon ^2\right] \left(\mathcal{F}_{ZL}\mathcal{F}_{Z'L}+
   \mathcal{F}_{ZR}\mathcal{F}_{Z'R}\right)+
   \nn \\ & ~~~~~ 
   \left[-2\epsilon ^2
   \left(\left(\epsilon _{h}^2-5\right) \epsilon _{\psi }^2+\epsilon
   _{h}^2\right)+\left(\epsilon_{h}^2-1\right)^2 \epsilon _{\psi
   }^2+\epsilon^4 \epsilon _{\psi }^2\right] \left(\mathcal{F}_{ZR}
   \mathcal{F}_{Z'L}+\mathcal{F}_{ZL}\mathcal{F}_{Z'R}\right)
\bigg\} \nn
    \\ & 
    \simeq 
    \frac{2}{3} f_{\phi} g^{\prime 2} \epsilon ^2 h_{\mathit{f}}^2
  \left[-\frac{h_{\mathit{f}}^2 f_{\phi }}{g^2 y_{\phi }^2}-\frac{4 \left(f_L
   y_L+f_R y_R\right)}{y_{\phi }}\right]\,, \nn
\end{align} 
where once again we remind that in the virtual $Z$ line we also sum over the Goldstone $\phi$, and 
make use of the Ward identities in eq.\,(\ref{fqe}) to rewrite its couplings. 
We find the identity 
\begin{align}
 V_{\psi\bar{\psi}}^{(\psi)} =  
 V_{Zh}^{(\psi)} & = 
 2\times 
 \raisebox{-12mm}{
	\begin{tikzpicture}
	\draw[thick,style={decorate, decoration={snake}}] (-0.75,0)--(0,0);
	\draw[->-=.6,>=Latex][thick] (0,0)--(1.25/2,1.25/2); 
	\draw[thick,style={decorate, decoration={snake}}] (1.25/2,1.25/2)--(1.25,1.25); 
        \draw[thick,dashed] (1.25,-1.25)--(0,0);
	\draw[->-=.8,>=Latex][thick] (1.25/2,-1.25/2)--(0,0);
	\draw[->-=.6,>=Latex][thick] (1.25/2,1.25/2)--(1.25/2,-1.25/2);
	\draw[thick,dash dot,persianblue][thick] (1.25,1.25)--(1.25,-1.25);
	\draw[thick,style={decorate, decoration={snake}}] (1.25,1.25)--(2.5,0); 
	\draw[thick,dashed] (2.5,0)--(1.25,-1.25); 	
	\draw[thick,dash dot,persianblue][thick] (1.25,1.25)--(1.25,-1.25); 
	\draw[thick,style={decorate, decoration={snake}}] (2.5,0)--(2.5+0.75,0);
	\draw[black,fill=tangerineyellow,thick] (0,0)circle(2.75pt);
	\draw[black,fill=tangerineyellow,thick] (1.25/2,1.25/2)circle(2.75pt);
	\draw[black,fill=tangerineyellow,thick] (1.25/2,-1.25/2)circle(2.75pt);
	\draw[black,fill=tangerineyellow,thick] (2.5,0)circle(2.75pt); 
	\node at (-0.2,-0.4) {\scalebox{0.85}{$\mathcal{F}_{Z'i}$}};
	\node at (2.7,-0.4) {\scalebox{0.85}{$\mathcal{F}_{Z'Zh}$}};
	\node at (0.35,0.93) {\scalebox{0.85}{$\mathcal{F}_{Z i}$}};
	\node at (0.35,-0.85) {\scalebox{0.85}{$\mathcal{F}_{h\psi}$}};
	\node at (1,-0.7) {\scalebox{0.85}{$h$}};
	\node at (1,+0.62) {\scalebox{0.85}{$Z$}};	
	\end{tikzpicture}} ~~~+~~~ h.c.
\end{align} 
where, in the language of fig.\,\ref{ASJ}, $V_{Zh}^{(\psi)}$ represents the bubble $A_1(d)$.

As in the other cases, the validity of the IR unitarity theorem for the $A_1$ bubble can be checked explicitly by adding diagrams that 
account for three-body decays.   
We have
\begin{align}
R_{Z\psi}^{(\psi)} & = \raisebox{-12mm}{\begin{tikzpicture}
	\draw[thick,style={decorate, decoration={snake}}] (-0.75,0)--(0,0);
	\draw[->-=.7,>=Latex][thick] (0,0)--(1.25/2,1.25/2); 
	\draw[thick,style={decorate, decoration={snake}}] (1.25/2,1.25/2)--(1.25,1.25); 
	\draw[->-=.6,>=Latex][thick] (1.25,-1.25)--(0,0);
	\draw[thick,style={decorate, decoration={snake}}] (1.25,1.25)--(2.5,0);
	\draw[->-=.7,>=Latex][thick] (1.25+1.25/2,-1.25/2)--(1.25,-1.25);
	\draw[thick,dashed] (2.5,0)--(1.25+1.25/2,-1.25/2);		
	\draw[thick,style={decorate, decoration={snake}}] (2.5,0)--(2.5+0.75,0);
	\draw[->-=.7,>=Latex][thick] (1.25/2,1.25/2)--(1.25+1.25/2,-1.25/2);
	\draw[thick,dash dot,persianblue][thick] (1.25,1.25)--(1.25,-1.25);
	\draw[black,fill=tangerineyellow,thick] (0,0)circle(2.75pt);
	\draw[black,fill=tangerineyellow,thick] (1.25+1.25/2,-1.25/2)circle(2.75pt); 
	\draw[black,fill=tangerineyellow,thick] (1.25/2,1.25/2)circle(2.75pt); 
	\draw[black,fill=tangerineyellow,thick] (2.5,0)circle(2.75pt);  
	\node at (-0.2,-0.4) {\scalebox{0.85}{$\mathcal{F}_{Z'i}$}};
	\node at (2.7,-0.4) {\scalebox{0.85}{$\mathcal{F}_{Z'Zh}$}};
	\node at (0.35,+0.95) {\scalebox{0.85}{$\mathcal{F}_{Zi}$}};
	\node at (2.1,-0.95) {\scalebox{0.85}{$\mathcal{F}_{h\psi}$}};
	\node at (1,+0.62) {\scalebox{0.85}{$Z$}};
	\node at (2.0,-0.2) {\scalebox{0.85}{$h$}};
	\end{tikzpicture}}
~~~+~~~
	\raisebox{-12mm}{
	\begin{tikzpicture}
	\draw[thick,style={decorate, decoration={snake}}] (-0.75,0)--(0,0);
	\draw[->-=.7,>=Latex][thick] (1.25/2,-1.25/2)--(0,0); 
	\draw[thick,style={decorate, decoration={snake}}] (1.25/2,-1.25/2)--(1.25,-1.25); 
	\draw[->-=.6,>=Latex][thick] (0,0)--(1.25,+1.25);
	\draw[thick,style={decorate, decoration={snake}}] (1.25,-1.25)--(2.5,0);
	\draw[->-=.7,>=Latex][thick] (1.25,+1.25)--(1.25+1.25/2,+1.25/2);
	\draw[thick,dashed] (2.5,0)--(1.25+1.25/2,+1.25/2);		
	\draw[thick,style={decorate, decoration={snake}}] (2.5,0)--(2.5+0.75,0);
	\draw[->-=.7,>=Latex][thick] (1.25+1.25/2,+1.25/2)--(1.25/2,-1.25/2);
	\draw[thick,dash dot,persianblue][thick] (1.25,1.25)--(1.25,-1.25);
	\draw[black,fill=tangerineyellow,thick] (0,0)circle(2.75pt);
	\draw[black,fill=tangerineyellow,thick] (1.25+1.25/2,+1.25/2)circle(2.75pt); 
	\draw[black,fill=tangerineyellow,thick] (1.25/2,-1.25/2)circle(2.75pt); 
	\draw[black,fill=tangerineyellow,thick] (2.5,0)circle(2.75pt);  
	\node at (-0.2,-0.4) {\scalebox{0.85}{$\mathcal{F}_{Z'i}$}};
	\node at (2.7,-0.4) {\scalebox{0.85}{$\mathcal{F}_{Z'Zh}$}};
	\node at (0.35,-0.95) {\scalebox{0.85}{$\mathcal{F}_{Zi}$}};
	\node at (2.1,+0.95) {\scalebox{0.85}{$\mathcal{F}_{h\psi}$}};
	\node at (1,-0.62) {\scalebox{0.85}{$Z$}};
	\node at (2.0,0) {\scalebox{0.85}{$h$}};
	\end{tikzpicture}} ~~~+~~~ h.c. \label{eq:HT1} \\
R_{\psi h}^{(\psi)} & =	\raisebox{-12mm}{
	\begin{tikzpicture}
	\draw[thick,style={decorate, decoration={snake}}] (-0.75,0)--(0,0);
	\draw[->-=.7,>=Latex][thick] (0,0)--(1.25/2,1.25/2); 
	\draw[thick,dashed] (1.25/2,1.25/2)--(1.25,1.25); 
	\draw[->-=.6,>=Latex][thick] (1.25,-1.25)--(0,0);
	\draw[thick,dashed] (1.25,1.25)--(2.5,0);
	\draw[->-=.7,>=Latex][thick] (1.25+1.25/2,-1.25/2)--(1.25,-1.25);
	\draw[thick,style={decorate, decoration={snake}}] (2.5,0)--(1.25+1.25/2,-1.25/2);		
	\draw[thick,style={decorate, decoration={snake}}] (2.5,0)--(2.5+0.75,0);
	\draw[->-=.7,>=Latex][thick] (1.25/2,1.25/2)--(1.25+1.25/2,-1.25/2);
	\draw[thick,dash dot,persianblue][thick] (1.25,1.25)--(1.25,-1.25);
	\draw[black,fill=tangerineyellow,thick] (0,0)circle(2.75pt);
	\draw[black,fill=tangerineyellow,thick] (1.25+1.25/2,-1.25/2)circle(2.75pt); 
	\draw[black,fill=tangerineyellow,thick] (1.25/2,1.25/2)circle(2.75pt); 
	\draw[black,fill=tangerineyellow,thick] (2.5,0)circle(2.75pt);  
	\node at (-0.2,-0.4) {\scalebox{0.85}{$\mathcal{F}_{Z'i}$}};
	\node at (2.7,-0.4) {\scalebox{0.85}{$\mathcal{F}_{Z'Zh}$}};
	\node at (0.35,+0.95) {\scalebox{0.85}{$\mathcal{F}_{h\psi}$}};
	\node at (2.1,-0.95) {\scalebox{0.85}{$\mathcal{F}^*_{Zi}$}};
	\node at (1,+0.62) {\scalebox{0.85}{$h$}};
	\node at (2.0,-0.2) {\scalebox{0.85}{$Z$}};
	\end{tikzpicture}}
	~~~+~~~
	\raisebox{-12mm}{
	\begin{tikzpicture}
	\draw[thick,style={decorate, decoration={snake}}] (-0.75,0)--(0,0);
	\draw[->-=.7,>=Latex][thick] (1.25/2,-1.25/2)--(0,0); 
	\draw[thick,dashed] (1.25/2,-1.25/2)--(1.25,-1.25); 
	\draw[->-=.6,>=Latex][thick] (0,0)--(1.25,+1.25);
	\draw[thick,dashed] (1.25,-1.25)--(2.5,0);
	\draw[->-=.7,>=Latex][thick] (1.25,+1.25)--(1.25+1.25/2,+1.25/2);
	\draw[thick,style={decorate, decoration={snake}}] (2.5,0)--(1.25+1.25/2,+1.25/2);		
	\draw[thick,style={decorate, decoration={snake}}] (2.5,0)--(2.5+0.75,0);
	\draw[->-=.7,>=Latex][thick] (1.25+1.25/2,+1.25/2)--(1.25/2,-1.25/2);
	\draw[thick,dash dot,persianblue][thick] (1.25,1.25)--(1.25,-1.25);
	\draw[black,fill=tangerineyellow,thick] (0,0)circle(2.75pt);
	\draw[black,fill=tangerineyellow,thick] (1.25+1.25/2,+1.25/2)circle(2.75pt); 
	\draw[black,fill=tangerineyellow,thick] (1.25/2,-1.25/2)circle(2.75pt); 
	\draw[black,fill=tangerineyellow,thick] (2.5,0)circle(2.75pt);  
	\node at (-0.2,-0.4) {\scalebox{0.85}{$\mathcal{F}_{Z'i}$}};
	\node at (2.7,-0.4) {\scalebox{0.85}{$\mathcal{F}_{Z'Zh}$}};
	\node at (0.35,-0.95) {\scalebox{0.85}{$\mathcal{F}_{h\psi}$}};
	\node at (2.1,+0.95) {\scalebox{0.85}{$\mathcal{F}^*_{Zi}$}};
	\node at (1,-0.62) {\scalebox{0.85}{$h$}};
	\node at (2.0,0) {\scalebox{0.85}{$Z$}};
	\end{tikzpicture}}~~~+~~~ h.c. \label{eq:HT2} 	
\end{align}
Notice that, both in eq.\,(\ref{eq:HT1}) and eq.\,(\ref{eq:HT2}),  the first diagram 
accounts for the emission of a soft fermion while the second diagram accounts for the emission of a soft antifermion. 
In eq.\,(\ref{eq:HT2}), the sum over the virtual Goldstone $\phi$ is understood.
Eq.\,(\ref{eq:HT1}) corresponds to the bubble $A_1(c)$ in the language of fig.\,\ref{ASJ} while 
eq.\,(\ref{eq:HT2}) to $A_1(b)$. 
We find
\begin{align}
R_{Z\psi}^{(\psi)} + R_{\psi h}^{(\psi)} = -  V_{\psi\bar{\psi}}^{(\psi)} =  
- V_{Zh}^{(\psi)}\,,
\end{align}
from which the IR unitarity theorem follows.
 
\subsection{$A_2$ diagram }

The virtual diagrams are
\begin{align}
V_{Zh}^{(h)} & =     
    \raisebox{-12mm}{
	\begin{tikzpicture}
	\draw[thick,style={decorate, decoration={snake}}] (-0.75,0)--(0,0);
	\draw[thick,style={decorate, decoration={snake}}] (0,0)--(1.25/2,1.25/2); 
	\draw[thick,dashed] (1.25/2,-1.25/2)--(0,0);
	\draw[thick,dashed] (1.25/2,1.25/2)--(1.25/2,-1.25/2);
	\draw[thick,style={decorate, decoration={snake}}] (1.25/2,1.25/2)--(1.25,1.25); 
    \draw[thick,dashed] (1.25,-1.25)--(0,0);
	\draw[thick,dash dot,persianblue][thick] (1.25,1.25)--(1.25,-1.25);
	\draw[thick,style={decorate, decoration={snake}}] (1.25,1.25)--(2.5,0); 
	\draw[thick,dashed] (2.5,0)--(1.25,-1.25); 	
	\draw[thick,dash dot,persianblue][thick] (1.25,1.25)--(1.25,-1.25); 
	\draw[thick,style={decorate, decoration={snake}}] (2.5,0)--(2.5+0.75,0);
	\draw[black,fill=tangerineyellow,thick] (0,0)circle(2.75pt);
	\draw[black,fill=tangerineyellow,thick] (1.25/2,1.25/2)circle(2.75pt);
	\draw[black,fill=tangerineyellow,thick] (1.25/2,-1.25/2)circle(2.75pt);
	\draw[black,fill=tangerineyellow,thick] (2.5,0)circle(2.75pt); 
	\node at (0.85,0) {\scalebox{0.85}{$h$}};
	\node at (0.15,0.5) {\scalebox{0.85}{$Z$}};
	\node at (0.43,-0.2) {\scalebox{0.85}{$h$}};
	\node at (-0.2,-0.4) {\scalebox{0.85}{$\mathcal{F}_{Z'Zh}$}};
	\node at (2.7,-0.4) {\scalebox{0.85}{$\mathcal{F}_{Z'Zh}$}};
	\node at (0.35,0.95) {\scalebox{0.85}{$\mathcal{F}_{ZZh}$}};
	\node at (0.35,-0.85) {\scalebox{0.85}{$\mathcal{F}_{hhh}$}};
	\node at (1,-0.7) {\scalebox{0.85}{$h$}};
	\node at (1,+0.62) {\scalebox{0.85}{$Z$}};	
	\end{tikzpicture}} ~~ + ~~ h.c. \label{eq:VirSqZh3}\\ \nn
	&  = -\frac{\epsilon _{h}^2}{6}   \mathcal{F}_{ZZh}  \mathcal{F}_{hhh} 
   \mathcal{F}_{Z^{\prime}Z h}^2  \left[\epsilon ^2 \left(10-\frac{3  \epsilon_{h}^2}{2}\right)+\frac{\epsilon_{h}^6+\epsilon _{h}^2}{2 
   \epsilon ^2}-5  \epsilon _{h}^2+\epsilon ^4+1\right]
    \simeq -8 g^{\prime 2} \lambda  \epsilon ^2 f_{\phi }^2 \left(\frac{\lambda }{g^2 y_{\phi
   }^2}+1\right)\,.\nn
   \end{align}
The three-body decay is described by 
\begin{align}\label{eq:ReZh2}
R_{Zh}^{(h)} =  
    \raisebox{-12mm}{
	\begin{tikzpicture}
	\draw[thick,style={decorate, decoration={snake}}] (-0.75,0)--(0,0);
	\draw[thick,style={decorate, decoration={snake}}] (0,0)--(1.25/2,1.25/2); 
	\draw[thick,style={decorate, decoration={snake}}] (1.25/2,1.25/2)--(1.25,1.25); 
	\draw[thick,dashed] (1.25,-1.25)--(0,0);
	\draw[thick,style={decorate, decoration={snake}}] (1.25,1.25)--(2.5,0);
	\draw[thick,dashed] (1.25+1.25/2,-1.25/2)--(1.25,-1.25);
	\draw[thick,dashed] (2.5,0)--(1.25+1.25/2,-1.25/2);
	\draw[thick,style={decorate, decoration={snake}}] (2.5,0)--(2.5+0.75,0);
	\draw[thick,dashed] (1.25/2,1.25/2)--(1.25+1.25/2,-1.25/2);
	\draw[thick,dash dot,persianblue][thick] (1.25,1.25)--(1.25,-1.25);
	\draw[black,fill=tangerineyellow,thick] (0,0)circle(2.75pt);
	\draw[black,fill=tangerineyellow,thick] (1.25+1.25/2,-1.25/2)circle(2.75pt); 
	\draw[black,fill=tangerineyellow,thick] (1.25/2,1.25/2)circle(2.75pt);   
	\draw[black,fill=tangerineyellow,thick] (2.5,0)circle(2.75pt);  
	\node at (0.85,-0.13) {\scalebox{0.85}{$h$}};
	\node at (0.12,0.5) {\scalebox{0.85}{$Z$}};
	\node at (-0.2,-0.4) {\scalebox{0.85}{$\mathcal{F}_{Z'Zh}$}};
	\node at (2.7,-0.4) {\scalebox{0.85}{$\mathcal{F}_{Z'Zh}$}};
	\node at (0.35,+0.95) {\scalebox{0.85}{$\mathcal{F}_{ZZh}$}};
	\node at (2.1,-0.95) {\scalebox{0.85}{$\mathcal{F}_{hhh}$}};
	\end{tikzpicture}} ~~ + ~~ h.c.
\end{align}   
and we find $R_{Zh}^{(h)} = - V_{Zh}^{(h)}$.




\begin{thebibliography}{99}




\bibitem{Agarwal:2021ais}
N.~Agarwal, L.~Magnea, C.~Signorile-Signorile and A.~Tripathi,
``The Infrared Structure of Perturbative Gauge Theories,''
[arXiv:2112.07099 [hep-ph]].
  \bibitem{Bloch:1937pw}
F.~Bloch and A.~Nordsieck,
``{\it Note on the Radiation Field of the electron},''
Phys. Rev. \textbf{52} (1937), 54-59
\bibitem{cc1}
P.~Ciafaloni and D.~Comelli,
``{\it Sudakov enhancement of electroweak corrections,}''
Phys. Lett. B \textbf{446} (1999), 278-284
[\hhref{hep-ph/9809321}]
;\\
M.~Beccaria, P.~Ciafaloni, D.~Comelli, F.~M.~Renard and C.~Verzegnassi,
``{\it Logarithmic expansion of electroweak corrections to four-fermion processes in the TeV region,}''
Phys. Rev. D \textbf{61} (2000), 073005
[\hhref{hep-ph/9906319}].
\\
P.~Ciafaloni and D.~Comelli,
``Electroweak Sudakov form-factors and nonfactorizable soft QED effects at NLC energies,''
Phys. Lett. B \textbf{476} (2000), 49-57
[arXiv:hep-ph/9910278 [hep-ph]].
\\
V.~S.~Fadin, L.~N.~Lipatov, A.~D.~Martin and M.~Melles,
``Resummation of double logarithms in electroweak high-energy processes,''
Phys. Rev. D \textbf{61} (2000), 094002
[arXiv:hep-ph/9910338 [hep-ph]].
\bibitem{BNV}
M.~Ciafaloni, P.~Ciafaloni and D.~Comelli,
``{\it Bloch-Nordsieck violating electroweak corrections to inclusive TeV scale hard processes,}''
Phys. Rev. Lett. \textbf{84} (2000), 4810-4813
[\hhref{hep-ph/0001142}];\\
M.~Ciafaloni, P.~Ciafaloni and D.~Comelli,
``{\it Electroweak Bloch-Nordsieck violation at the TeV scale: 'Strong' weak interactions?,}''
Nucl. Phys. B \textbf{589} (2000), 359-380
[\hhref{hep-ph/0004071}];\\
M.~Ciafaloni, P.~Ciafaloni and D.~Comelli,
``{\it Electroweak double logarithms in inclusive observables for a generic initial state,}''
Phys. Lett. B \textbf{501} (2001), 216-222
[\hhref{hep-ph/0007096}];\\
M.~Ciafaloni, P.~Ciafaloni and D.~Comelli,
``{\it Bloch-Nordsieck violation in spontaneously broken Abelian theories,}''
Phys. Rev. Lett. \textbf{87} (2001), 211802
[\hhref{hep-ph/0103315}].
\\
M.~Ciafaloni, P.~Ciafaloni and D.~Comelli,
``Enhanced electroweak corrections to inclusive boson fusion processes at the TeV scale,''
Nucl. Phys. B \textbf{613} (2001), 382-406
[arXiv:hep-ph/0103316 [hep-ph]].
\bibitem{MS}
A.~A.~Penin,
``{\it High-Energy Limit of Quantum Electrodynamics beyond Sudakov Approximation,}''
Phys. Lett. B \textbf{745} (2015), 69-72
[erratum: Phys. Lett. B \textbf{751} (2015), 596-596; erratum: Phys. Lett. B \textbf{771} (2017), 633-634]
[\hhref{1412.0671}];\\
T.~Liu and A.~Penin,
``{\it High-Energy Limit of Mass-Suppressed Amplitudes in Gauge Theories,}''
JHEP \textbf{11} (2018), 158
[\hhref{1809.04950}]; \\
  A.~A.~Penin,
  ``{\it Two-loop corrections to Bhabha scattering},''
  Phys.\ Rev.\ Lett.\  {\bf 95} (2005) 010408
  [\hhref{hep-ph/0501120}];\\
M.~Beneke,
``{\it Renormalons,}''
Phys. Rept. \textbf{317} (1999), 1-142
[\hhref{hep-ph/9807443}].

 \bibitem{Ak}
 M.~Beneke, V.~M.~Braun and V.~I.~Zakharov,
``Bloch-Nordsieck cancellations beyond logarithms in heavy particle decays,''
Phys. Rev. Lett. \textbf{73} (1994), 3058-3061
[arXiv:hep-ph/9405304 [hep-ph]].
\\
R.~Akhoury, L.~Stodolsky and V.~I.~Zakharov,
``{\it Power corrections and KLN cancellations,}''
Nucl. Phys. B \textbf{516} (1998), 317-332
[\hhref{hep-ph/9609368}];\\
A.~Sinkovics, R.~Akhoury and V.~I.~Zakharov,
``{\it Cancellation of 1 / m(Q) corrections to the inclusive decay width of a heavy quark,}''
Phys. Rev. D \textbf{58} (1998), 114025
[\hhref{hep-ph/9804401}].
\bibitem{Ciafaloni:2009tf}
M.~Ciafaloni, P.~Ciafaloni and D.~Comelli,
``Anomalous Sudakov Form Factors,''
JHEP \textbf{03} (2010), 072
[arXiv:0909.1657 [hep-ph]].
\bibitem{Passarino:1978jh}
  G.~Passarino and M.~J.~G.~Veltman,
  ``{\it One Loop Corrections for $e^+ e^-$ Annihilation Into $\mu^+ \mu^-$ in the Weinberg Model,}''
  Nucl.\ Phys.\ B {\bf 160} (1979) 151.
  
 \bibitem{Frenkel:1976bj}
J.~Frenkel and J.~C.~Taylor,
``{\it Exponentiation of Leading Infrared Divergences in Massless Yang-Mills Theories,}''
Nucl. Phys. B \textbf{116} (1976), 185-194
\\
D.~Amati, R.~Petronzio and G.~Veneziano,
``Relating Hard QCD Processes Through Universality of Mass Singularities,''
Nucl. Phys. B \textbf{140} (1978), 54-72.
\bibitem{g/h}
M.~Ciafaloni, P.~Ciafaloni and D.~Comelli,
``{\it Enhanced Electroweak Corrections to Inclusive Boson Fusion Processes at the TeV Scale,}''
Nucl.Phys. B\textbf{613} (2001), 382 
[\hhref{hep-ph/0103315}].





 \bibitem{Kinoshita:1962ur}
T.~Kinoshita,
 ``{\it Mass singularities of Feynman amplitudes},''
J. Math. Phys. \textbf{3} (1962), 650-677
\bibitem{Lee:1964is}
T.~D.~Lee and M.~Nauenberg,
 ``{\it Degenerate Systems and Mass Singularities},''
Phys. Rev. \textbf{133} (1964), B1549-B1562

\bibitem{Frye:2018xjj}
C.~Frye, H.~Hannesdottir, N.~Paul, M.~D.~Schwartz and K.~Yan,
``{\it Infrared Finiteness and Forward Scattering},''
Phys. Rev. D \textbf{99} (2019) no.5, 056015
[\hhref{1810.10022}].
 
  

\bibitem{equivalence}
B.~W.~Lee, C.~Quigg and H.~B.~Thacker,
Phys. Rev. D \textbf{16} (1977), 1519;
\\
M.~S.~Chanowitz and M.~K.~Gaillard,
Nucl. Phys. B \textbf{261} (1985), 379-431;
\\
J.~M.~Cornwall, D.~N.~Levin and G.~Tiktopoulos,
Phys. Rev. D \textbf{10} (1974), 1145
[erratum: Phys. Rev. D \textbf{11} (1975), 972].





\bibitem{Muta:1987mz}
T.~Muta,
``{\it Foundations of quantum chromodynamics: An Introduction to perturbative methods in gauge theories},''
World Sci. Lect. Notes Phys. \textbf{5} (1987), 1-409
%

\bibitem{Contopanagos:1991yb}
H.~F.~Contopanagos and M.~B.~Einhorn,
``{\it Theory of the asymptotic S matrix for massless particles},''
Phys. Rev. D \textbf{45} (1992), 1291-1321.
\\
J.~C.~Taylor,
``{\it Scattering of very light charged particles},''
Phys. Rev. D \textbf{54} (1996), 2975-2977
[\hhref{hep-ph/9306234}].
\\
M.~Lavelle and D.~McMullan,
``{\it Collinearity, convergence and cancelling infrared divergences},''
JHEP \textbf{03} (2006), 026
[\hhref{hep-ph/0511314}].
\\
M.~Ciafaloni,
``INFRARED SINGULARITIES AND COHERENT STATES IN GAUGE THEORIES,''
Adv. Ser. Direct. High Energy Phys. \textbf{5} (1989), 491-572.
\\
V.~Del Duca, L.~Magnea and G.~F.~Sterman,
``Collinear Infrared Factorization and Asymptotic Evolution,''
Nucl. Phys. B \textbf{324} (1989), 391-411
\\
H.~Hannesdottir and M.~D.~Schwartz,
``$S$ -Matrix for massless particles,''
Phys. Rev. D \textbf{101} (2020) no.10, 105001
[arXiv:1911.06821 [hep-th]].

 \bibitem{Doria:1980ak}
R.~Doria, J.~Frenkel and J.~C.~Taylor,
 ``Counter Example to Nonabelian Bloch-Nordsieck Theorem,''
Nucl. Phys. B \textbf{168} (1980), 93-110.
\\
C.~E.~Carneiro, M.~Day, J.~Frenkel, J.~C.~Taylor and M.~T.~Thomaz,
``{\it Leading non-cancelling infrared divergences in peerturbative QCD},''
Nucl. Phys. B \textbf{183} (1981), 445-470.
\\
C.~Di'Lieto, S.~Gendron, I.~G.~Halliday and C.~T.~Sachrajda,
 ``{\it A Counter Example to the Bloch-Nordsieck Theorem in Nonabelian Gauge Theories},''
Nucl. Phys. B \textbf{183} (1981), 223-250.
\\
  A.~Andrasi, M.~Day, R.~Doria, J.~Frenkel and J.~C.~Taylor,
  ``{\it Soft Divergences in Perturbative {QCD},}''
  Nucl.\ Phys.\ B {\bf 182} (1981) 104.
 \\
  N.~Yoshida,
  ``{\it Diagrammatical Display of the Counter Example to Nonabelian Bloch-nordsieck Conjecture,}''
  Prog.\ Theor.\ Phys.\  {\bf 66} (1981) 269.
\\
  S.~Catani, M.~Ciafaloni and G.~Marchesini,
  ``{\it Noncancelling Infrared Divergences In Qcd Coherent State,}''
  Nucl.\ Phys.\ B {\bf 264} (1986) 588.
\\
F.~Caola, K.~Melnikov, D.~Napoletano and L.~Tancredi,
``{\it On the non-cancellation of infrared singularities in collisions of massive quarks},''
[\hhref{2011.04701}].




\bibitem{Ito:1981}
  I.~Ito,
  ``{\it Cancellation of Infrared Divergence and Initial Degenerate State in {QCD}},''
  Prog.\ Theor.\ Phys.\  {\bf 65} (1981) 1466.
 \\
  T.~Muta and C.~A.~Nelson,
  ``{\it Role of Quark - Gluon Degenerate States in Perturbative {QCD}},''
  Phys.\ Rev.\ D {\bf 25} (1982) 2222.

\bibitem{Ciafaloni:2010ti}
  P.~Ciafaloni, D.~Comelli, A.~Riotto, F.~Sala, A.~Strumia and A.~Urbano,
  ``{\it Weak Corrections are Relevant for Dark Matter Indirect Detection},''
  JCAP {\bf 1103} (2011) 019
  [\hhref{1009.0224}].\\
  P.~Ciafaloni, M.~Cirelli, D.~Comelli, A.~De Simone, A.~Riotto and A.~Urbano,
  ``{\it On the Importance of Electroweak Corrections for Majorana Dark Matter Indirect Detection},''
  JCAP {\bf 1106} (2011) 018
  [\hhref{1104.2996}].



\bibitem{tHooft:1978jhc}
G.~'t Hooft and M.~J.~G.~Veltman,
Nucl. Phys. B \textbf{153} (1979), 365-401
\bibitem{Ellis:2007qk}
R.~K.~Ellis and G.~Zanderighi,
``{\it Scalar one-loop integrals for QCD,}''
JHEP \textbf{02} (2008), 002
[\hhref{arXiv:0712.1851} [hep-ph]].
\bibitem{Roth:1996pd}
M.~Roth and A.~Denner,
``{\it High-energy approximation of one loop Feynman integrals,}''
Nucl. Phys. B \textbf{479} (1996), 495-514
[\hhref{hep-ph/9605420}].
\bibitem{Scharf:1993ds}
R.~Scharf and J.~B.~Tausk,
``{\it Scalar two loop integrals for gauge boson selfenergy diagrams with a massless fermion loop,}''
Nucl. Phys. B \textbf{412} (1994), 523-552
[\hhref{hep-ph/9512336}].
  
 


 




\end{thebibliography}
\end{document}